\documentclass[10pt,twocolumn,letterpaper]{article}

\usepackage{cvpr}
\usepackage{times}
\usepackage{epsfig}
\usepackage{graphicx}
\usepackage{amsmath}
\usepackage{amssymb}

\cvprfinalcopy 

\ifcvprfinal\fi

\begin{document}

\title{Motion Deblurring using Spatiotemporal Phase Aperture Coding}

\author{Shay Elmalem, Raja Giryes and Emanuel Marom\\
Tel Aviv University\\
}

\maketitle

\begin{abstract}
   
Motion blur is a known issue in photography, as it limits the exposure time while capturing moving objects. Extensive research has been carried to compensate for it. In this work, a computational imaging approach for motion deblurring is proposed and demonstrated. Using dynamic phase-coding in the lens aperture during the image acquisition, the trajectory of the motion is encoded in an intermediate optical image. This encoding embeds both the motion direction and extent by coloring the spatial blur of each object. The color cues serve as prior information for a blind deblurring process, implemented using a convolutional neural network (CNN) trained to utilize such coding for image restoration. We demonstrate the advantage of the proposed approach over blind-deblurring with no coding and other solutions that use coded acquisition, both in simulation and real-world experiments.

\end{abstract}

\section{Introduction}

Finding the proper exposure setting is a well-known challenge in photography. In general, one has to balance between aperture size, exposure time and the gain to achieve a good image (the trade-off between these factors is sometimes referred to as the 'exposure triangle'). This balancing process involves many trade-offs, and therefore requires complex skills and rich experience. In many cases, a large exposure is necessary to allow a sufficient amount of light to reach the sensor in order to achieve a good image with respect to the lighting condition, which usually is not controllable. To increase the amount of light in the sensor plane, one may increase the aperture size. However, large aperture results in a shallow depth-of-field and increased sensitivity to optical aberrations. Increasing the sensor gain can intensify the image signal, with the price of a higher noise level. Increasing the exposure time allows more light to be integrated into the image sensor, but introduces motion blur. 

\begin{figure}[ht]
\begin{center}
\includegraphics[width=1\linewidth]{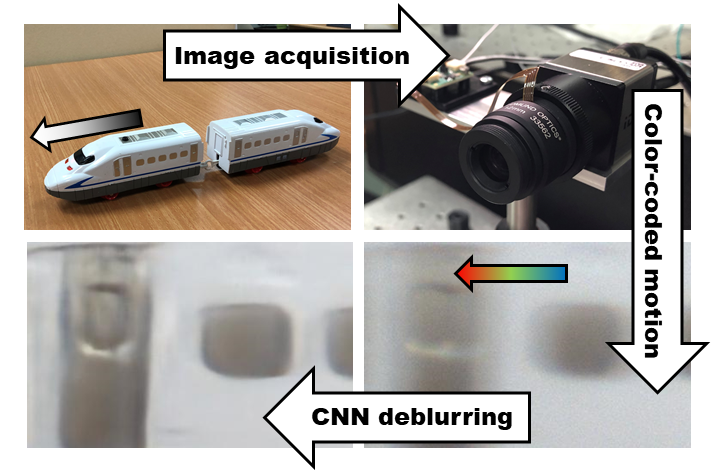}
\end{center}
   \caption{\textbf{Motion deblurring using spatiotemporal phase aperture coding:} A moving scene is captured using a camera with spatiotemporal phase aperture coding, which generates a motion-color coded PSF. The PSF coding serves as a prior for the CNN that performs blind spatially varying motion deblurring.}
\label{fig:teaser}
\end{figure}

Various efforts have been dedicated to balance automatically the exposure parameters \cite{photographyTextbook}. Yet, such solutions are either very specific to the scenario or provide medium performance. A different approach tries to eliminate one (or more) of the exposure triangle vertices, by developing methods that can restore the artifacts introduced by a non-balanced exposure. For example, one may apply a high gain and then perform a denoising operation \cite{Lefkimmiatis_denoising_2017_CVPR,Schwartz2018_DeepISP_TL,seeInDark,liba2019handheld}; increase the aperture size and restore the blurred image using out-of-focus deblurring algorithms \cite{Zhang_2017_CVPR}; or take long exposures and revert the motion-related blur \cite{motBlurComparison}, which is the focus of this work. 

In addition to the pure post-processing methods, solutions based on computational imaging \cite{compImg_rev} attempt to globally analyze the scenario, and then re-design the whole imaging system. In such an approach, the image acquisition is manipulated in a way that (generally) leads to an intermediate image with low-quality. However, this image is distorted in a very specific way such that it encodes information acquired during the exposure. Such information encoding is designed so that it can be employed in the post-processing stage for the final image restoration. Such methods have been demonstrated for various applications, including extended depth-of-field \cite{Dowski:95,Levin_coded_aper,Zalevsky:06,EDOF_spr,EDOF_DL}, hyper/multi-spectral imaging \cite{CASSI,Jonathan_hyper}, depth estimation \cite{Levin_coded_aper,Nayr_codedAper_dpt,IEEE_depth} and motion deblurring \cite{Raskar_Flut_1,Agrawal2009_Flutt_2,Levin_motionInvariant,Levin_orthParab,nayar_hybrid,lightFld_deb}, to name a few. 

Computational imaging methods for motion deblurring had been presented before, either using a temporal shutter coding \cite{Raskar_Flut_1} or a parabolic motion of the camera \cite{Levin_motionInvariant}. In both methods, some assumption/prior knowledge on the motion direction is needed, which limits their performance, as discussed in Section~\ref{prevWork}. 

{\bf{Contribution:}} In this work, a computational imaging approach for motion deblurring is proposed and analyzed. The innovation in the proposed approach is the encoding method that embeds dynamic cues for the motion trajectory and extent in the intermediate image (see Fig.~\ref{fig:teaser}), which serve as a strong prior for both shift-variant Point-Spread Function (PSF) estimation and deblurring operation. 

\begin{figure}
    \def\psfWd{0.283}
	\centering
	\begin{tabular}{c c c}
		\includegraphics[width = \psfWd\columnwidth]{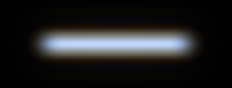}&
		\includegraphics[width = \psfWd\columnwidth]{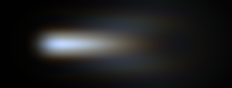}&
		\includegraphics[width = \psfWd\columnwidth]{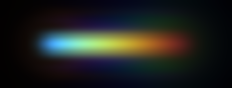}\\

	\end{tabular}
	\caption{\textbf{Motion blurred PSF simulation:} (left) conventional camera, (middle) gradual focus variation in conventional camera and (right) the proposed camera- gradual focus variation with phase aperture coding.}
	\label{fig:PSFs}
\end{figure}

Our encoding is achieved by performing spatiotemporal phase-coding in the lens aperture plane during the image acquisition. The PSF of the coded system induces a specific chromatic-temporal coupling, which (unlike in a conventional camera) results in a color-varying spatial blur (see Fig.~\ref{fig:PSFs}). Such a PSF encodes the different motion trajectory of each object. A convolutional neural network (CNN) is trained to analyze the embedded cues and use them to reconstruct a deblurred image. 

The encoding design is performed for a general case, whereby objects move in different directions/velocities. Such encoding allows blind deblurring of the intermediate image, since the required prior information for both PSF estimation and image restoration is embedded in it. The deblurring CNN is trained to estimate the spatially varying PSF using the encoded cues, and reconstruct a sharp image. An experiential setup of a camera performing the designed spatiotemporal blur is presented. We demonstrate its ability to perform motion deblurring both in the presence of uniform and non-uniform motion.


\section{Related work} \label{prevWork}

{\textbf{Blind Motion deblurring:}} Motion deblurring is a vastly studied challenge in image processing. Various image priors have been tested for motion deblurring, e.g. image statistics \cite{Levin_motBlur_NIPS} and sparsity \cite{Krishnan} to name a few. In recent years, deep models are used to implicitly learn the transformation from a blurred to a sharp image, such that the model encapsulates both the non-uniform PSF estimation and the image deblurring operation. This approach was demonstrated using various networks: recurrent and scale-recurrent networks \cite{DynamicSceneRNN_CVPR18,Tao_SRN_CVPR_18}, adversarial training \cite{Nah_deepDeblur_gopro,DeblurGAN_CVPR18} and frame burst to sharp image \cite{Wieschollek_deepDblur_multiframe}. For a recent review and comparison of existing methods, see \cite{motBlurComparison}.

{\textbf{Computational imaging-based motion deblurring:}} 
Various works proposed to manipulate the imaging process during exposure to allow better motion deblurring. 
Such approaches include the use of hybrid imaging \cite{nayar_hybrid}, light field camera \cite{lightFld_deb} or usage of the rolling shutter effects \cite{rollingShtr_deb}. In this work, our focus is on spatiotemporal schemes. We detail now two such prior frameworks.

Raskar \etal developed a temporal amplitude coding of the aperture to counteract motion blur \cite{Raskar_Flut_1}. The continuous exposure is analyzed as a wide temporal box filter with narrow frequency response, which limits the motion deblurring performance. Using this analysis, the authors propose a temporal amplitude coding of the aperture (referred to as 'fluttered shutter'), by interchangeably closing and opening it during exposure in some pre-determined timing (i.e. in some 'temporal code'). Such coding generates a much wider frequency response, which in turn is utilized for improved motion deblurring results. While achieving very good results, it requires prior knowledge of the motion direction and extent. In addition, it suffers from reduced light efficiency due to the (interchangeable) closing of the aperture during half of the exposure. A follow-up work analyzes the case of fluttered shutter PSF estimation as a part of the deblurring process, thus, avoiding the requirement for prior knowledge of the motion parameters \cite{Agrawal2009_Flutt_2}. However, such code design requires a compromise in the deblurring performance, to achieve both PSF invertibility and estimation abilities. Moreover, since it also relies on temporal amplitude coding, light efficiency is still decreased. A rigorous model analyzing the design and implementation of a fluttered shutter camera is presented in \cite{flt_prdx,flt_prdx_followup}. Jeon \etal extended the method to multi-image photography using complementary fluttering patterns \cite{flt_multiFrm}.

Another approach by Levin \etal searches for a sensor motion that leads to a motion invariant PSF \cite{Levin_motionInvariant}. The motivation is that with such a PSF, one may perform non-blind deconvolution using the known kernel on the entire image at once, without the requirement to estimate each object motion trajectory. After a rigorous analysis, it is shown that parabolic motion of the image sensor during exposure leads to the desired motion invariant PSF. Intuitively, one may think of this image acquisition technique as a process in which every moving object, at least for a fraction of the exposure, is in the same velocity of the sensor (assuming the velocity is inside a predefined range). Since each object is 'tracked' by the camera for one brief moment, and in the rest of the exposure it is moving relative to the camera, the blur of all objects turns out to be similar (for the full analysis see \cite{Levin_motionInvariant}). This allows applying a conventional deblurring approach that assumes a uniform blur. While this is a major advantage, this approach has one serious limitation: The PSF encoding is limited to the axis in which the parabolic motion took place. If an object moves in different directions, the motion invariant PSF assumption no longer holds, and the performance degrades. In the case of movement to an orthogonal direction, the deblurring ability is completely lost. To solve this issue, follow-up work \cite{Levin_orthParab} proposed an advanced solution based on two images taken with two orthogonal parabolic motions. Such a solution allows deblurring of motion in all directions. Yet, it requires a more complex setup and acquisition of two images. The rigorous model presented in \cite{flt_prdx} for the fluttered shutter camera can also be applied to the parabolic motion camera.

{\textbf{Spatiotemporal coding for other applications:}} Spatiotemporal coded cameras were investigated also for other tasks.

In \cite{Nayar_codedExp_1} the authors suggest a rolling shutter mechanism that can assist computational imaging applications such as optical flow estimation and high-speed photography. It has been shown that using such a modified rolling shutter, one may extract a video sequence just from a single coded exposure photograph \cite{Nayar_codedExposure_2,Nayar_codedExposure_3}. In \cite{P2C2}, the authors present an approach to convert low resolution and low frame-rate video sequences to higher resolution and higher frame-rate, using a dynamic mask designed as a spatiotemporal shutter. A similar problem is approached in \cite{flt2vid} using a fluttered shutter camera. Llull \etal \cite{Llull:13} use compressed sensing techniques to extract more than ten frames from a single snapshot, where a moving coded aperture mask is used to generate the required spatiotemporal coding.

\section{Spatiotemporal aperture coding}  \label{mask}
    In order to achieve blind deblurring of motion blurred images, the blur kernel has to be estimated and thereafter inverted (even if both of these operations are jointly performed). In a general scene, objects move in different directions and velocities, making the blur kernel shift-dependent. Therefore, linear shift invariant deconvolution operations cannot be used. Yet, one may encode cues in the acquired image to mitigate some of the hurdles. To this end, we aim at encoding in the intermediate image enough information that allows both estimating and inverting the spatially-varying PSF of the acquired image, such that improved motion deblurring of a general scene is achieved. 

{\bf The spatiotemporal PSF design.} The design-goal of such a PSF is to encode the object trajectory during the image acquisition. To achieve this task, the PSF has to vary along the trajectory in some way that provides cues for both the motion direction and extent. One may suggest spatial variations of the PSF along the motion trajectory (i.e. during the exposure time), however, such a variation introduces a spatial blur. Since trade-offing motion blur with spatial blur is not desired, the PSF variation has to take place in another dimension.

In our proposed design, the motion variations are projected onto the color space. If the PSF can change in color during the motion, a motion-variant encoding with cues to the motion direction and velocity can be achieved. Generally, color-coding requires color filtering, which results in loss of light and requires some mechanism for filter replacement (either mechanically or electronically); both of these issues are not desired. Therefore, in order to achieve motion-color coding, a phase-mask is used. In various works \cite{DowskiCath,cossairt2010diffusion,Zalevsky:06,EDOF_spr,EDOF_DL,IEEE_depth,nagahara2008flexible}, phase-masks are used for PSF engineering. The advantage in light throughput of phase over amplitude aperture coding is significant (in many amplitude-coding based systems the light throughput is reduced by $\sim$50\%; see for example \cite{Levin_coded_aper,mitra2017CodedAperture}). 

In several previous works \cite{Zalevsky:06,EDOF_spr,EDOF_DL,IEEE_depth}, phase-masks formed of several concentric rings are used for extended depth-of-field (EDOF) and depth estimation. Such masks act as a circularly-symmetric diffraction grating, and therefore introduce a predesigned and controlled axial chromatic aberration. This controlled aberration engineer the PSF to have a joint defocus-color dependency. As opposed to a conventional corrected lens (in which the response is designed to be the same for all colors), incorporating such a phase-mask in the lens aperture introduces a discrepancy in the lens response to the different colors. For example, a phase-mask can be designed to generate an in-focus PSF, which is narrow for the blue wavelength band, wider for green, and even wider for red. To generate such joint defocus-color dependency, the phase-mask is designed in a way that defocus-variations change the width of the PSFs at different colors, such that in another focus plane the narrow PSF color is either green or red, and the 'order' of the PSFs width is interchanging. 

This joint dependency can be used for both EDOF \cite{EDOF_spr,EDOF_DL} and depth estimation \cite{IEEE_depth}, by focusing the lens to a specific plane in a scene with objects located at various depths. In such a configuration, each object is blurred by a different blur kernel, according to its defocus condition. This color-depth encoding of the blur kernels allows high quality EDOF (which requires blind shift-variant convolution in the general case) and single image monocular depth estimation. 

We suggest using a similar phase-mask for the spatiotemporal encoding required for motion deblurring. We assume a scene with objects located relatively far from the lens (in relation to the focal length), in a way that all objects can be considered as located practically in infinity (such a setting, known as infinite-conjugate imaging, is commonly used in various applications, for example security cameras and smartphone cameras). By adding a mask containing the proper phase rings, the PSF is modulated to be 'colored' (by narrow PSF in a certain color band and the opposite in the other bands, and not by chromatic filtering). When the focus setting varies, the PSF also changes to a different color. Therefore, if the focus changes gradually during the exposure, the desired spatiotemporal dependency is achieved: The 'color' of the PSF (i.e. the ratio between the PSF width in the different color channels) varies during the exposure, and as every object moves, its motion is blurred differently (in the chromatic dimension) along the trajectory. 

In \cite{EDOF_DL,EDOF_spr}, the color differences serve as cues to estimate the correct PSF, and deblurring can be done using the sharp color channel (i.e. the color in which the PSF is narrow) that 'carries' the image information (as in most natural images objects always have some color content in all channels, and pure monochromatic objects are rare). Therefore, the color-dependent blur can be deblurred by transferring resolution from channel to channel. In our case of motion blur, the color cues are designed to indicate the motion trajectory for shift variant PSF estimation, and thereafter the PSF information is used for deblurring. Since strong defocus-color dependency is desired, a similar mask to the one presented in \cite{IEEE_depth} is used. 

As described above, a proper focus variation should be performed during the exposure to achieve the desired colored motion trajectory. Focus/defocus condition is quantified using the $\psi$ measure, defined as: 
\begin{equation}
		\psi = \frac{{\pi {R^2}}}{\lambda }\left( {\frac{1}{{{z_\mathrm{o}}}} + 
			\frac{1}{{{z_{\mathrm{img}}}}} - \frac{1}{f}} \right),
\end{equation}
where $R$ is the lens exit pupil radius, $\lambda$ is the illumination wavelength, $z_\mathrm{o}$ is the object distance, $z_\mathrm{img}$ is the sensor distance and $f$ is the lens' focal length. When the lens is focused properly, $\psi=0$. If a focus variation is introduced, then $\psi$ changes. By examining the phase-coded lens response, it seems that the defocus variation domain providing the strongest separation is between $0<\psi<8$ (calculated for blue wavelength), and therefore it is taken as the domain for the gradual focus variation during exposure.

{\bf The PSF encoding simulation.} The proposed encoding is illustrated in Fig.~\ref{fig:PSFs}. Fig.~\ref{fig:PSFs}(left) presents blur of a moving point source captured by a conventional camera. If gradual focus variation is performed to a clear aperture lens during exposure (Fig.~\ref{fig:PSFs}(middle)), the PSF gets wider in all the colors simultaneously, and thus introduces a considerable spatial blur in the last parts of the motion. However, if the same focus variation is performed to a lens equipped with a ring phase mask (Fig.~\ref{fig:PSFs}(right)), the PSF colors change along the motion line, from blue through green to red.

\begin{figure}
    \def\dotSz{0.40}
    
	\centering
	\begin{tabular}{c c}
		\includegraphics[width = \dotSz\columnwidth]{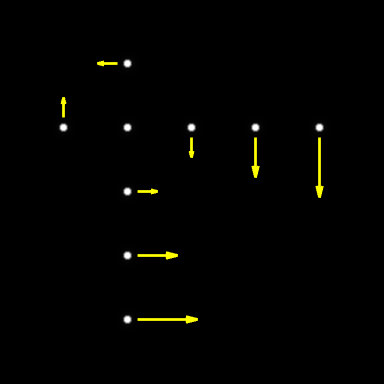} &
		\includegraphics[width = \dotSz\columnwidth]{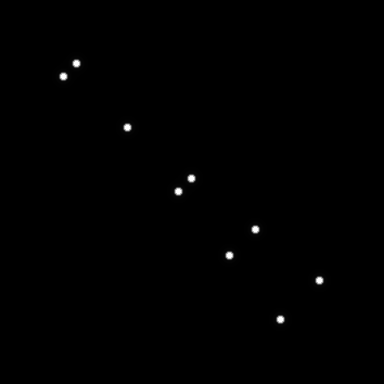}\\
		{\small{(a) First frame}}&
		{\small{(b) Last frame}}\\
		
		\includegraphics[width = \dotSz\columnwidth]{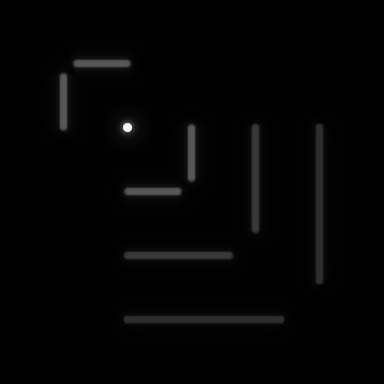}&
		\includegraphics[width = \dotSz\columnwidth]{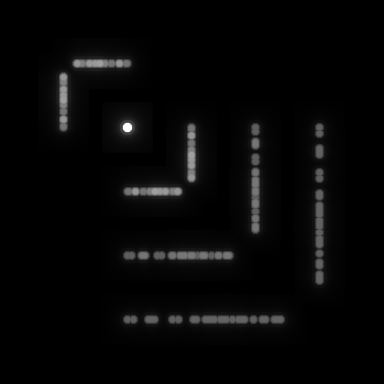}\\
	    
		{\small{(c) Conventional camera}}&
		{\small{(d) Fluttered Shutter camera}}\\
		
		\includegraphics[width = \dotSz\columnwidth]{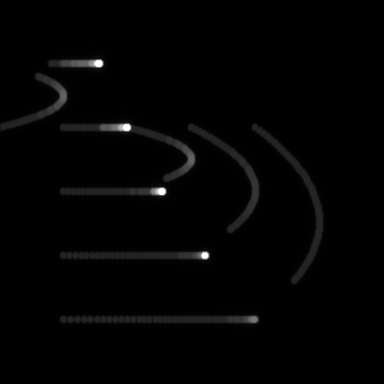}&
		\includegraphics[width = \dotSz\columnwidth]{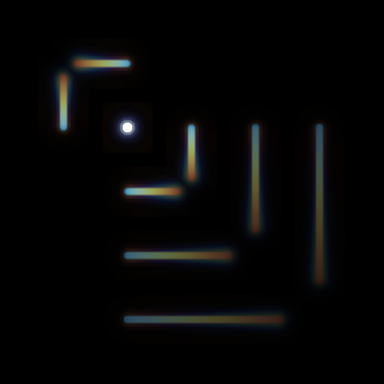}\\
		
		{\small{(e) Parabolic motion camera}} &
		{\small{(f) Our camera}}\\

	\end{tabular}
	\caption{\textbf{Simulation of the different coding methods:} (a) first frame (arrows indicate dots path and velocity), (b) last frame, (c) conventional static camera, (d) fluttered shutter camera \cite{Raskar_Flut_1}, (e) Parabolic motion camera \cite{Levin_motionInvariant} and (f) our proposed camera. (The imaging is performed on single pixel dots to simulate point sources. 
	For visualization purposes dilation and gamma correction are applied.)}
	\label{fig:dots}
\end{figure}

To further illustrate the motion encoding of our method, we simulate imaging of moving point sources using our method, and compare it to a conventional camera, the fluttered-shutter camera \cite{Raskar_Flut_1} and the parabolic motion camera \cite{Levin_motionInvariant}. Fig.~\ref{fig:dots} presents the PSF encoding performed by the different methods (this is an extension of a similar comparison shown in Fig.~3 of \cite{Levin_motionInvariant}). The original scene is formed of two sets of point sources arranged in two orthogonal lines. While the joint dot stays in place, all the other dots are moving in different velocities, as illustrated by the arrows in Fig.~\ref{fig:dots}(a). 

Imaging simulation of this scene is performed using the four methods. In the conventional case, the stationary dot stays 'as-is', and all the other dots are blurred according to their motion trajectory. Using fluttered shutter camera, parts of the dots' trace is blocked, and the code can be clearly seen. As suggested in \cite{Raskar_Flut_1}, such a code generates an easy to invert PSF, assuming the motion direction and extent is known. Indeed, some PSF estimation can be done for blind deblurring, but an inherent invertibility/estimation trade-off exists, as discussed in \cite{Agrawal2009_Flutt_2}. In addition, the light throughput loss caused by fluttered shutter is clearly seen (for the code proposed in \cite{Raskar_Flut_1} the loss is 50\%).

Using the parabolic motion camera (with parabolic motion in the horizontal direction), the PSF is roughly motion invariant in the direction of the sensor motion, as clearly seen in all horizontal dots. Yet, in any other direction, and most significantly in the orthogonal (in this case, vertical) one, each dot linear motion and the sensor parabolic motion are composed, making the PSFs highly motion-variant. 

In the proposed joint phase-mask and focus variations coding, each PSF is colored according to the different motion trajectory. The direction is encoded by the blue-green-red transition, and the extent of the transition indicates the velocity of the motion.

{\bf Spectral analysis.} To analyze the motion encoding ability of our scheme, a spectral analysis of the PSF is carried using the spatiotemporal Fourier analysis model proposed in \cite{Levin_motionInvariant}. In this model, a single spatial dimension is examined vs. the temporal dimension, and a 2D Fourier Transform (FT) is carried on this $(x,t)$ plane (which is a slice of the full $(x,y,t)$ space). In such setting, different velocities of a point source form lines at different angles in the $(x,t)$ plane. The analysis in \cite{Levin_motionInvariant} included only the spectrum amplitude, but in our analysis we include also its phase since our encoding is also phase dependent as we show next.  

\begin{figure}[tb]
    \def\pltWd{0.28}
	\centering
	\begin{tabular}{c c c}
		\includegraphics[width = \pltWd\columnwidth]{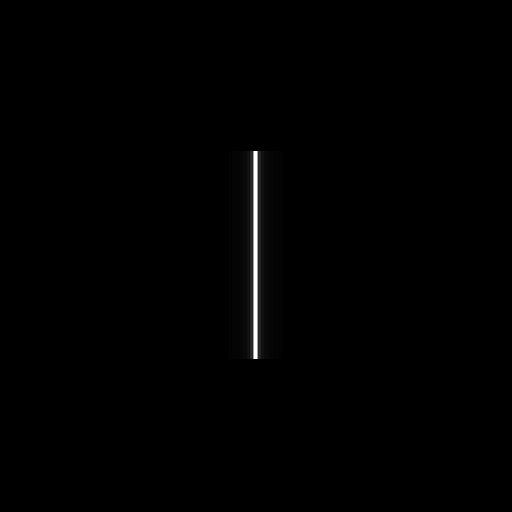} &
		\includegraphics[width = \pltWd\columnwidth]{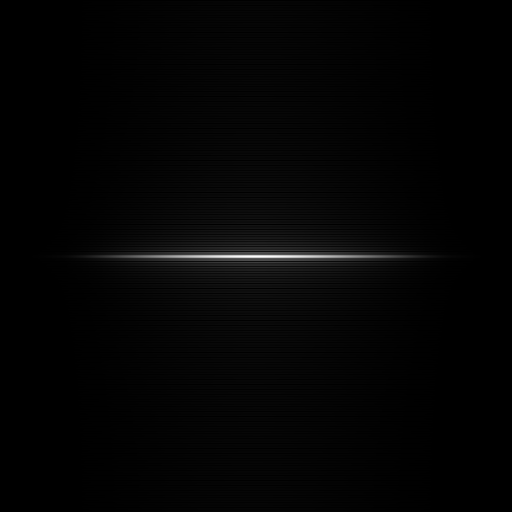} &
		\includegraphics[width = \pltWd\columnwidth]{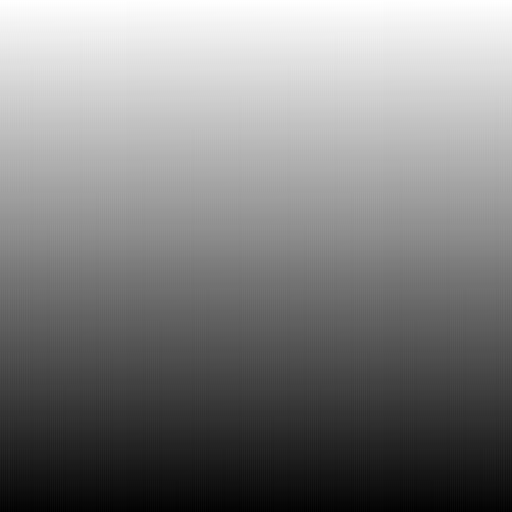}\\
		
		\includegraphics[width = \pltWd\columnwidth]{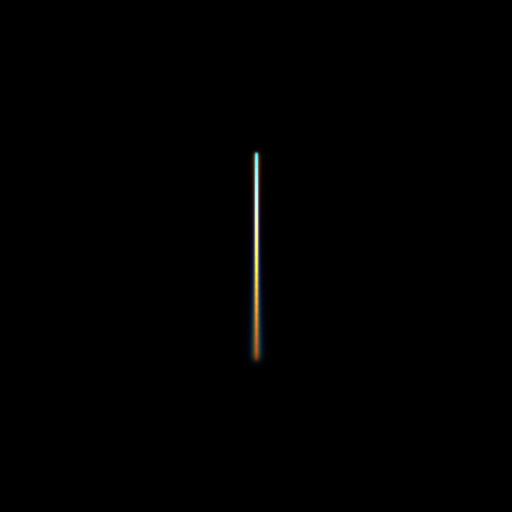} &
		\includegraphics[width = \pltWd\columnwidth]{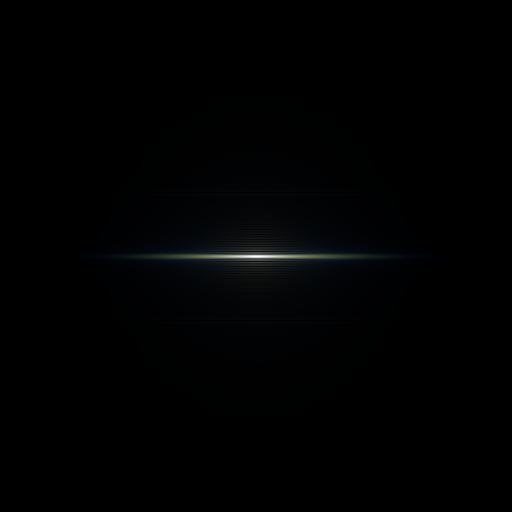} &
		\includegraphics[width = \pltWd\columnwidth]{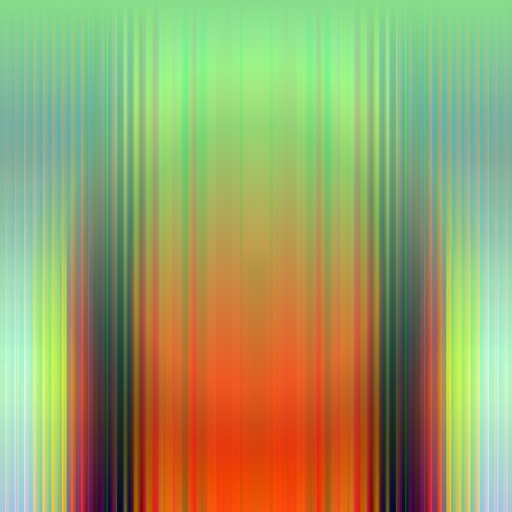}\\
		
		{\small{(a) $(x,t)$ PSF}}&
		{\small{(a) FT amp.}}&
		{\small{(b) FT ph.}}\\

	\end{tabular}
	\caption{\textbf{PSF spectral analysis.} PSFs and the corresponding spectra of a (top) static camera and (bottom) our method. (a) $(x,t)$ slice of PSF and its (b) amplitude and (c) phase in Fourier domain.}
	\label{fig:spctAnal}
\end{figure}

We compare our method with a conventional camera in Fig.~\ref{fig:spctAnal} (a full analysis including the fluttered shutter and parabolic motion cameras appears in the supplementary material). For the conventional static camera, the $(x,t)$ slice of the PSF has a Sinc spectrum amplitude, which allows good reconstruction of object at this velocity (represented by the angle of the $(x,t)$ PSF). Since the PSF is 'gray' (i.e. has no chromatic shift along its trajectory), its spectrum phase is also gray. This 'gray phase' feature is common also to fluttered-shutter and parabolic motion cameras, as can bee seen in the full analysis in the supplementary material. 

Our proposed PSF can be considered as an infinite sequence of smaller PSFs, each one of a different color. As all PSFs have a similar spatial shape, but each has a different color and different location in the $(x,t)$ plane, the spectrum amplitude is 'white' and similar to the spectrum amplitude of the conventional PSF. Yet, the phase (which holds the shift information) is colored, according to the shift (i.e. spatiotemporal location) of each color. Our spatiotemporal chromatic coupling can be considered as utilization of the spectrum phase as a degree of freedom for the coding. The color variations in the phase indicate the coupling between the color and the trajectory, as can be seen in Fig.~\ref{fig:spctAnal}. 

\section{The color-coded motion deblurring network} \label{CNN}
    As described in the previous section, the dynamic phase aperture coding generates color variations in the spatiotemporal blur kernel. These chromatic cues encode the different motion trajectories, without a limitation on the motion direction. These cues serve as prior information both for PSF estimation and image deblurring, thus allowing shift-variant motion deblurring. 

Traditionally, spatially varying deblurring is performed in two stages: PSF estimation for the different objects/segments, and then deblurring is applied to each of them. As presented in \cite{EDOF_DL,Nah_deepDeblur_gopro}, this task can be solved using a single CNN, trained with a dataset containing the various possibilities of the shift-variant blur. One may treat the CNN operation as an end-to-end process that extracts the cues, which allow the PSF estimation, and then utilizes the acquired PSF information for image deblurring. 

{\bf Training data.} To train such a CNN for our motion deblurring process, images containing moving objects blurred with our spatiotemporal varying blur kernel (and their corresponding sharp images) are needed. Since experimentally acquiring a motion-blurred image and its pixel-wise accurate sharp image is very complex (even without the dynamic aperture coding), an imaging simulation is used. Using the GoPro dataset \cite{Nah_deepDeblur_gopro}, which contains high frame-rate videos of various scenes, we simulate images with the motion-color coded blur by blurring (using the coded kernel) consecutive frames and then summing them up. Sequences of 9 frames are used, and a dataset containing 2,500 images is created; 80\% of it is used for training and the rest for validation and testing. Since our deblurring process is based on local cues encoded by our spatiotemporal kernel and not the image statistics, as we show hereafter, a CNN trained on this synthetic data generalizes well real-world images. 

{\bf The deblurring network architecture.} Since image restoration is sought, a fully-convolutional network (FCN) architecture is considered. As shown in the work of Nah \etal \cite{Nah_deepDeblur_gopro}, multiscale processing is an efficient tool to grasp the structure of motion-blurred objects. Therefore, the network architecture we use is based on the known U-Net structure \cite{Unet}, as it is one of the leading multiscale FCN architectures. A skip-connection is added between the output and the input, leaving the 'U' structure to estimate the residual correction for the input image. Empirically, this simplifies the convergence (the full structure and details of the network are presented in the supplementary material). 

The U-Net architecture is trained using patches of size 128x128 taken from the dataset described above. Since the final goal is to present the performance on images taken with a real camera, noise augmentation is used, with similar noise to the one observed in real images taken using the target camera (AWGN with $\sigma=9$). The network is trained using the Huber loss \cite{huber1964}, and the average reconstruction results on the test set are $PSNR=29.5, SSIM=0.93$. Examples of the reconstruction performance achieved on images from the test set in different cases are presented in the supplementary material. 

{\bf Ablation study.} As an ablation study, we generated a version of the same dataset without our spatiotemporal coding, and trained the same architecture on it. In this case, we get a significant over-fitting and poor results on the test set ($PSNR=24.6, SSIM=0.84$). 

In another ablation test, we evaluate another network structure, which is similar to the one presented in \cite{EDOF_DL}. Consecutive blocks of Conv-BN-ReLU (no pooling) with direct skip connection from the input to the output are used. Such an architecture is also designed to learn the residual correction needed to the image for the deblurring operation, but without multiscale operation. Nominal performance is achieved with this network structure ($PSNR=27.5, SSIM=0.9$), probably because the multiscale operation is important for this task. However, this architecture is much more shallow and with just 2\% of the weights of the full U-net model, and it still achieves comparable results to the model of \cite{Nah_deepDeblur_gopro} (see Section~\ref{exp} for the comparison). This comparison demonstrates the benefit of the aperture coding- the encoded cues are such strong guidance for the deblurring operation, that a very shallow model achieves comparable performance to a much larger one. The full details on this model and test appear in the supplementary material. 


\section{Experiments} \label{exp}
    We start by evaluating our proposed method in simulation. Two different comparisons are presented; the first is to  other computational imaging methods: the fluttered shutter camera \cite{Raskar_Flut_1} and the parabolic motion camera \cite{Levin_motionInvariant}, demonstrating the advantages of our dynamic aperture phase coding vs. other coding methods. The second comparison is to the deblurring CNN presented by Nah \etal \cite{Nah_deepDeblur_gopro}, which is designed for conventional cameras. Such a comparison illustrates the benefits of coded aperture. Following that, we present real-world results from our designed prototype. 

\subsection{Comparison to other coding methods}
In order to demonstrate our PSF estimation ability in the motion deblurring process vs. the motion direction sensitivity of the other methods, a scene with rotating spoke resolution target is simulated. Such scene contains motion in all directions and in various velocities (according to the distance from the center of the spoke target) simultaneously.


\begin{figure}[tb]
    \def\spkSz{0.4}
	\centering
	\begin{tabular}{ c c c}
		\includegraphics[width = \spkSz\columnwidth]{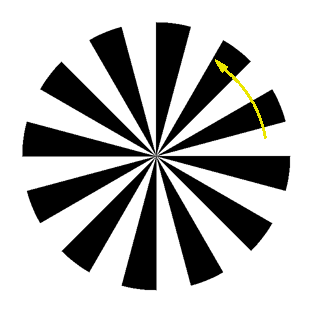} &
		\includegraphics[width = \spkSz\columnwidth]{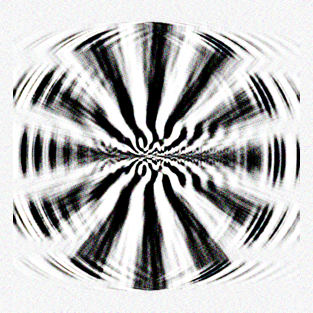}\\
		{\small{(a) Rotating target}}&
		{\small{(b) Flutter-shutter rec.}}\\
		
		\includegraphics[width = \spkSz\columnwidth]{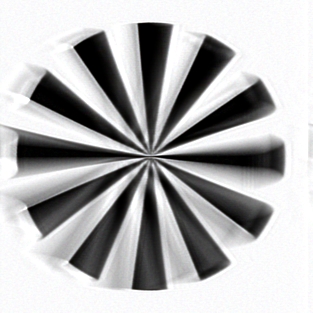} &
		\includegraphics[width = \spkSz\columnwidth]{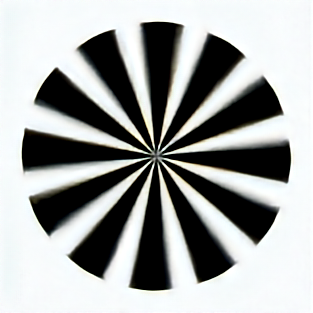}\\
		
		{\small{(c) Parabolic motion rec.}}&
		{\small{(d) Our rec.}}\\

	\end{tabular}
	\caption{\textbf{Simulation results of rotating target:} (a) rotating target and the reconstruction results for (b) fluttered-shutter, (c) parabolic motion camera and (d) our method.}
	\label{fig:sim_res}
\end{figure}

The synthetic scene serves as an input to the imaging simulation for the three different methods (fluttered shutter, parabolic motion and ours). The fluttered shutter code being used (both in the imaging and reconstruction) is for motion to the right, in the extent of the linear motion of the outer parts of the spoke target. The parabolic motion takes place on the horizontal direction. Each imaging result is noised using AWGN with $\sigma=3$ to simulate a real imaging scenario in good lighting conditions (since the fluttered shutter coding blocks 50\% of the light throughput, the noise level of its image is practically doubled). Fig.~\ref{fig:sim_res} presents the deblurring results of the three different techniques.\footnote{The imaging simulations for the fluttered shutter and parabolic motion cameras were implemented by us, following the descriptions in \cite{Raskar_Flut_1,Levin_motionInvariant}. The fluttered shutter reconstruction is performed using the code released by the authors. The parabolic motion reconstruction is performed using the Lucy-Richardson deconvolution algorithm \cite{Richardson:72,Lucy:1974yx}, as suggested by the authors in Section 4.1 of \cite{Levin_motionInvariant}. As the authors stated, a little better performance can be achieved using the original algorithm used in \cite{Levin_motionInvariant}, but its implementation is not available. Moreover, probably much better results can be achieved for both methods using a CNN based reconstruction. However, the main issue in the current comparison is the sensitivity of the other coding methods to the motion direction, which is not related to the used reconstruction algorithm.}

The fluttered shutter based reconstruction restores the general form of the  area with the corresponding motion coding (outer lower part, moving right), and some of the opposite direction (outer upper part, moving left), and fails on all other directions/velocities. This can be partially solved using a different coding that allows both PSF estimation and inversion. Yet, this introduces an estimation-invertibility trade-off. Note also that a rotating target is a challenging case for shift-variant PSF estimation, and thus, as can be seen, a restoration with incorrect PSF leads to poor results. Moreover, the noise sensitivity of this approach is apparent, as it blocks 50\% of the light throughput. 

The parabolic-motion method achieves good reconstruction for the horizontal motion (both left and right) as can be seen in the upper and lower parts of the spoke (which move horizontally). Yet, notice that its performance are not the same for left/right (as any practical finite parabolic motion cannot generate a true motion invariant PSF). Also, both vertical motions are not coded properly, and therefore are not reconstructed well. Using our method, motion in all directions can be estimated, which allows a shift variant blind deblurring of the scene. 

\subsection{Comparison to blind deblurring} 
To analyze the advantages in motion-cues coding, our method is compared to the multiscale motion deblurring CNN presented by Nah \etal \cite{Nah_deepDeblur_gopro}. The test set of the GoPro dataset is used as the input. Since Nah \etal trained their model on sequences of between 7-13 frames, similar scenes were created using both our coding method and simple frame summation (as used in \cite{Nah_deepDeblur_gopro}, with the proper gamma-related transformations). Note that in our case, a spatial (diffraction related) blur is added with the motion blur, so our model is handling a more challenging task. 

The reconstruction results are compared for several noise levels- $\sigma=[0,3]$ on a $[0,255]$ scale (the reference method was trained with $\sigma=2$). The measures on each motion length are averaged over the different noise levels, and the results are displayed in Table~\ref{Tab:goProStats}. As can be clearly seen, our method provides an advantage in the recovery error over the method of Nah \etal \cite{Nah_deepDeblur_gopro} in both PSNR and SSIM (visual reconstruction results are presented in the supplementary material). In small motion lengths, both methods provide visually pleasing restorations (though our method is more accurate in terms of PSNR/SSIM). Yet, as the motion length increases our improvement becomes more significant. This can be explained by the fact that the architecture used in \cite{Nah_deepDeblur_gopro} is trained using adversarial loss, and therefore inherent data hallucination occurs in their reconstruction. As the motion length gets larger, such data hallucination is less accurate, and therefore the reduction in their PSNR/SSIM performance is more significant. Our method employs the encoded motion cues for the reconstruction, therefore providing more accurate results.

Note also that our model is trained only on images generated using sequences of 9 frames, and the results for shorter/longer sequences are still better than the model from \cite{Nah_deepDeblur_gopro} that is trained on sequences in all this range. This clearly shows that our model has learned to extract the color-motion cues and utilize them for the image deblurring, beyond the specific extent present in the training data. 

Note that in our dataset an additional diffraction-related spatial blur is added (as mentioned above), so in a case that a similar spatial blur is added to the original GoPro dataset (without the motion-color cues), our advantage over \cite{Nah_deepDeblur_gopro} is expected to be even larger. Note also that our network converges well and provides good performance for higher noise levels (as presented in the following), however, for a fair comparison to \cite{Nah_deepDeblur_gopro} we limit the noise level here to $\sigma=3$. 

\begin{table}
\begin{center}
\begin{tabular}{|l|c|c|}
\hline
$N_{frames}$ & Nah \etal & Ours \\
\hline\hline

$N=7$ & $28.1/0.93$ & $\bf{30.9}/\bf{0.95}$ \\
$N=9$ & $27/0.91$ & $\bf{30}/\bf{0.94}$ \\
$N=11$ & $25.9/0.89$ & $\bf{28.9}/\bf{0.92}$ \\
$N=13$ & $24.9/0.87$ & $\bf{28}/\bf{0.91}$ \\

\hline
\end{tabular}
\end{center}
\caption{\textbf{Quantitative comparison to blind deblurring:} PSNR/SSIM comparison between the method presented in \cite{Nah_deepDeblur_gopro} and our method, for various lengths of motion ($N_{frames}$). Results are the average measures for various scenes and different noise levels. See the supplementary material for per noise level statistics and additional information.} \label{Tab:goProStats}
\end{table}

\begin{figure}[h]
\begin{center}
\includegraphics[width=0.6\linewidth]{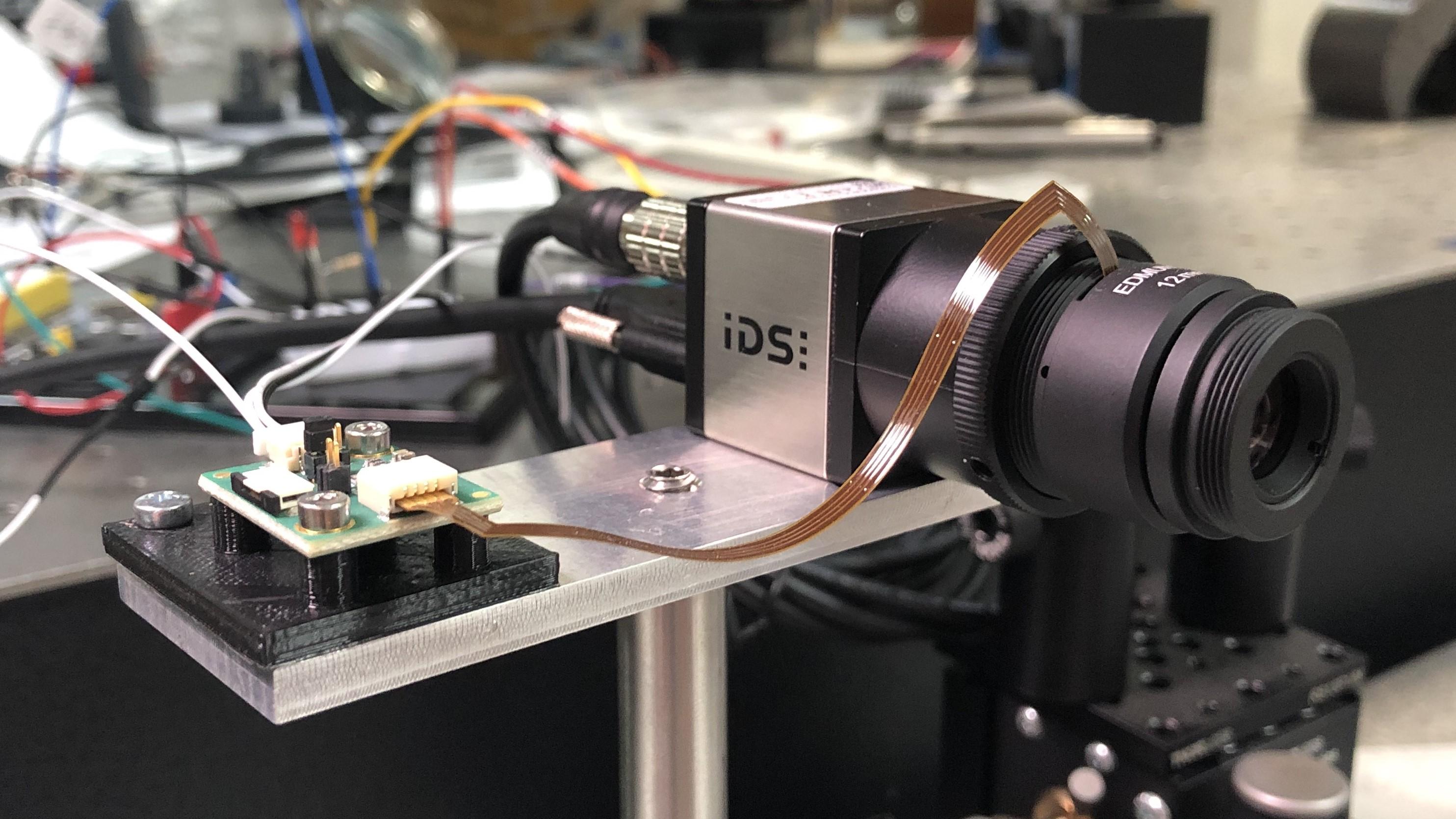}
\end{center}
   \caption{\textbf{The table-top experimental setup:} The liquid-lens and our phase-mask are incorporated in the C-mount lens. The micro-controller synchronizes the focus variation to the frame exposure using the camera flash signal.}
\label{fig:setup}
\end{figure}

\subsection{Table-top experiment}
 
 \begin{figure}[tb]
 \begin{center}
        \includegraphics[width = 0.8\columnwidth]{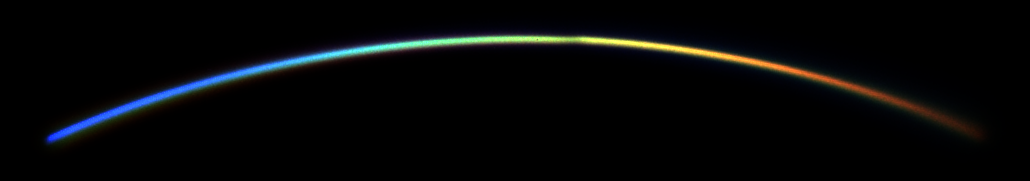}
    \end{center}
	\caption{\textbf{Experimental validation of PSF coding:} a moving white LED captured with our camera, validates the required PSF encoding.}
	\label{fig:exp_LEDs}
\end{figure}

Following the simulation results, a real-world setup is built (see Fig.~\ref{fig:setup}). A C-mount lens with $f=12[mm]$ is mounted on a 18MP camera with pixel size of $1.25[\mu m]$. A similar phase-mask to the one used in \cite{IEEE_depth} and a liquid focusing lens are incorporated in the aperture plane of the main lens. A signal from the camera indicating the start of the exposure (originally designed for flash activation) is used to trigger the liquid lens to perform the required focus variation (a detailed description of the experimental setup is presented in the supplementary material). The liquid lens is calibrated to introduce a focus variation equivalent to $\psi=[0,8]$ during exposure.

The first real-world test validates the desired PSF spatiotemporal encoding. Two white LEDs are mounted on a spinning wheel, and acts as point-sources, similar to the point sources simulated in Fig.~\ref{fig:dots}. A motion blurred image of the spinning LEDs is acquired, with the phase-mask incorporated in the lens and the proper focus variation during exposure. Zoom-in on one of the LEDs is presented in Fig.~\ref{fig:exp_LEDs}. The gradual color changes along the motion trajectory is clearly visible. The full image including both LEDs is presented in the supplementary material.  

\begin{figure}[tb]
     \def\stlSz{0.45}
	\centering
	\begin{tabular}{c c}
		\includegraphics[width = 0.428\columnwidth]{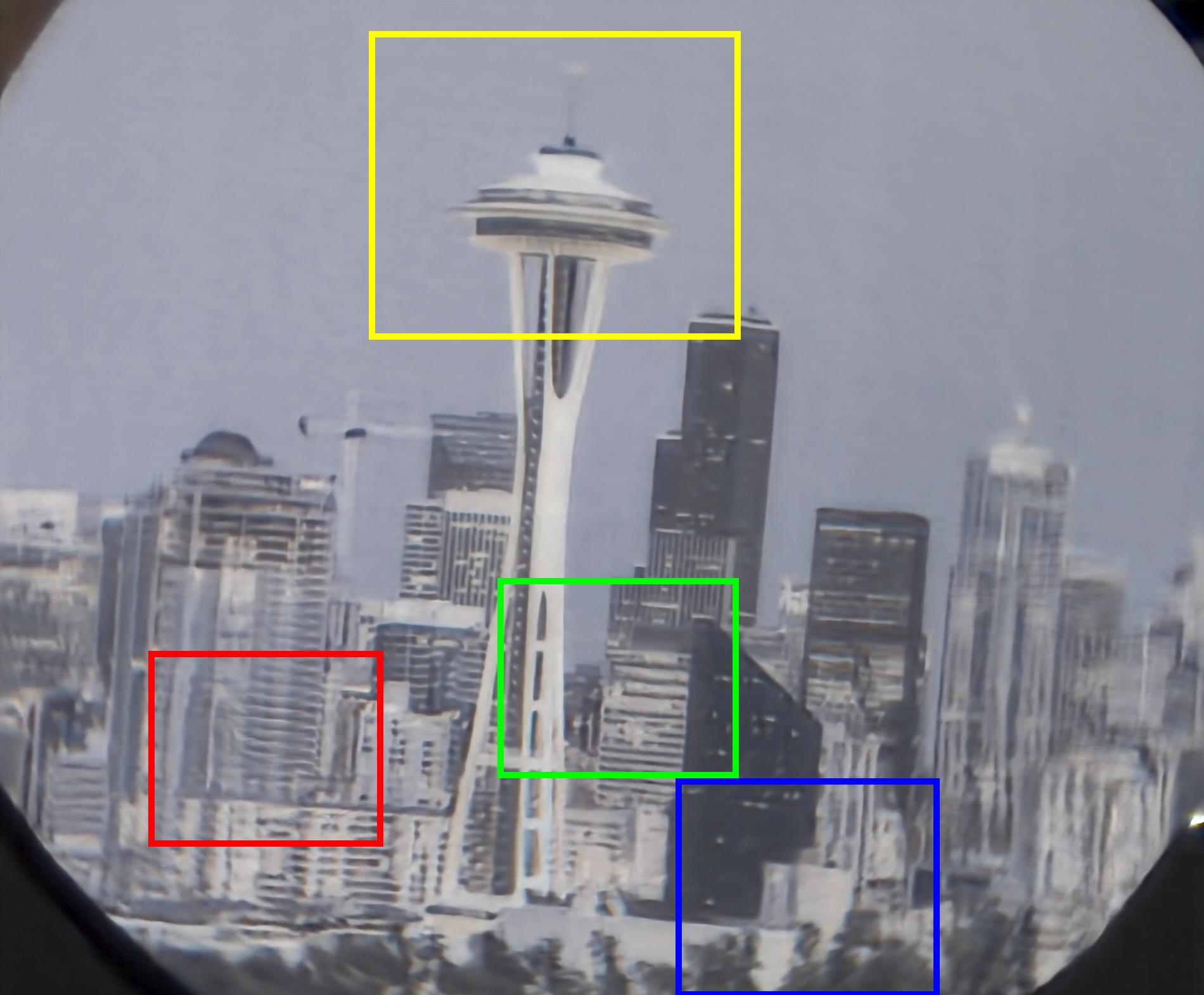} &
		\includegraphics[width = \stlSz\columnwidth]{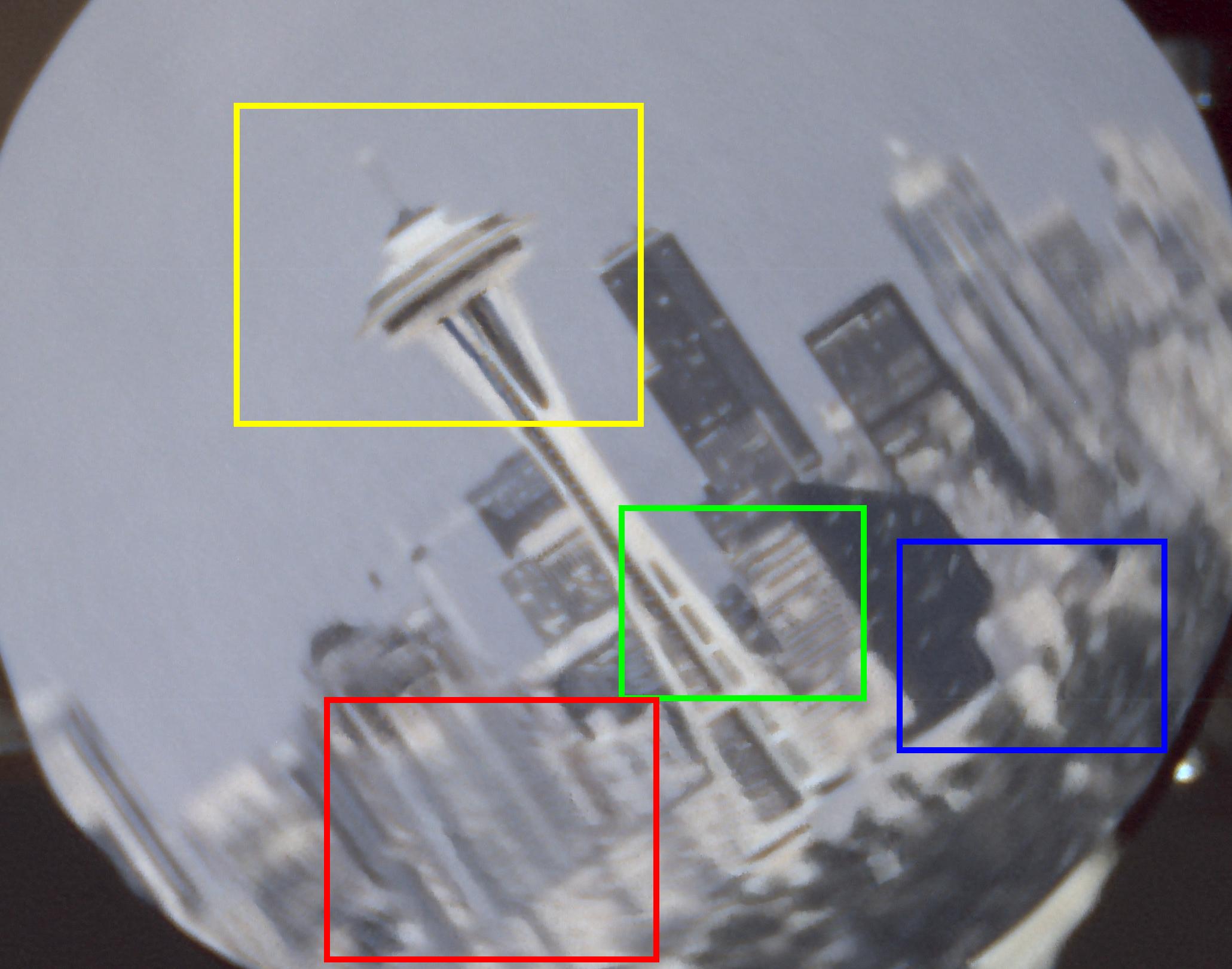}\\
		
		\includegraphics[width = 0.428\columnwidth]{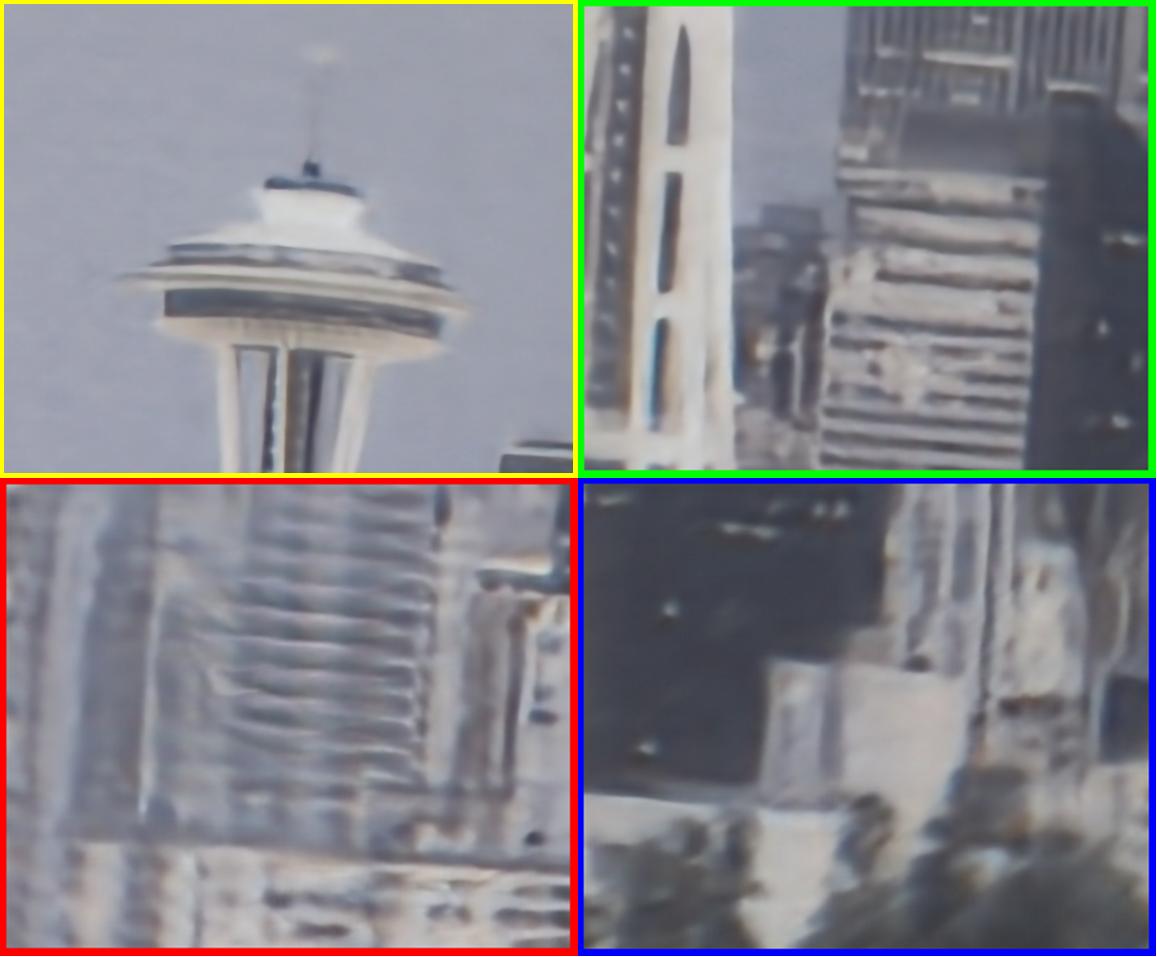} &
		\includegraphics[width = \stlSz\columnwidth]{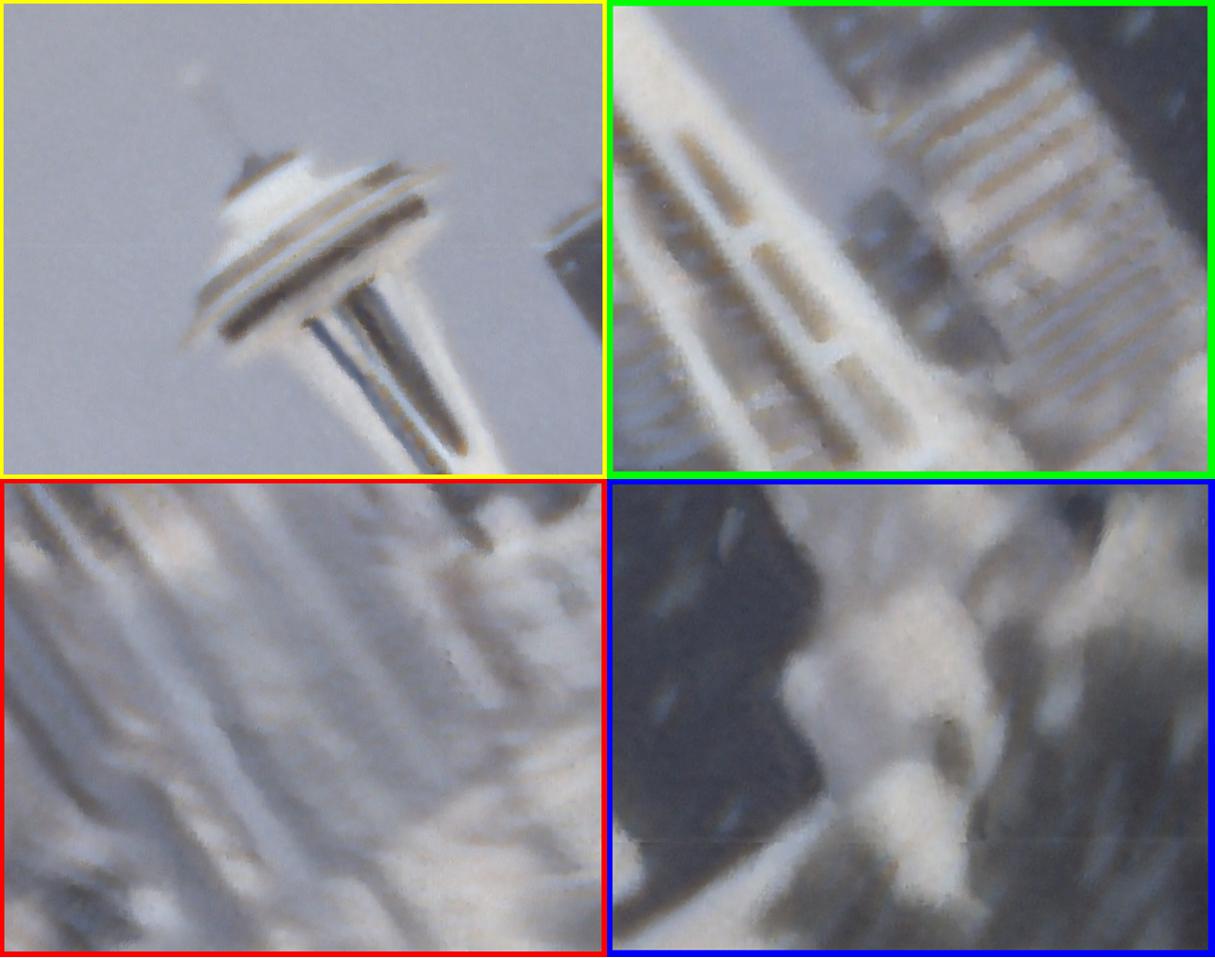}\\

		{\small{(a) Our reconstruction}}&
		{\small{(b) Nah \etal reconstruction \cite{Nah_deepDeblur_gopro}}}\\

	\end{tabular}
	\caption{\textbf{Seattle's view experiment:} reconstruction results of (top) rotating view of Seattle and (bottom) zoom-ins, using (a) out method and (b)  Nah \etal reconstruction \cite{Nah_deepDeblur_gopro}.}
	\label{fig:stl}
\end{figure}

Following the PSF validation experiment, a deblurring experiment on moving objects is carried. In order to examine various motion directions and velocities at once, a rotating object is used. For reference, image of the same object is captured with a conventional camera (i.e. the same camera with a fixed focus and without the phase-mask), and deblurred using the multiscale motion blur CNN (Nah \etal \cite{Nah_deepDeblur_gopro}). Results on a rotating photo of Seattle's view are presented in Fig.~\ref{fig:stl}. 

\begin{figure}[tb]
    \def\trnSz{0.9}
	\centering
	\begin{tabular}{c c}
		
		\includegraphics[width = \trnSz\columnwidth]{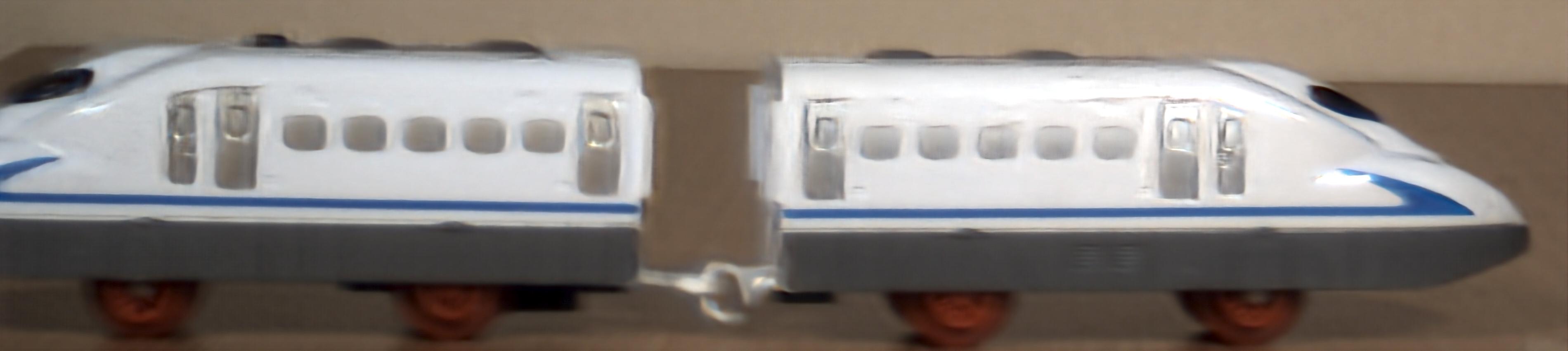}\\
		\includegraphics[width = \trnSz\columnwidth]{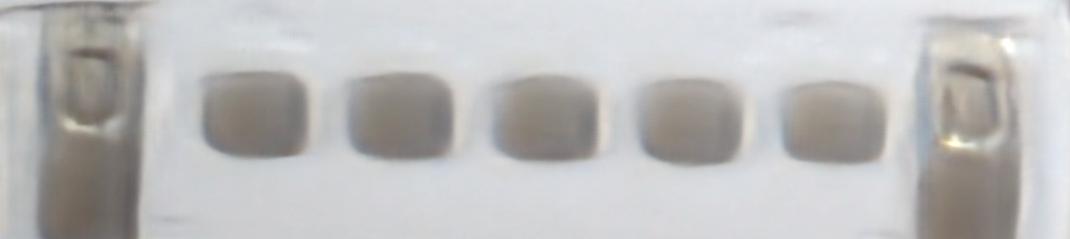}\\
		{\small{(a) Our reconstruction}}\\
		
		\includegraphics[width = \trnSz\columnwidth]{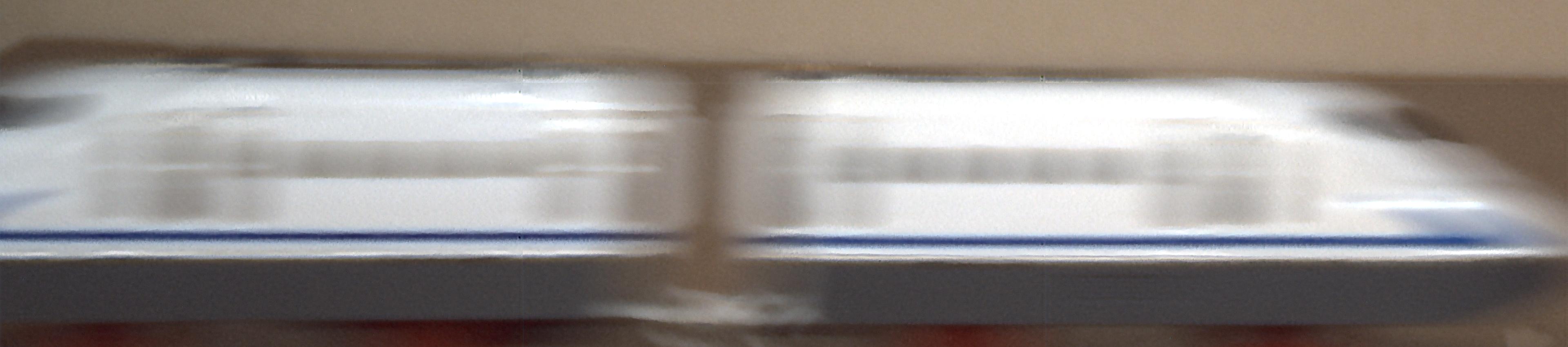}\\
		\includegraphics[width = \trnSz\columnwidth]{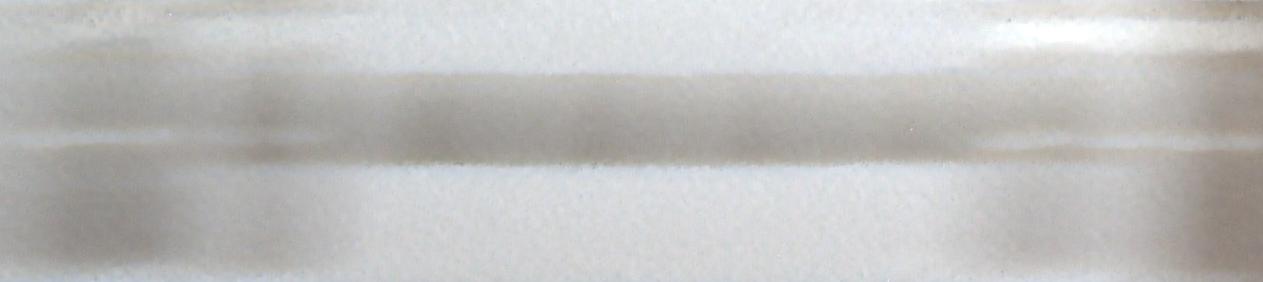}\\
		{\small{(b) Nah \etal reconstruction (\cite{Nah_deepDeblur_gopro})}} \\

	\end{tabular}
	\caption{\textbf{Train experiment:} Recovery results of a moving train using (a) our method and (b)  Nah \etal reconstruction \cite{Nah_deepDeblur_gopro}.}
	\label{fig:trn}
\end{figure}

In addition to the rotating target test, a linearly moving object (toy train) is also captured, in the same configuration as the previous example. The results are presented in Fig.~\ref{fig:trn}. As can be clearly seen, our camera provides much better results in both cases. The full results of the experiments above along with additional experiments and demonstrations are provided in the supplementary material. 

\section{Conclusion} \label{sum}
    A computational imaging approach for motion deblurring is presented. The method is based on spatiotemporal phase coding of the lens aperture, to achieve a motion-variant PSF. The phase coding is achieved using two components: (i) the static/spatial part- a phase-mask designed to code the PSF to have a joint color-defocus dependency; and (ii) the dynamic/temporal part- a gradual variation of the focus setting performed during the image exposure. Jointly, these coding mechanisms achieve motion variant PSF, which is exhibited in a gradual color change of the blur along the motion trajectory. Such a PSF encodes cues to the motion extent and velocity in the acquired image. These cues are in turn utilized for the motion deblurring process, implemented using a CNN model. The CNN operation encapsulates both the PSF estimation and the spatially-variant motion deblurring. 

Our approach is compared to blind deblurring methods and computational imaging based strategies. Its shift-variant PSF estimation ability and generalization potential to real-world scenes are analyzed and discussed. Our technique achieves better performance compared to the other solutions in various scenarios, without imposing a limitation on the motion direction. An experimental setup implementing the proposed method is presented, and the spatiotemporal PSF color encoding is validated in a real world experiment. In addition, as our encoding provides cues to the entire motion trajectory, our approach holds potential for video-from-motion and temporal super-resolution applications, similar to \cite{vidFromMot,Nayar_codedExposure_2,Nayar_codedExposure_3,flt2vid,Llull:13}.


{\small
\bibliographystyle{ieee_fullname}
\bibliography{egbib}
}

\appendix
\section{PSF spectral analysis} \label{psf_spect}
    As presented in Section~3 of the paper, a PSF spectral analysis is performed to analyze the differences between the different coding methods. The comparison is performed using the spatiotemporal Fourier analysis model proposed in \cite{Levin_motionInvariant}. In this model, a single spatial dimension is examined vs. the temporal dimension, and a 2D Fourier Transform (FT) is carried on this $(x,t)$ plane (which is a slice of the full $(x,y,t)$ space). In such setting, different velocities of a point source form lines at different angles in the $(x,t)$ plane. The analysis in \cite{Levin_motionInvariant} included only the spectrum amplitude, but in our analysis we include also its phase since our encoding is also phase dependent, as presented in the paper and further examined next.

\begin{figure}[tb]
    \def\pltWd{0.28}
	\centering
	\begin{tabular}{c c c c}

		\includegraphics[width = \pltWd\columnwidth]{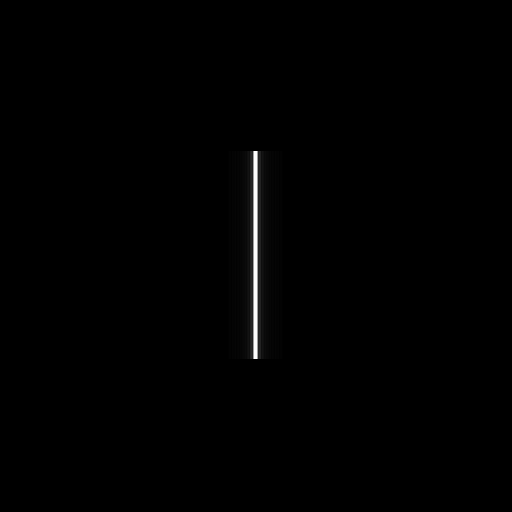} &
		\includegraphics[width = \pltWd\columnwidth]{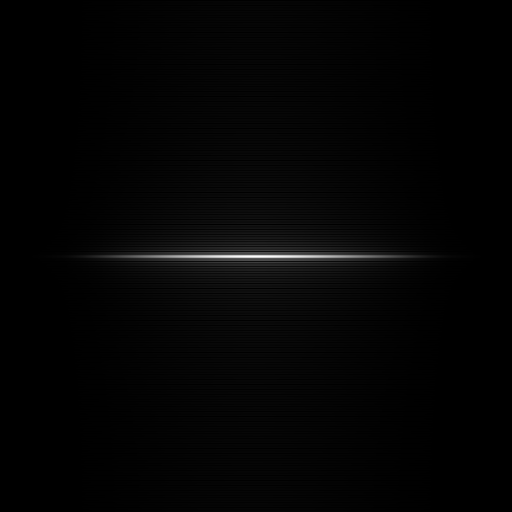} &
		\includegraphics[width = \pltWd\columnwidth]{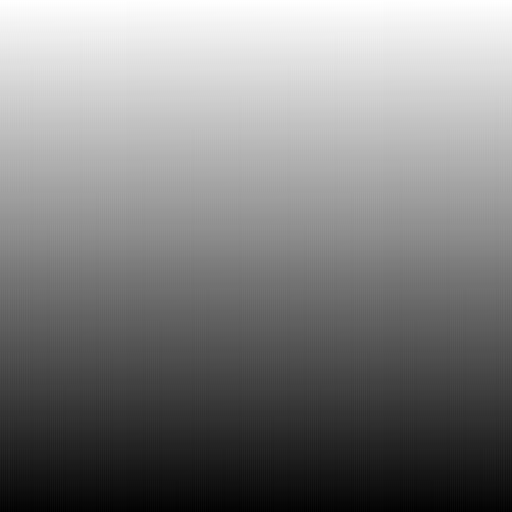}\\

		\includegraphics[width = \pltWd\columnwidth]{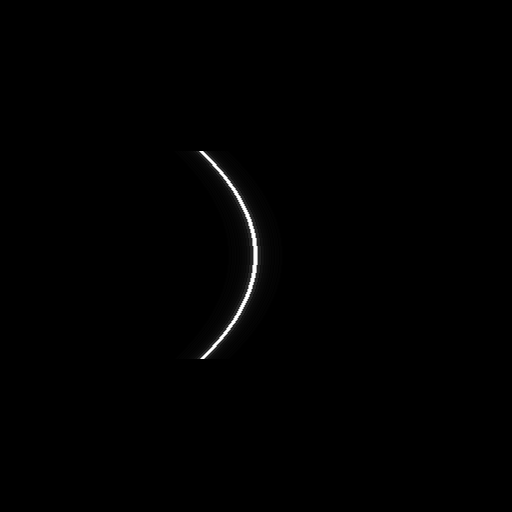} &
		\includegraphics[width = \pltWd\columnwidth]{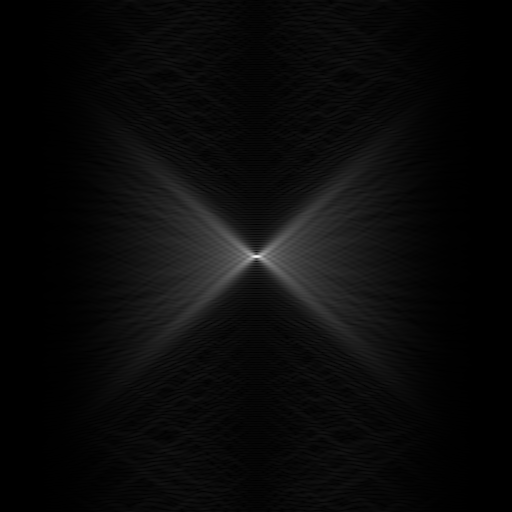} &
		\includegraphics[width = \pltWd\columnwidth]{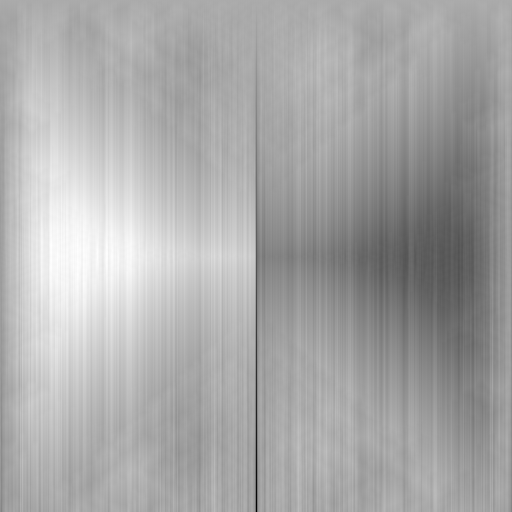}\\

		\includegraphics[width = \pltWd\columnwidth]{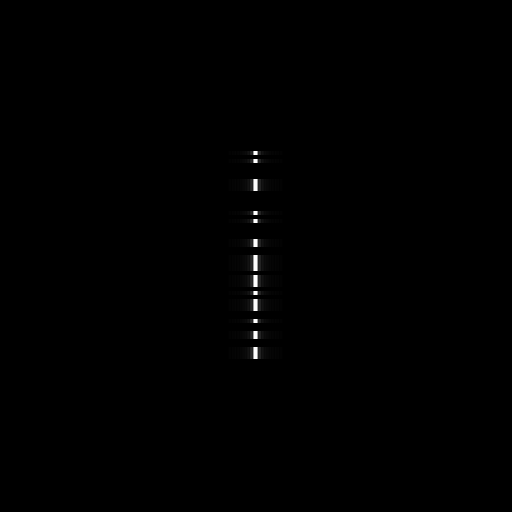} &
		\includegraphics[width = \pltWd\columnwidth]{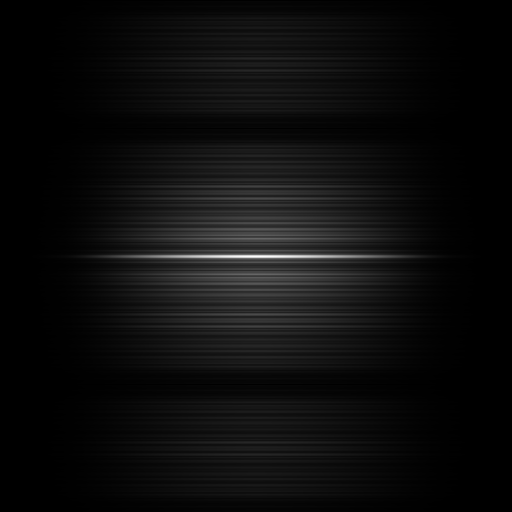} &
		\includegraphics[width = \pltWd\columnwidth]{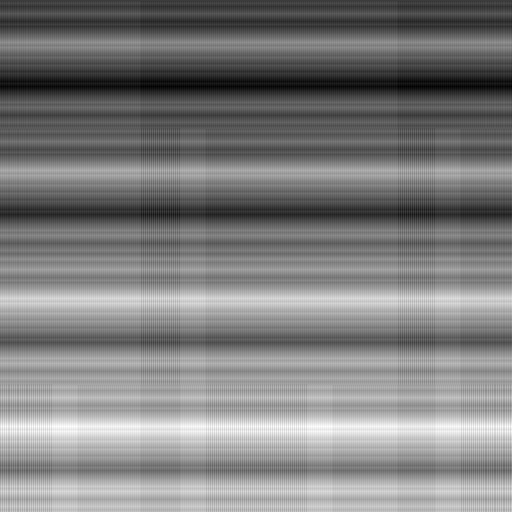}\\

		\includegraphics[width = \pltWd\columnwidth]{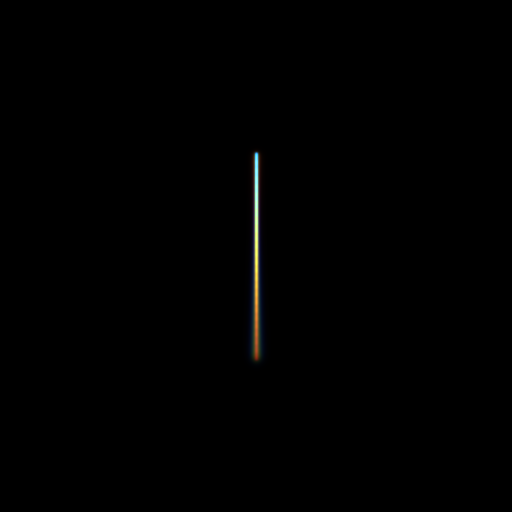} &
		\includegraphics[width = \pltWd\columnwidth]{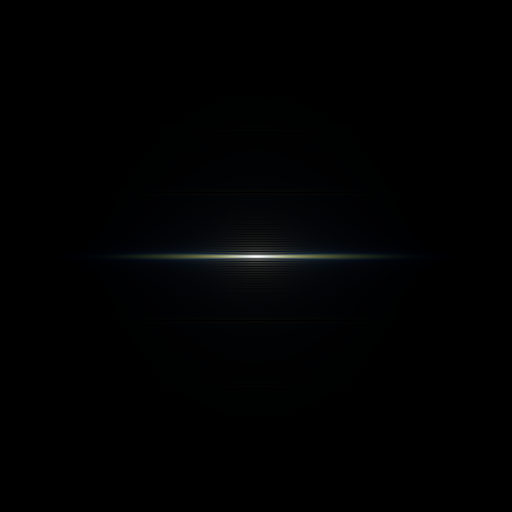} &
		\includegraphics[width = \pltWd\columnwidth]{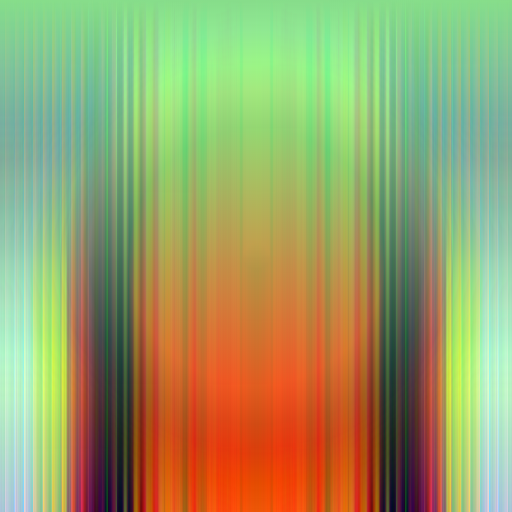}\\
		
		{\small{(a) $(x,t)$ PSF}}&
		{\small{(b) FT amp.}}&
		{\small{(c) FT ph.}}\\

	\end{tabular}
	\caption{\textbf{PSF spectral analysis.} PSFs and the corresponding spectra for a static point source captured using (top to bottom) static camera, parabolic motion camera, fluttered-shutter camera and our method. (a) $(x,t)$ slice of PSF and its (b) amplitude and (c) phase in Fourier domain.}
	\label{fig:spctAnal_1}
\end{figure}

We start by comparing all methods on a static point, represented by a vertical line in the $(x,t)$ plane (see Fig.~\ref{fig:spctAnal_1}, which is the full version of the comparison presented in the paper). In the three reference methods, the PSF is 'gray' (i.e. has no chromatic shift along its trajectory), and therefore the spectrum phase is also gray. 

\begin{figure}[tb]
    \def\pltWd{0.28}
	\centering
	\begin{tabular}{c c c c}

		\includegraphics[width = \pltWd\columnwidth]{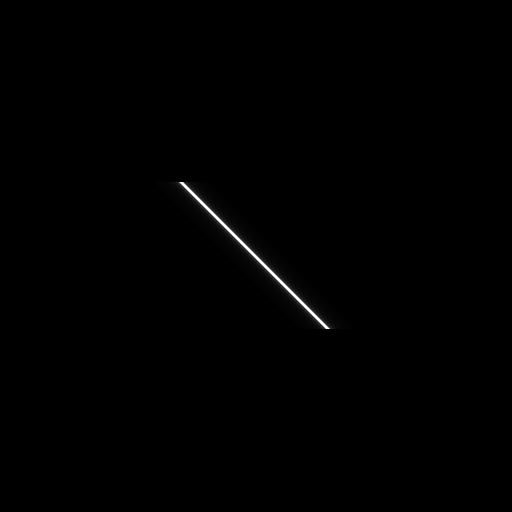} &
		\includegraphics[width = \pltWd\columnwidth]{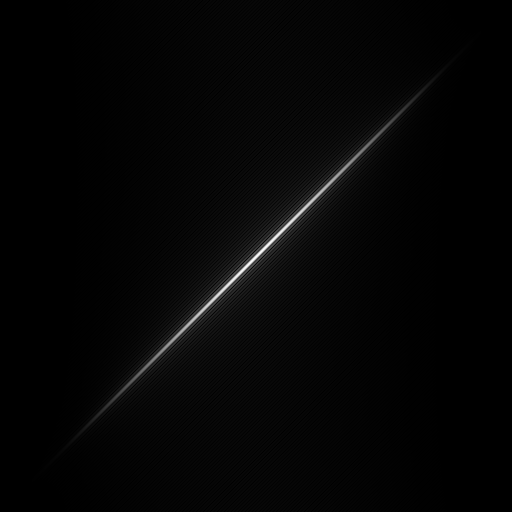} &
		\includegraphics[width = \pltWd\columnwidth]{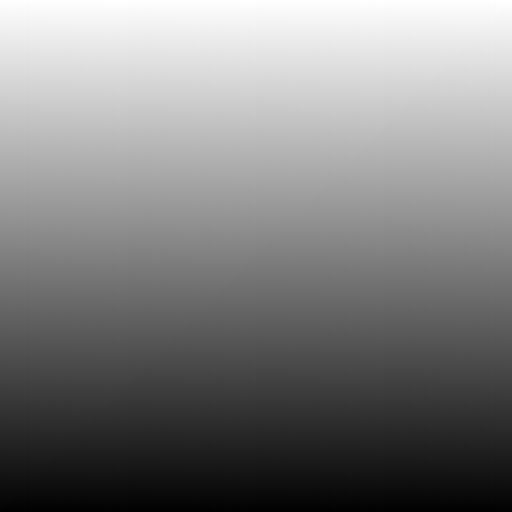}\\

		\includegraphics[width = \pltWd\columnwidth]{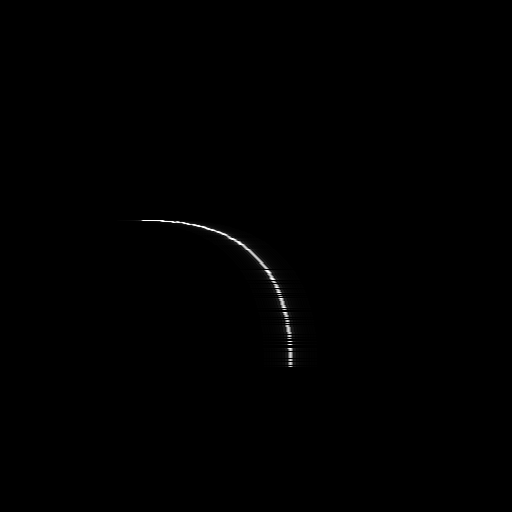} &
		\includegraphics[width = \pltWd\columnwidth]{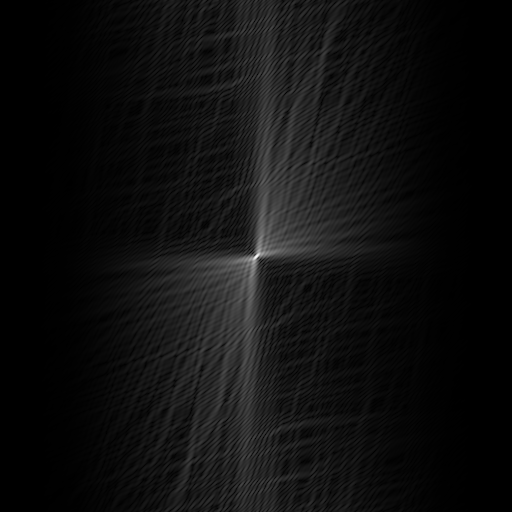} &
		\includegraphics[width = \pltWd\columnwidth]{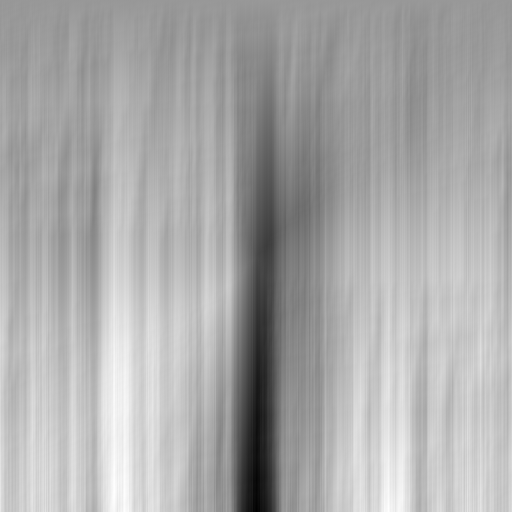}\\

		\includegraphics[width = \pltWd\columnwidth]{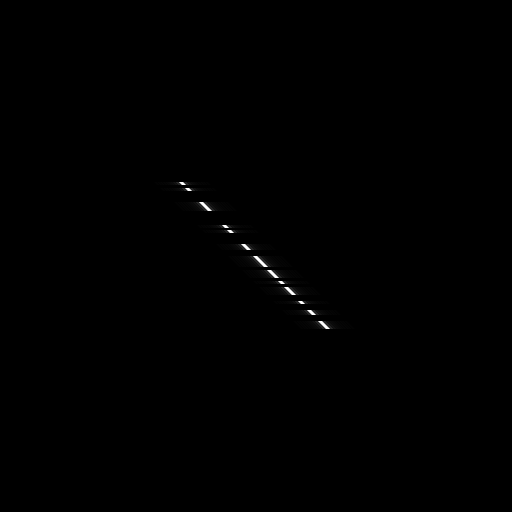} &
		\includegraphics[width = \pltWd\columnwidth]{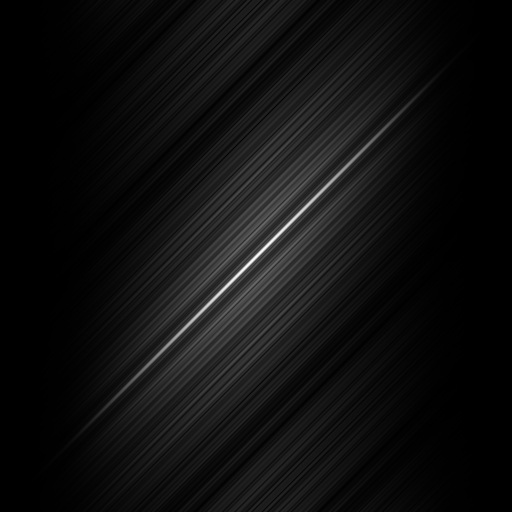} &
		\includegraphics[width = \pltWd\columnwidth]{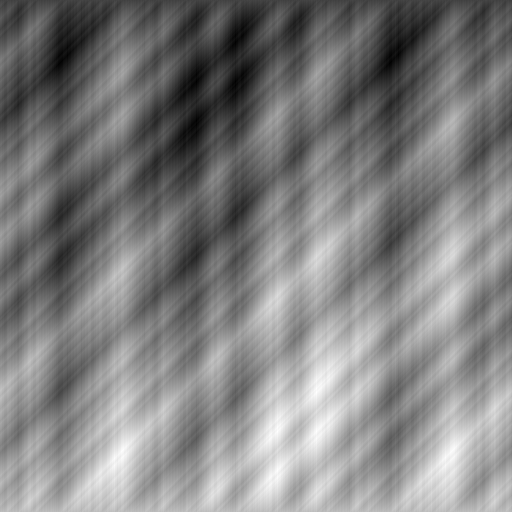}\\

		\includegraphics[width = \pltWd\columnwidth]{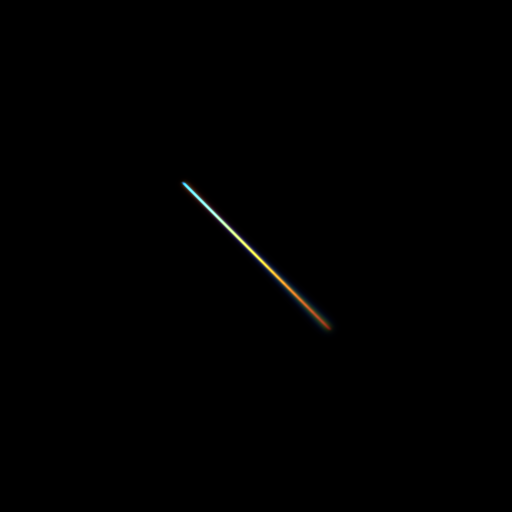} &
		\includegraphics[width = \pltWd\columnwidth]{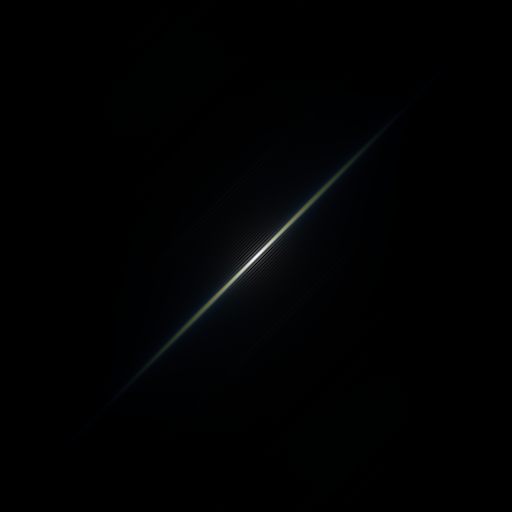} &
		\includegraphics[width = \pltWd\columnwidth]{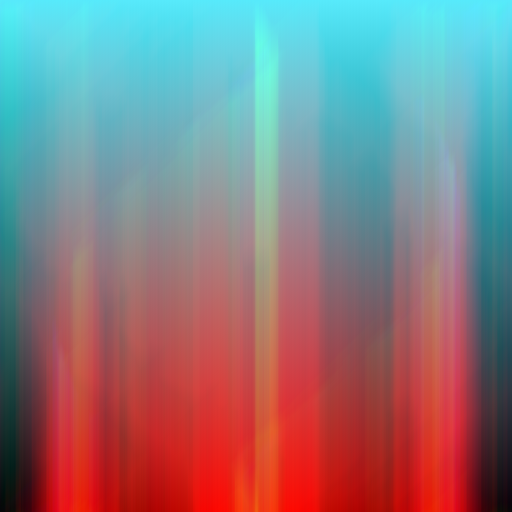}\\
		
		{\small{(a) $(x,t)$ PSF}}&
		{\small{(b) FT amp.}}&
		{\small{(c) FT ph.}}\\

	\end{tabular}
	\caption{\textbf{PSF spectral analysis.} PSFs and the corresponding spectra for a moving point source (the velocity is indicated by the angle at the ($(x,t)$ plane) captured using (top to bottom) static camera, parabolic motion camera, fluttered-shutter camera and our method. (a) $(x,t)$ slice of PSF and its (b) amplitude and (c) phase in Fourier domain.}
	\label{fig:spctAnal_2}
\end{figure}

\begin{figure}[tb]
    \def\pltWd{0.28}
	\centering
	\begin{tabular}{c c c c}

		\includegraphics[width = \pltWd\columnwidth]{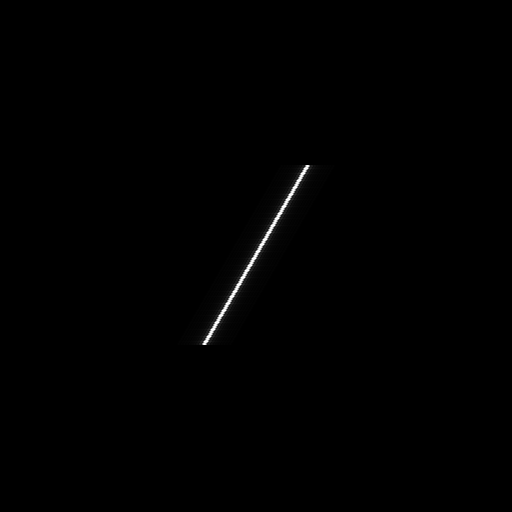} &
		\includegraphics[width = \pltWd\columnwidth]{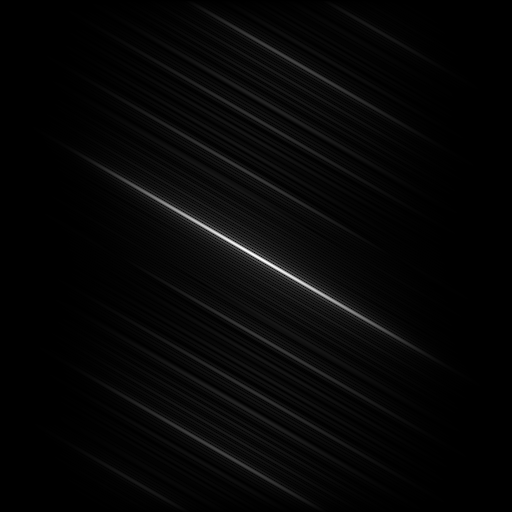} &
		\includegraphics[width = \pltWd\columnwidth]{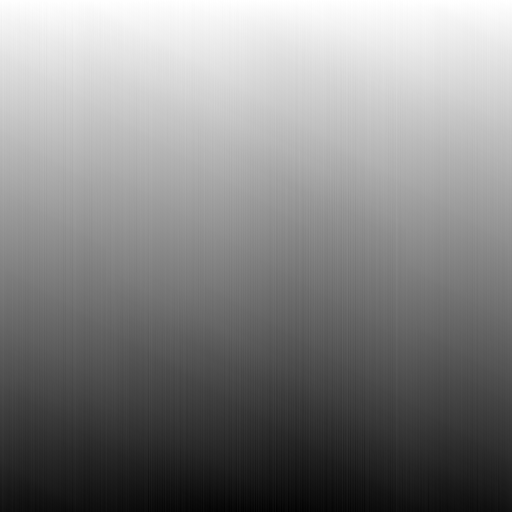}\\

		\includegraphics[width = \pltWd\columnwidth]{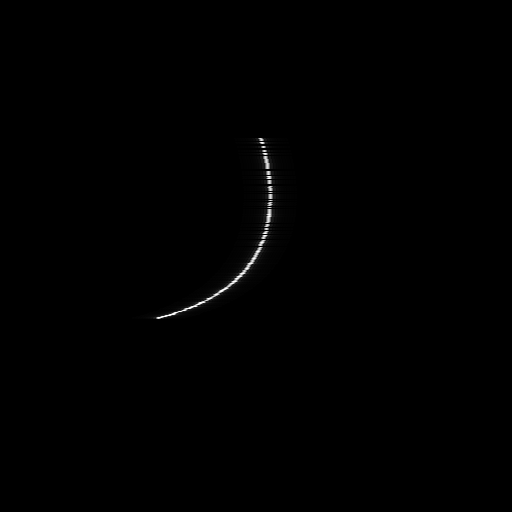} &
		\includegraphics[width = \pltWd\columnwidth]{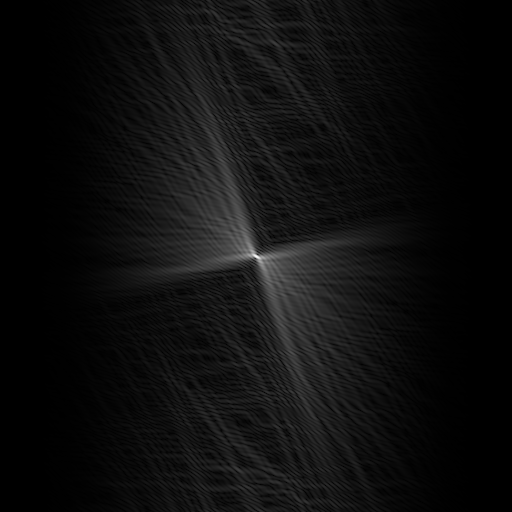} &
		\includegraphics[width = \pltWd\columnwidth]{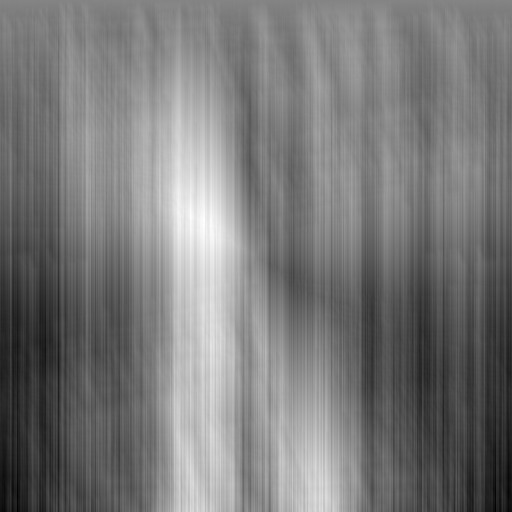}\\

		\includegraphics[width = \pltWd\columnwidth]{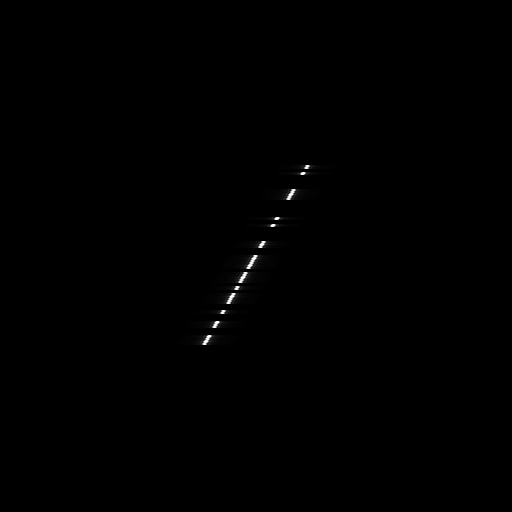} &
		\includegraphics[width = \pltWd\columnwidth]{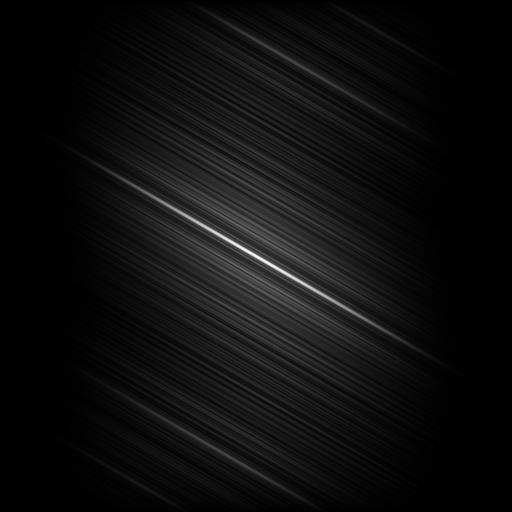} &
		\includegraphics[width = \pltWd\columnwidth]{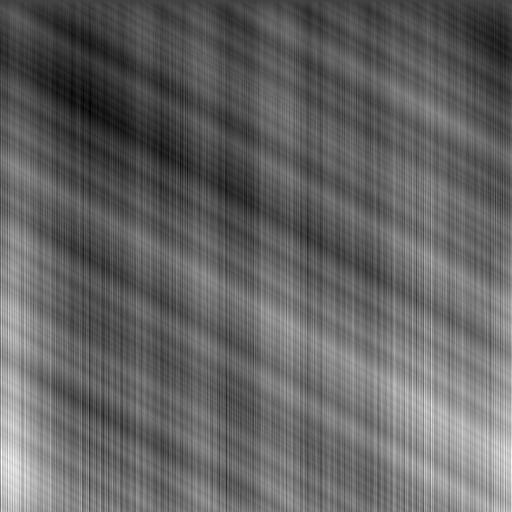}\\

		\includegraphics[width = \pltWd\columnwidth]{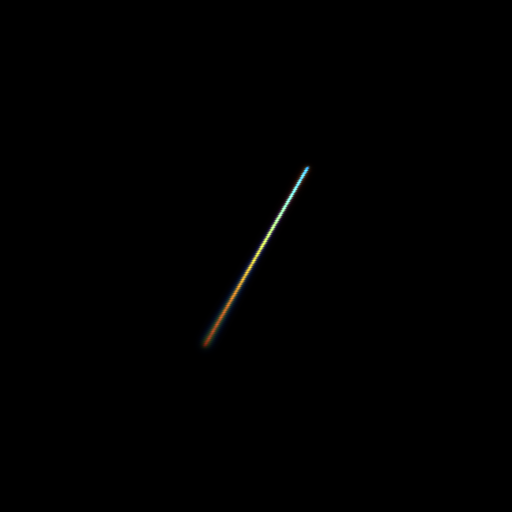} &
		\includegraphics[width = \pltWd\columnwidth]{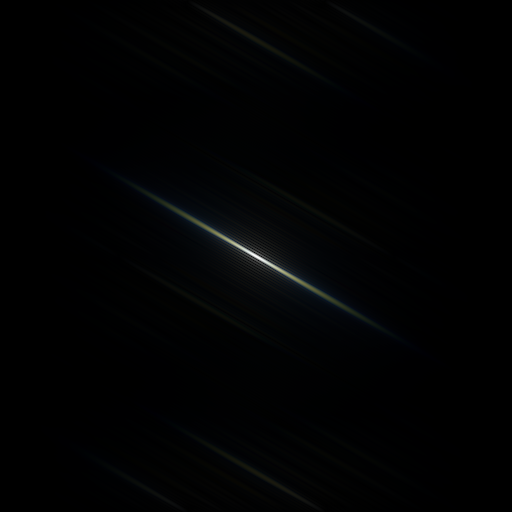} &
		\includegraphics[width = \pltWd\columnwidth]{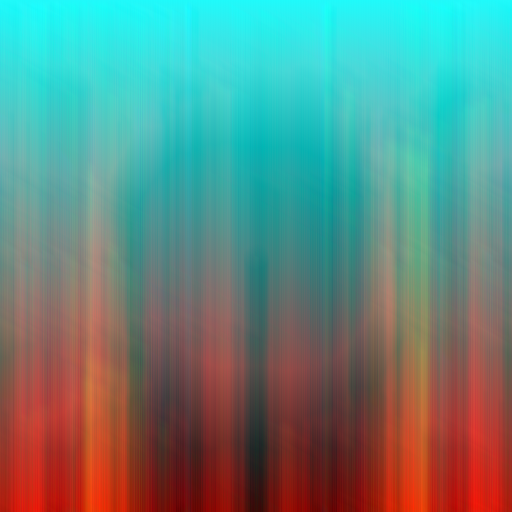}\\
		
		{\small{(a) $(x,t)$ PSF}}&
		{\small{(b) FT amp.}}&
		{\small{(c) FT ph.}}\\

	\end{tabular}
	\caption{\textbf{PSF spectral analysis.} PSFs and the corresponding spectra for a moving point source (the velocity is indicated by the angle at the ($(x,t)$ plane) captured using (top to bottom) static camera, parabolic motion camera, fluttered-shutter camera and our method. (a) $(x,t)$ slice of PSF and its (b) amplitude and (c) phase in Fourier domain.}
	\label{fig:spctAnal_3}
\end{figure}

As discussed in the paper, our proposed PSF can be considered as an infinite sequence of smaller PSFs, each one of a different color. As all PSFs have a similar spatial shape, but each has a different color and different location in the $(x,t)$ plane, the spectrum amplitude is 'white' and similar to the spectrum amplitude of the conventional PSF. Yet, the phase (which holds the shift information) is colored, according to the shift (i.e. spatiotemporal location) of each color. Our spatiotemporal chromatic coupling can be considered as utilization of the spectrum phase as a degree of freedom for the coding. The color variations in the phase indicate the coupling between the color and the trajectory, as can be seen in Fig.~\ref{fig:spctAnal_1} (note that vertical artifacts in the phase plots are due to errors of the phase unwrapping method used in the process). Additional comparisons presenting different velocities (which correspond to different angles in the $(x,t)$) space are presented in Figs.~\ref{fig:spctAnal_2}-\ref{fig:spctAnal_3}. In the conventional camera, no information is encoded in the phase (as can be seen, the three phases are almost the same). The parabolic motion camera is designed to generate a motion invariant PSF, therefore its phase also holds little information (the minor differences are due to the fact that the PSF is not fully motion invariant, due to the finite parabolic motion). Using our method, the different velocities of the source is coded in the different colored pattern of the spectrum phase. Note that the phase of the fluttered-shutter camera PSF indeed contains some motion estimation information (i.e. the temporal code generate phase variations), but this ability holds an estimation-invertibility  trade-off, as mentioned in the paper and discussed in \cite{Agrawal2009_Flutt_2}. In our method, where the color space is utilized for encoding, the motion cues are much stronger and allow improved PSF estimation while preserving PSF invertibility as in each part of the motion at least part of the spectrum is sharp, and can serve as a guide to reconstruct the blurred colors.

\section{CNN structure and details} \label{cnn_det}
    \subsection{U-Net model}
As discussed in the paper, our proposed architecture is based on the known U-net architecture \cite{Unet}. The U-net model includes several downsampling blocks, with their corresponding upsampling operations, which concatenate to their output the input of the same scale downsampling block, thus allowing multiscale processing. We use the U-net model available in \url{https://github.com/milesial/Pytorch-UNet}, with several modifications. A skip-connection is added between the input and the output, thus, letting the 'U' structure to estimate a 'residual' correction to the input blurred image (empirically, this change allows a much faster convergence).

The net contains four downsampling blocks and their corresponding four upsampling blocks, with additional convolutions in the input and output. Each convolution operation consists of $3\times3$ filters (with proper padding to keep the original input size, and without bias), followed by BatchNorm layer and a Leaky-ReLU activation. Each CONV block contains double Conv-BN-ReLU sequence. Downsampling blocks perform convolution and then a $2\times2$ max-pooling operation. Upsampling blocks perform a $\times2$ trainable upsampling (transpose-Conv layer), CONV block, and a concatenation with the corresponding downsampling block's output. The number of filters in each layer appears in Table~\ref{Tab:CNN_det}.

\begin{table}
\begin{center}
\begin{tabular}{|l|c|c|}
\hline
Layer & $N_{in}$ & $N_{out}$ \\
\hline\hline

$CONV_{in}$ & 3 & 64 \\
$DOWN_1$ & 64 & 128 \\
$DOWN_2$ & 128 & 256 \\
$DOWN_3$ & 256 & 512 \\
$DOWN_4$ & 512 & 512 \\
$UP_1$ & 1024 & 256 \\
$UP_2$ & 512 & 128 \\
$UP_3$ & 256 & 64 \\
$UP_4$ & 128 & 64 \\
$CONV_{out}$ & 64 & 3 \\

\hline
\end{tabular}
\end{center}
\caption{\textbf{CNN layers:} The number of filters in each convolution operation.} \label{Tab:CNN_det}
\end{table}

\subsection{Shallow model for ablation study}
As mentioned in the Section 4 in the paper, as part of the ablation study, a shallow model was trained using the same dataset. This test checked the impact of the model depth on the reconstruction performance, and specifically the multiscale operation effect. As mentioned in the paper, nominal performance is achieved with this network structure. However, it still achieves comparable results to a pure-computational model (like \cite{Nah_deepDeblur_gopro}), demonstrating the benefit of the aperture coding- the encoded cues are such strong guidance for the deblurring operation, that a very shallow model achieves comparable performance to a much larger one. 

The shallow model is made of ten consecutive blocks, where each one contains Conv-BN-ReLU layers (without any pooling). A skip connection is made from the input to the output (in similarity to our U-Net structure), what leaves the inner layers to estimate only the residual correction the blurred image. The filter size in each block is $3\times3$, and each layer has 32 filters (besides the output layer, which has only three, to generate the final residual image).


\section{Test-set results} \label{testSet}
    Several examples from the test set of Nah \etal \cite{Nah_deepDeblur_gopro} are presented (Note that the scenes are created using the full GoPro high FPS scenes according to the process presented in the following). For comparison, the performance of Nah \etal \cite{Nah_deepDeblur_gopro} blind deblurring are presented as well. Note that our input images contain both frame summation (for motion simulation) and coded spatial blur (for the aperture phase coding simulation), in contrast to the input of \cite{Nah_deepDeblur_gopro}, which simulates only motion blur (by summing up the same frame sequence without any additional blur). The sharp image used as the deblurring target is the middle frame of each sequence. 

\begin{figure*}[tb]
	\centering
	\begin{tabular}{c c c}
		\includegraphics[width = 0.6\columnwidth]{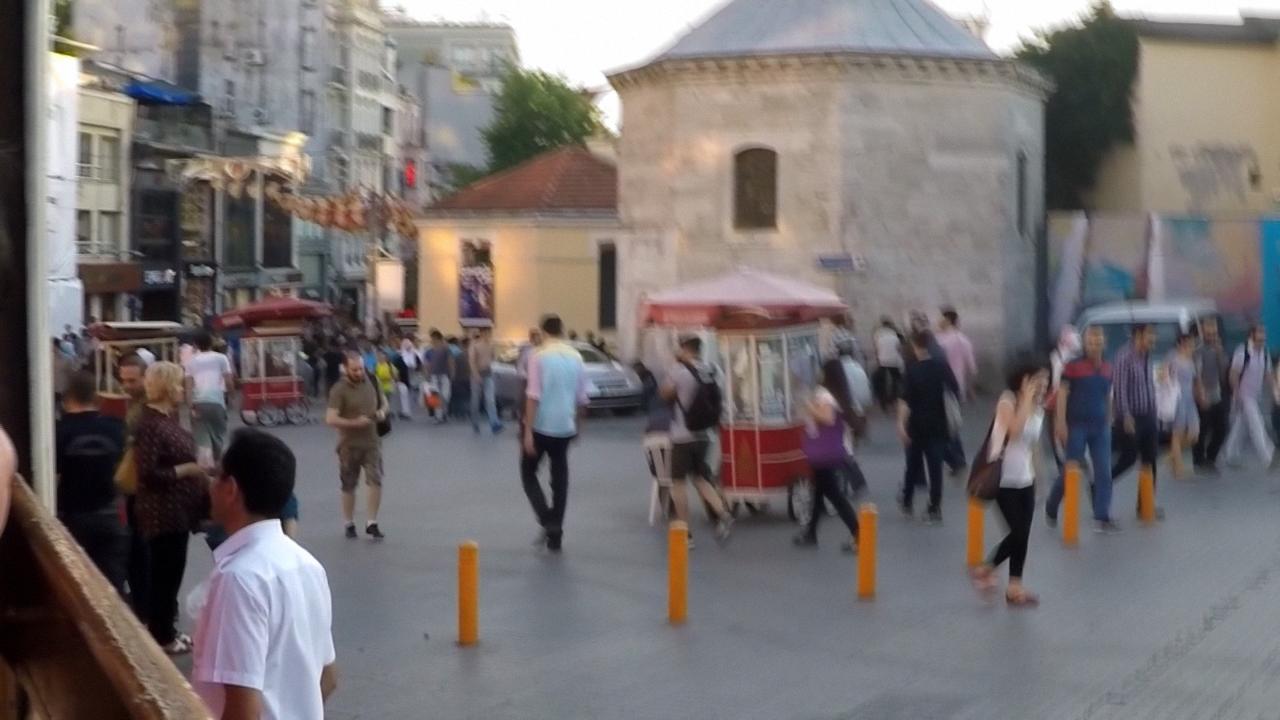} &
		\includegraphics[width = 0.6\columnwidth]{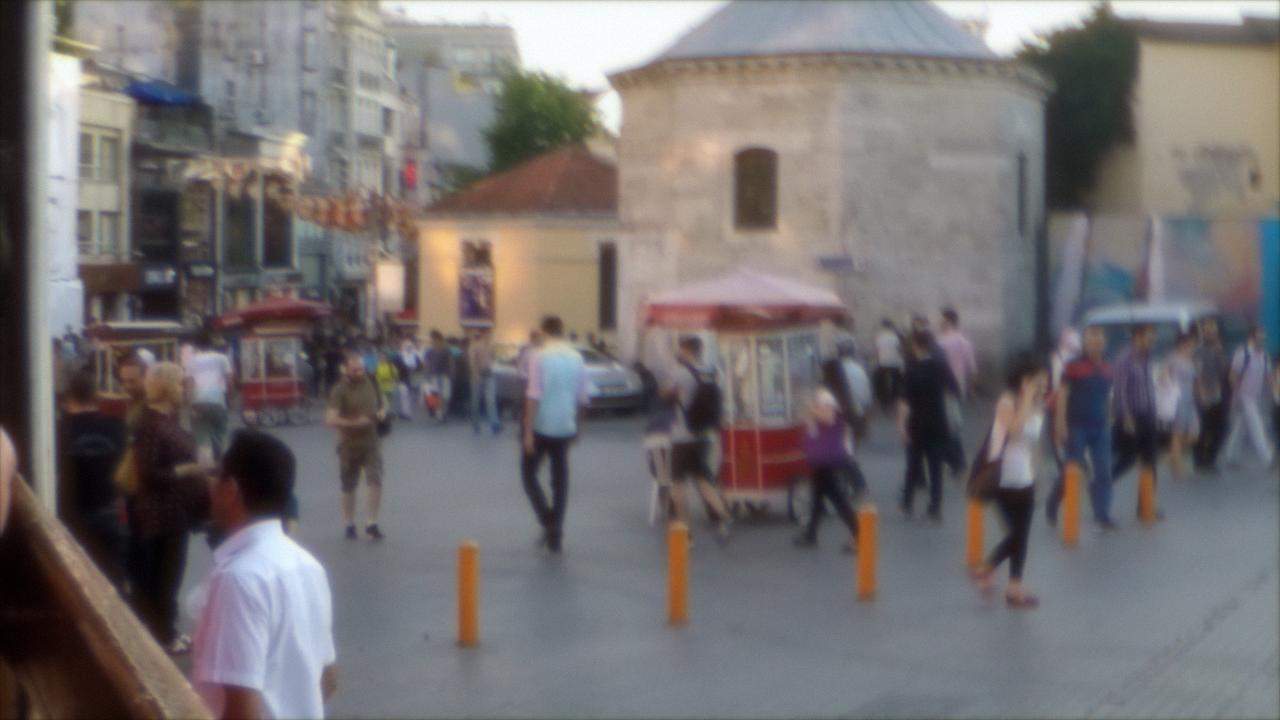} &
		\\
		
	    {\small{(a) Conventional image}}&
		{\small{(b) Our coded image}} &
		\\
		
		\includegraphics[width = 0.6\columnwidth]{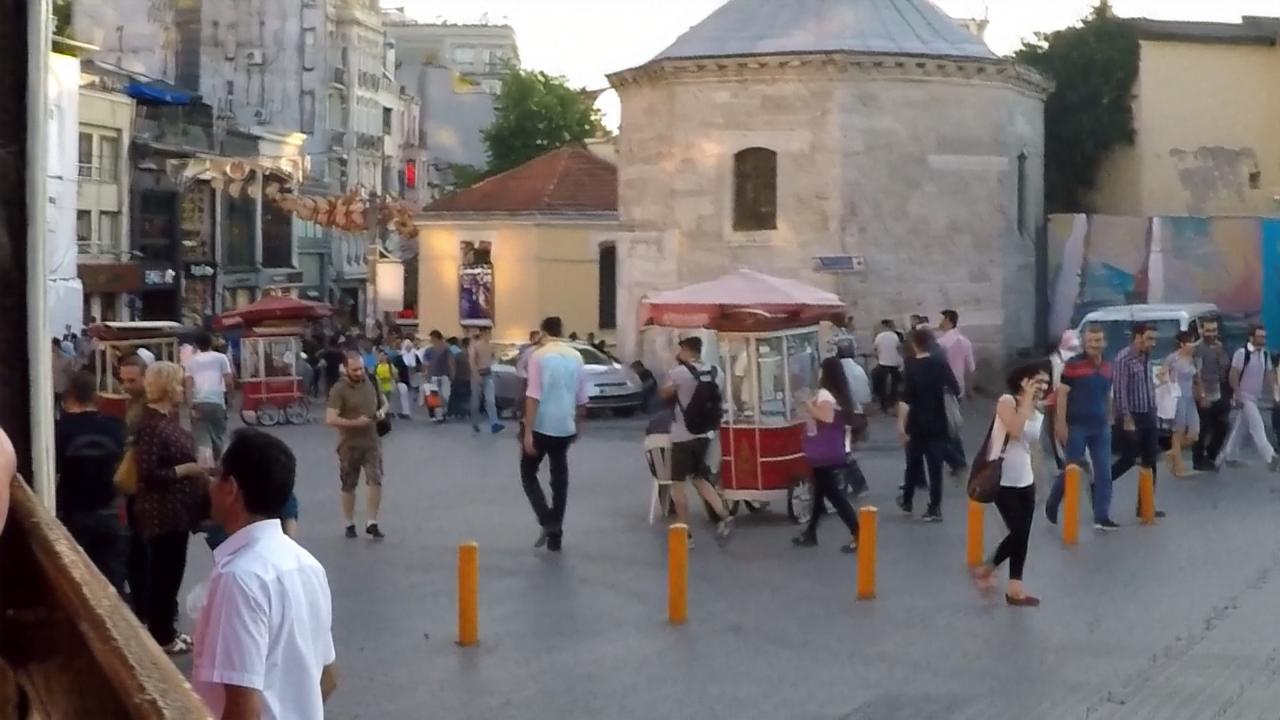} &
		\includegraphics[width = 0.6\columnwidth]{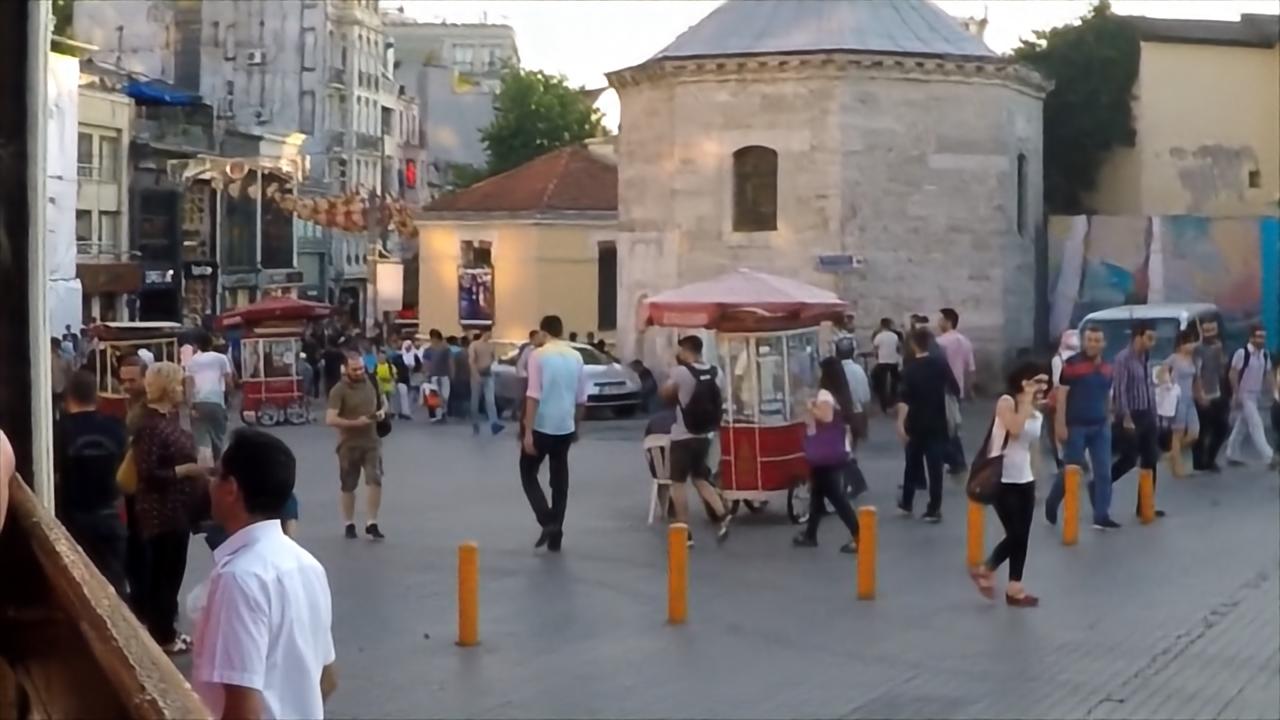} &
		\includegraphics[width = 0.6\columnwidth]{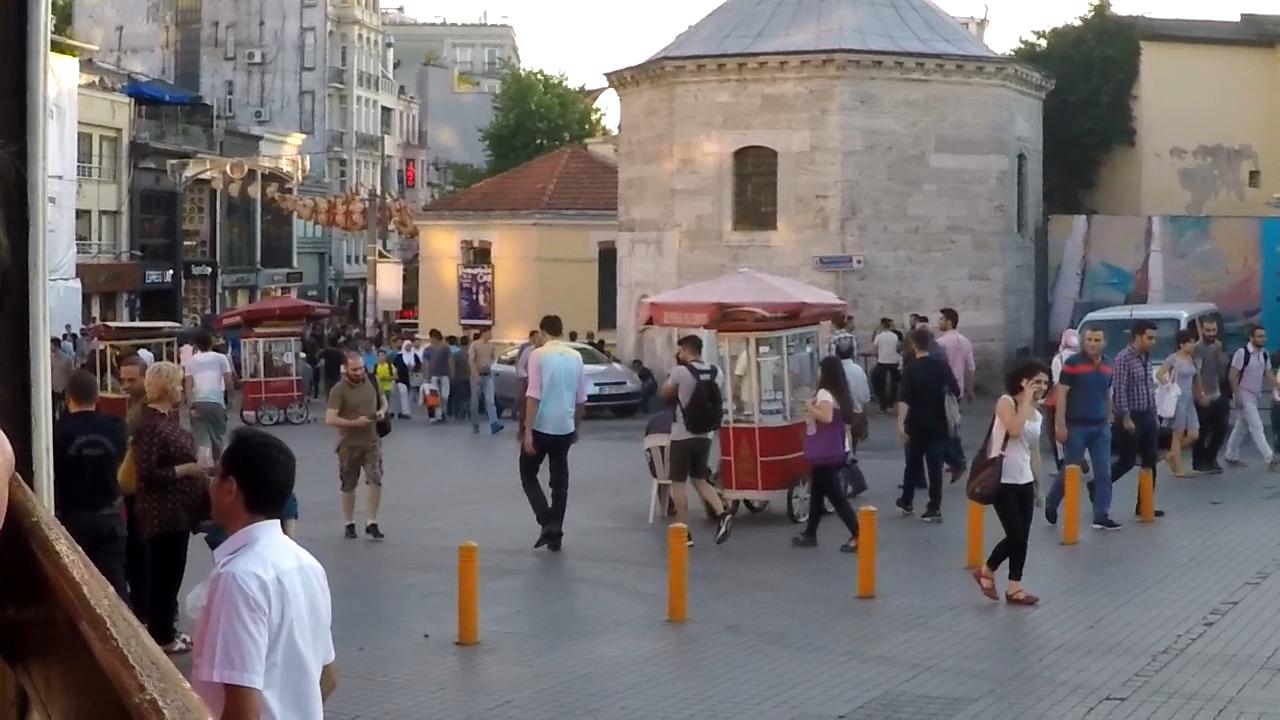}\\
		
		{\small{(c) Nah \etal rec. $(27.5, 0.91)$}}&
		{\small{(d) Our rec., $(31.4, 0.95)$}} &
		{\small{(e) GT}}\\

	\end{tabular}
	\caption{\textbf{Test set results:} A visual motion deblurring example with a sequence of 11 frames and a noise level of $\sigma=3$. (a) Conventional motion blur, (b) Our image (color-coded motion), (c) Deblurring results using (a) and \cite{Nah_deepDeblur_gopro}, (d) our deblurring results and (e) Ground truth image.}
	\label{fig:testSet_1}
\end{figure*}

Several representing examples for different sequence lengths and noise levels are presented in Figs.~\ref{fig:testSet_1}-\ref{fig:testSet_4}. Our reconstruction results achieve higher PSNR/SSIM in all cases. Note that the additional diffraction-related blur that exists in our images causes our CNN to deblur a more difficult task then the reference task solved in \cite{Nah_deepDeblur_gopro}. In a case where a similar spatial blur (without the motion-color cues) is introduced to the input images of \cite{Nah_deepDeblur_gopro} (with a corresponding re-training of their CNN), the advantage of our method is expected to be even larger. 

\begin{figure*}[tb]
	\centering
	\begin{tabular}{c c c}
		\includegraphics[width = 0.6\columnwidth]{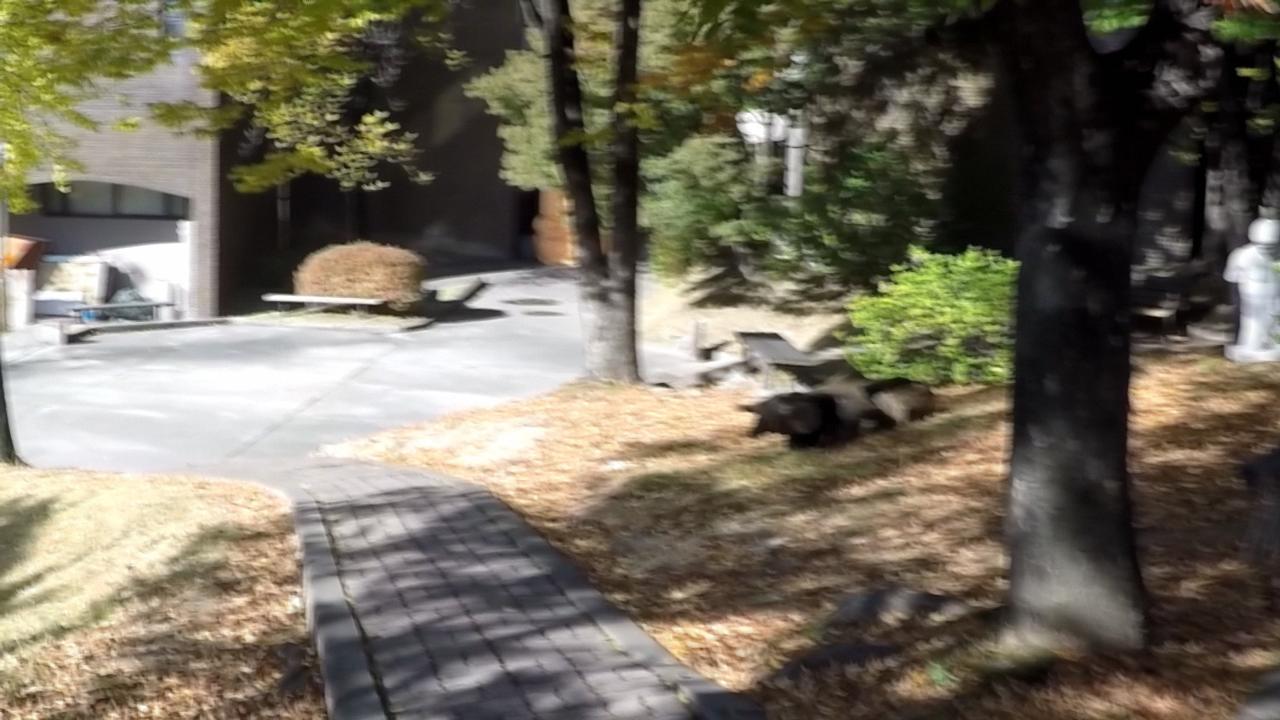} &
		\includegraphics[width = 0.6\columnwidth]{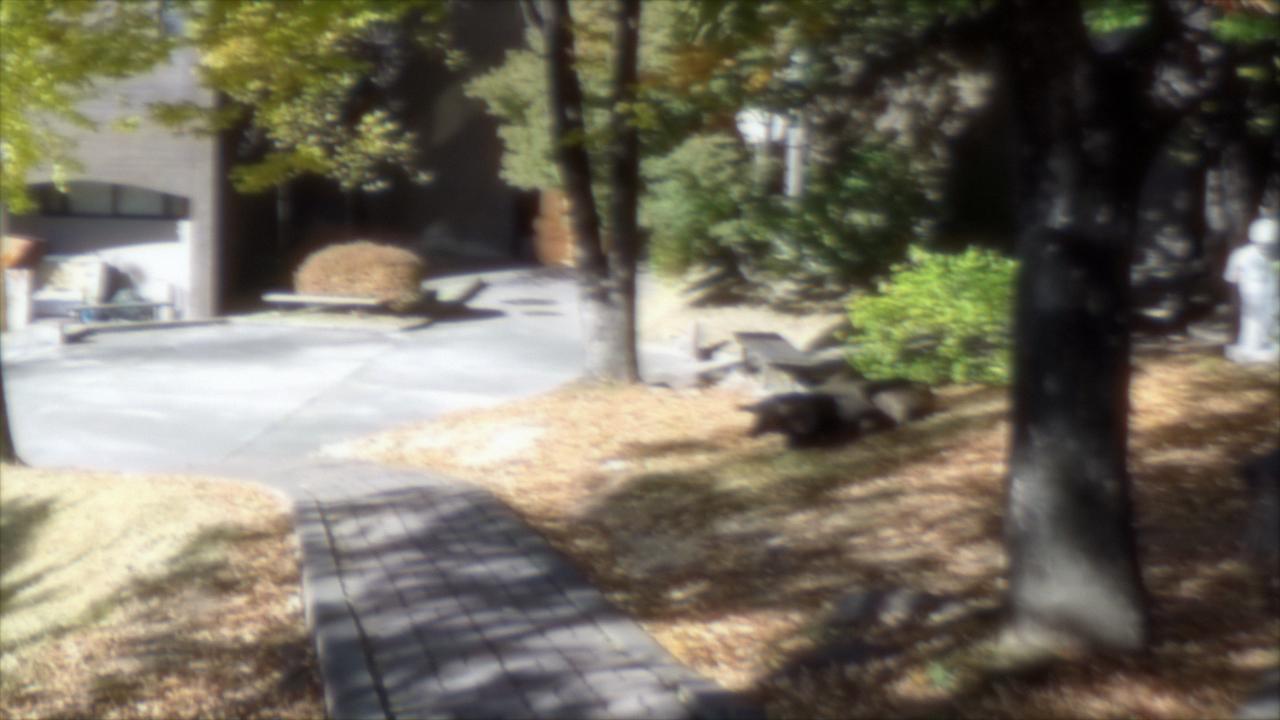} &
		\\
		
	    {\small{(a) Ref. image}}&
		{\small{(b) Our image}} &
		\\
		
		\includegraphics[width = 0.6\columnwidth]{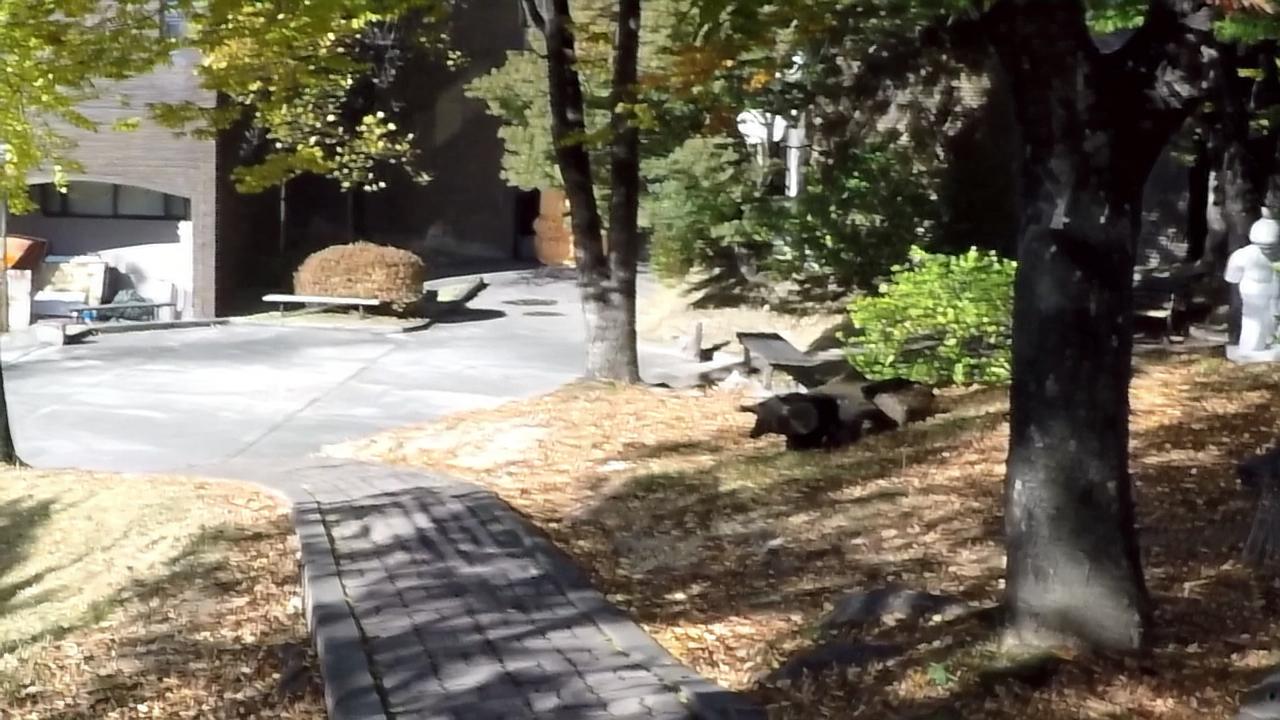} &
		\includegraphics[width = 0.6\columnwidth]{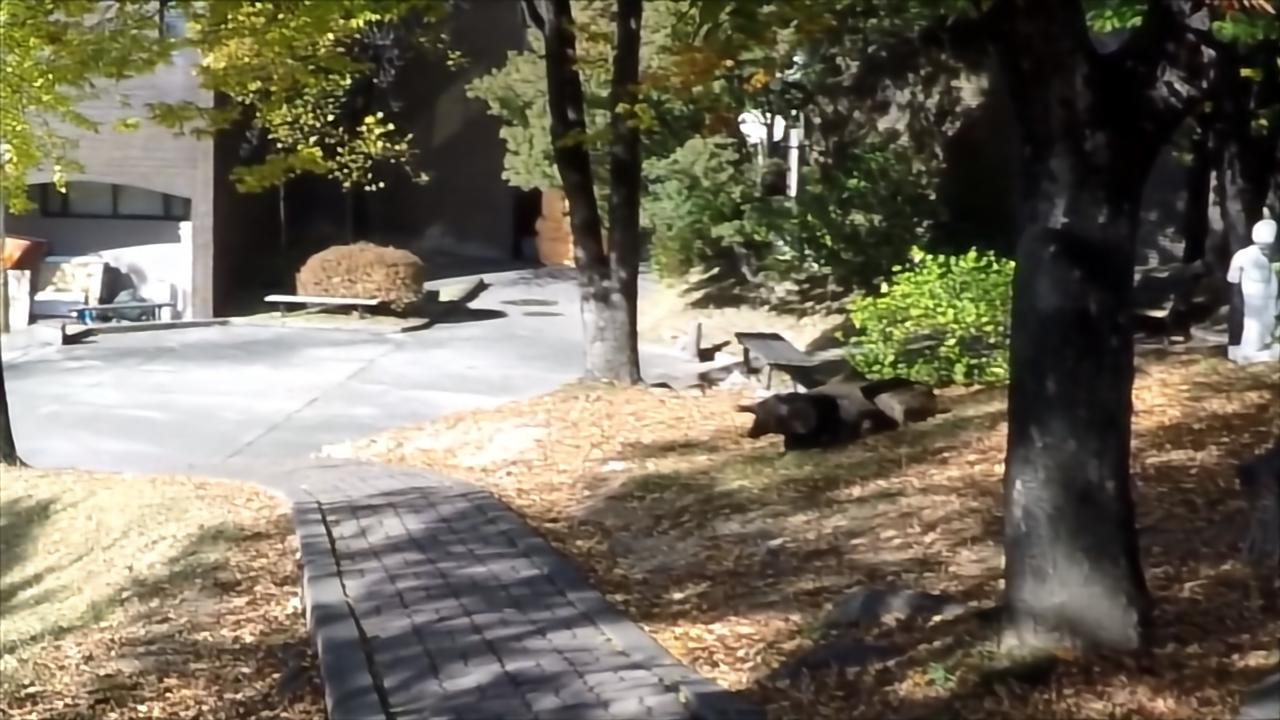} &
		\includegraphics[width = 0.6\columnwidth]{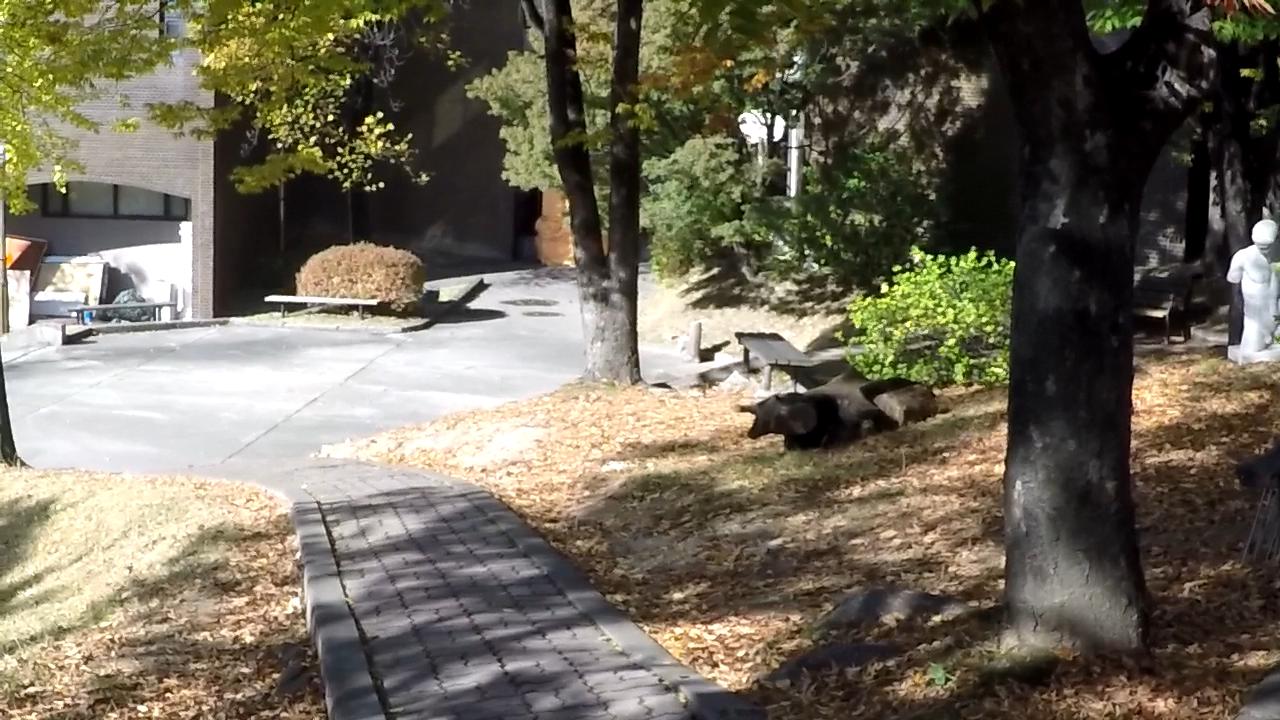}\\
		
		{\small{(c) Ref rec., $(24.2, 0.85)$}}&
		{\small{(d) Our rec., $(28.45, 0.92)$}} &
		{\small{(e) GT}}\\

	\end{tabular}
	\caption{\textbf{Test set results:} A visual motion deblurring example with a sequence of 13 frames and a noise level of $\sigma=2$. ((a) Conventional motion blur, (b) Our image (color-coded motion), (c) Deblurring results using (a) and \cite{Nah_deepDeblur_gopro}, (d) our deblurring results and (e) Ground truth image.}
\end{figure*}

\begin{figure*}[tb]
	\centering
	\begin{tabular}{c c c}
		\includegraphics[width = 0.6\columnwidth]{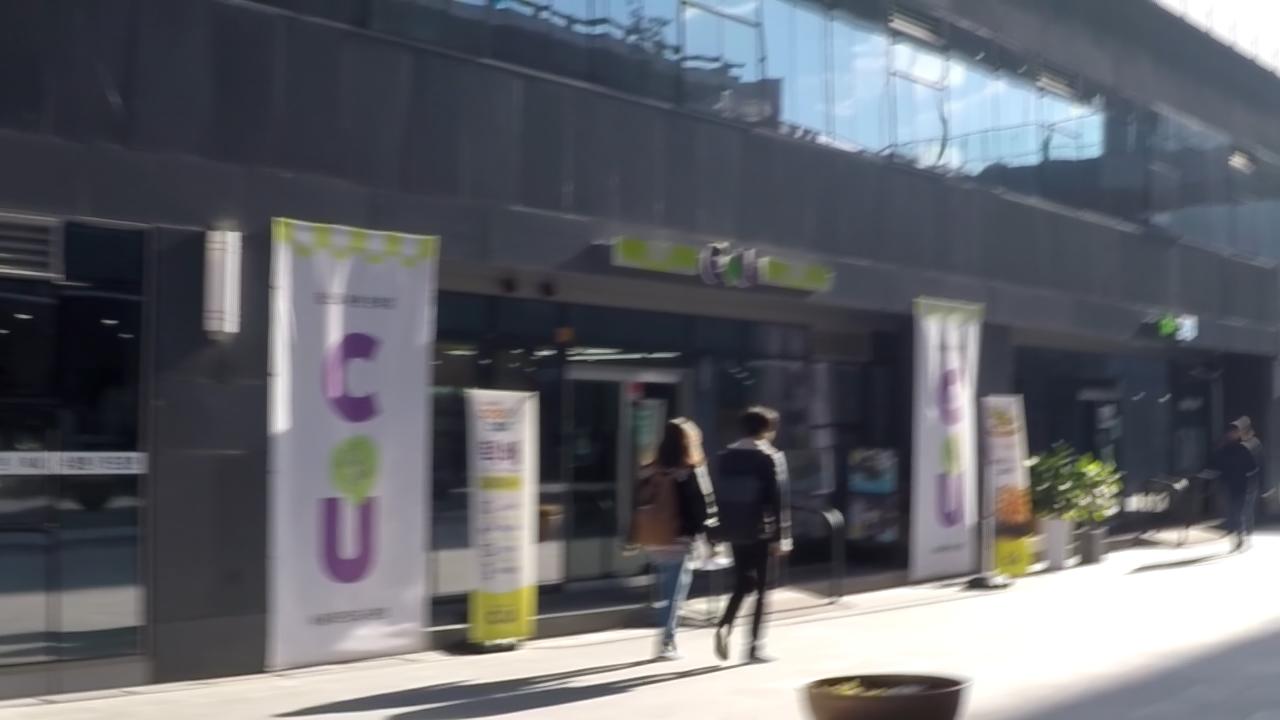} &
		\includegraphics[width = 0.6\columnwidth]{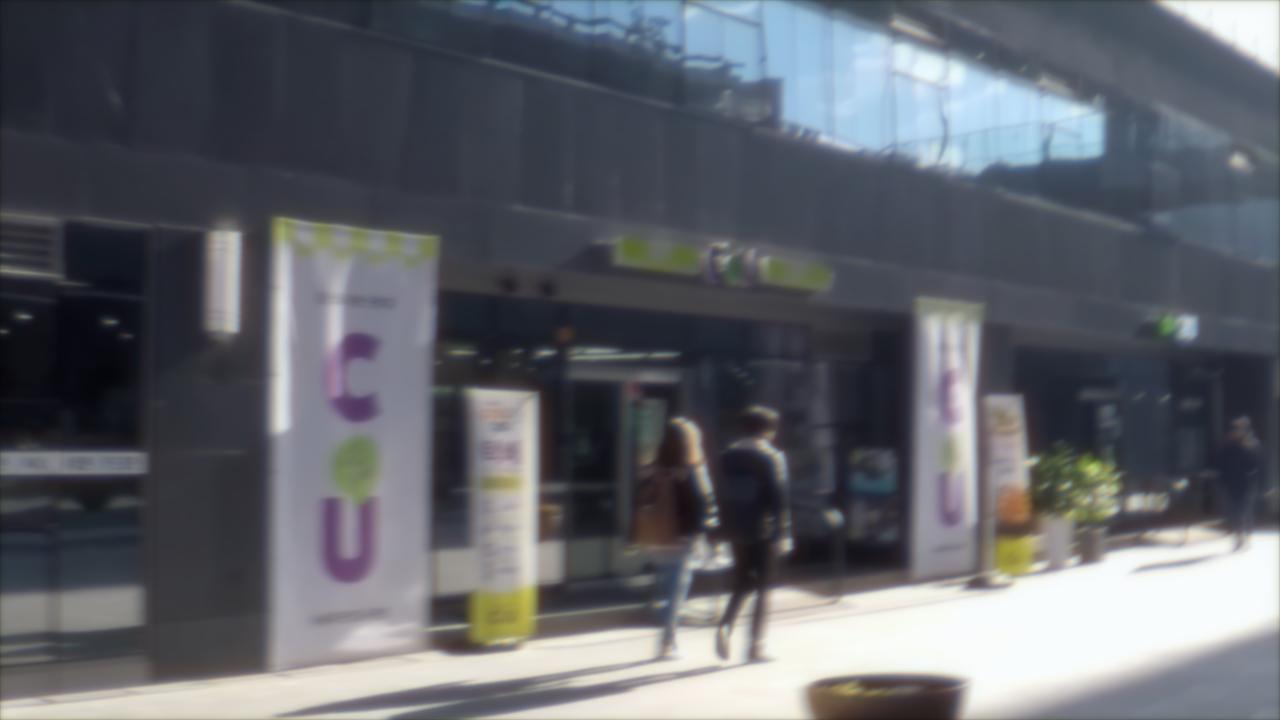} &
		\\
		
	    {\small{(a) Ref. image}}&
		{\small{(b) Our image}} &
		\\
		
		\includegraphics[width = 0.6\columnwidth]{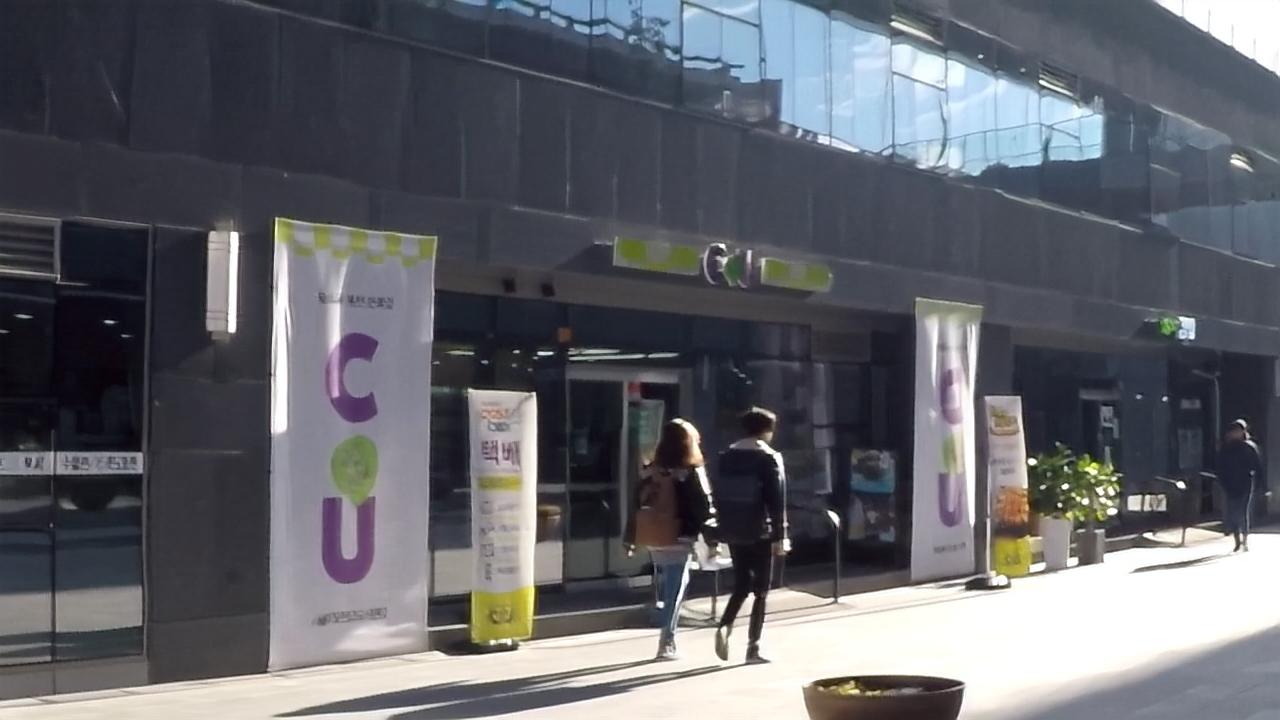} &
		\includegraphics[width = 0.6\columnwidth]{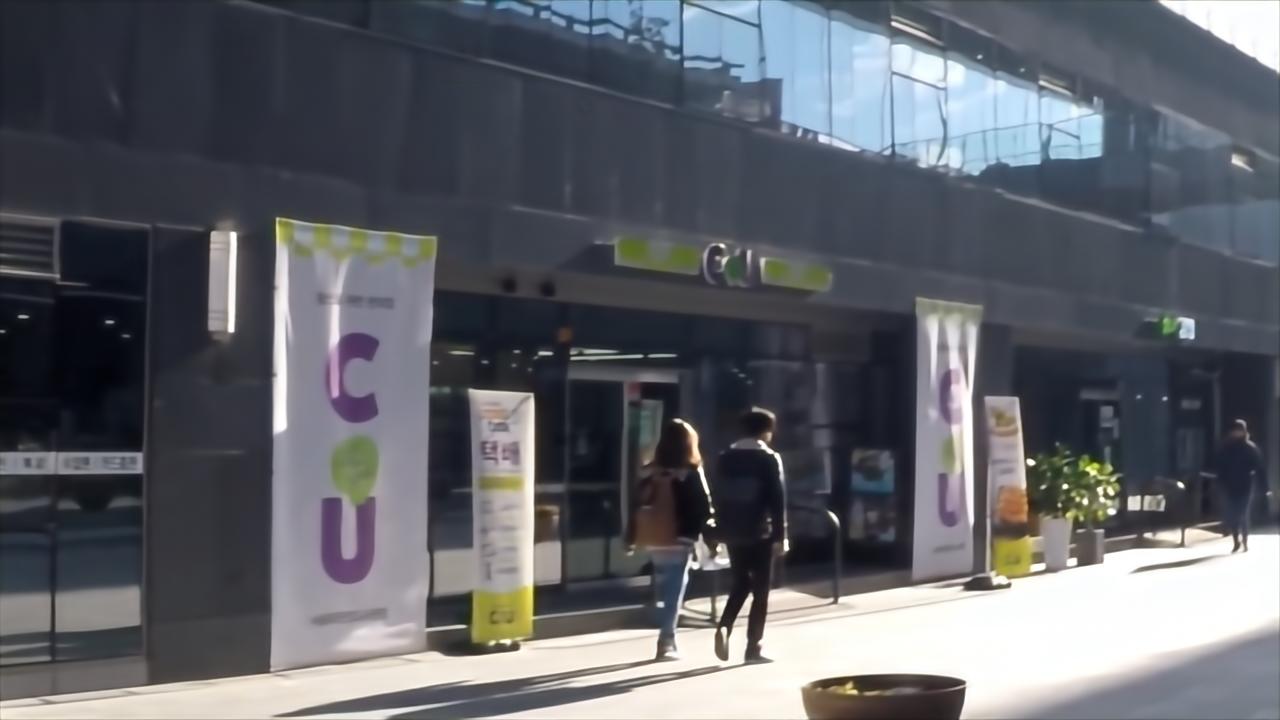} &
		\includegraphics[width = 0.6\columnwidth]{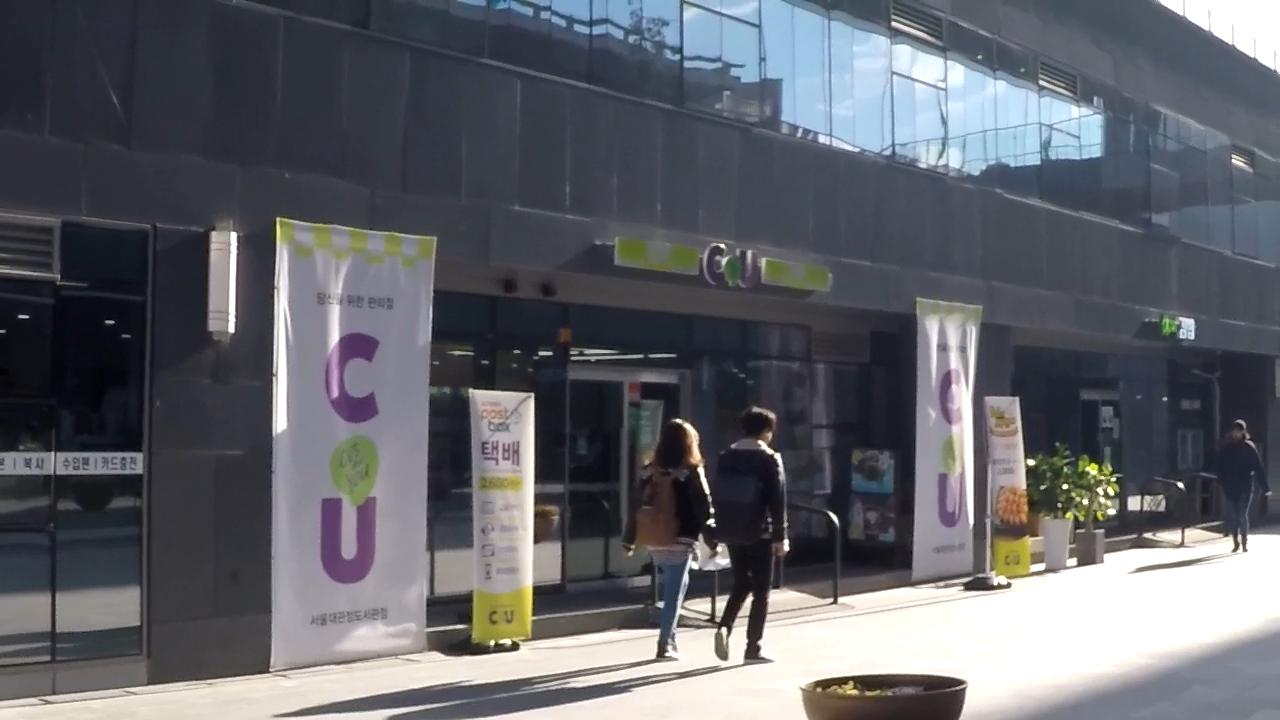}\\
		
		{\small{(c) Ref rec., $(27.4, 0.95)$}}&
		{\small{(d) Our rec., $(33.1, 0.97)$}} &
		{\small{(e) GT}}\\

	\end{tabular}
	\caption{\textbf{Test set results:} A visual motion deblurring example with a sequence of 7 frames and no noise. (a) Conventional motion blur, (b) Our image (color-coded motion), (c) Deblurring results using (a) and \cite{Nah_deepDeblur_gopro}, (d) our deblurring results and (e) Ground truth image.}
	\label{fig:testSet_3}
\end{figure*}

\begin{figure*}[tb]
	\centering
	\begin{tabular}{c c c}
		\includegraphics[width = 0.6\columnwidth]{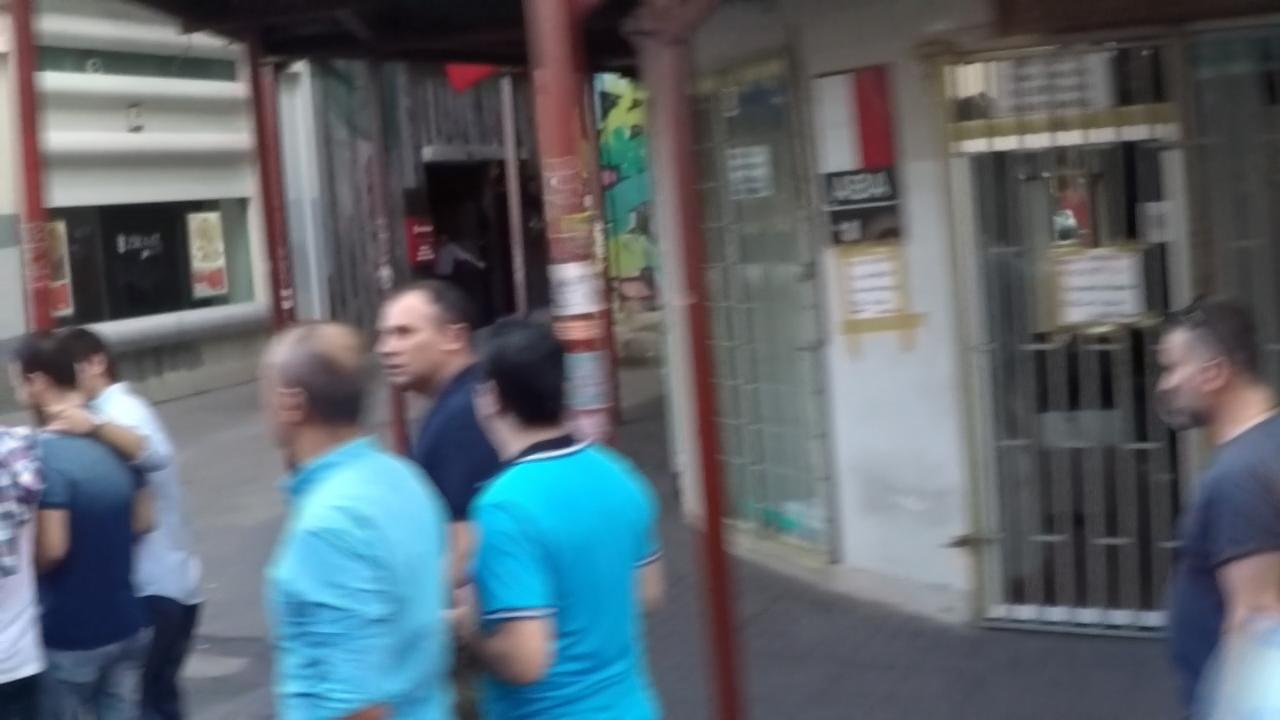} &
		\includegraphics[width = 0.6\columnwidth]{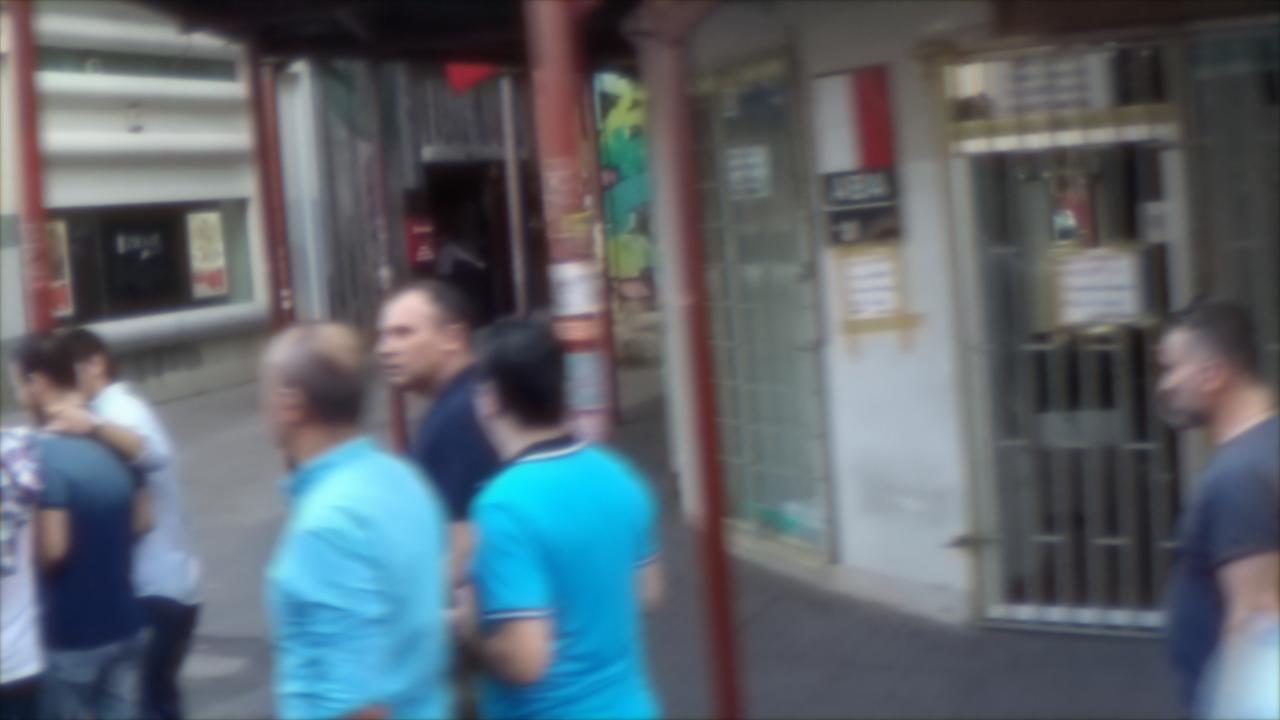} &
		\\
		
	    {\small{(a) Ref. image}}&
		{\small{(b) Our image}} &
		\\
		
		\includegraphics[width = 0.6\columnwidth]{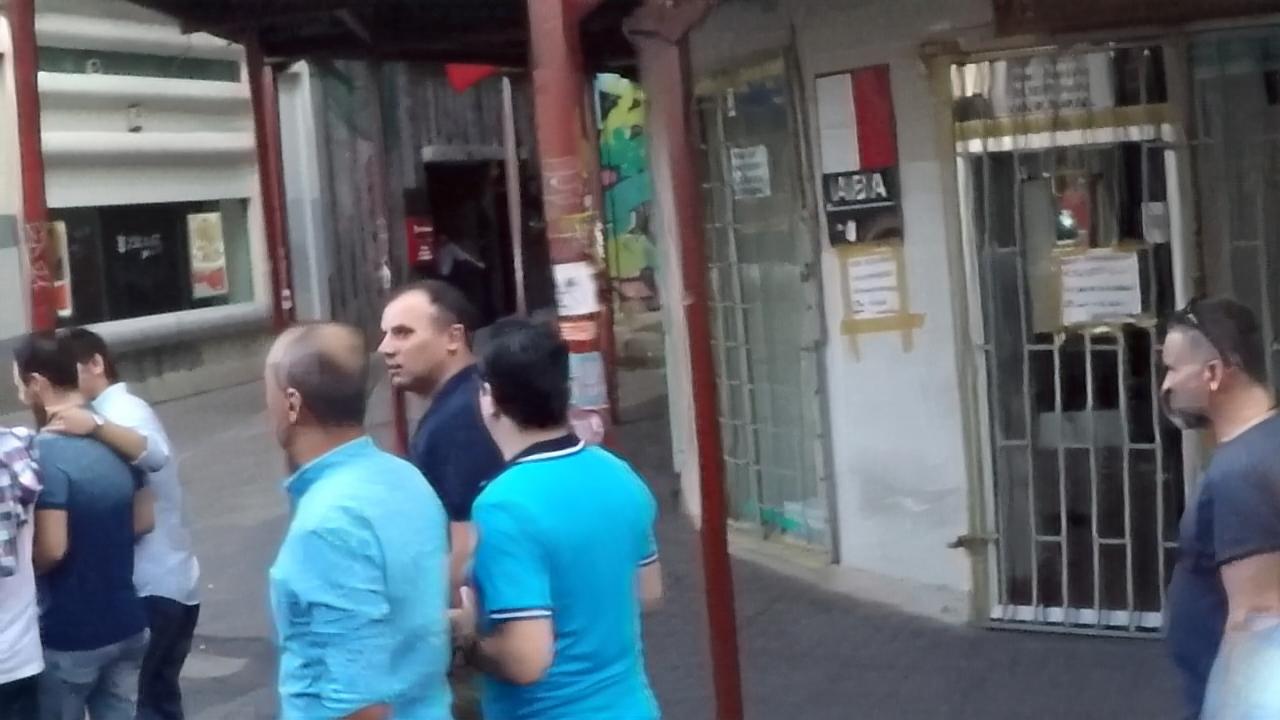} &
		\includegraphics[width = 0.6\columnwidth]{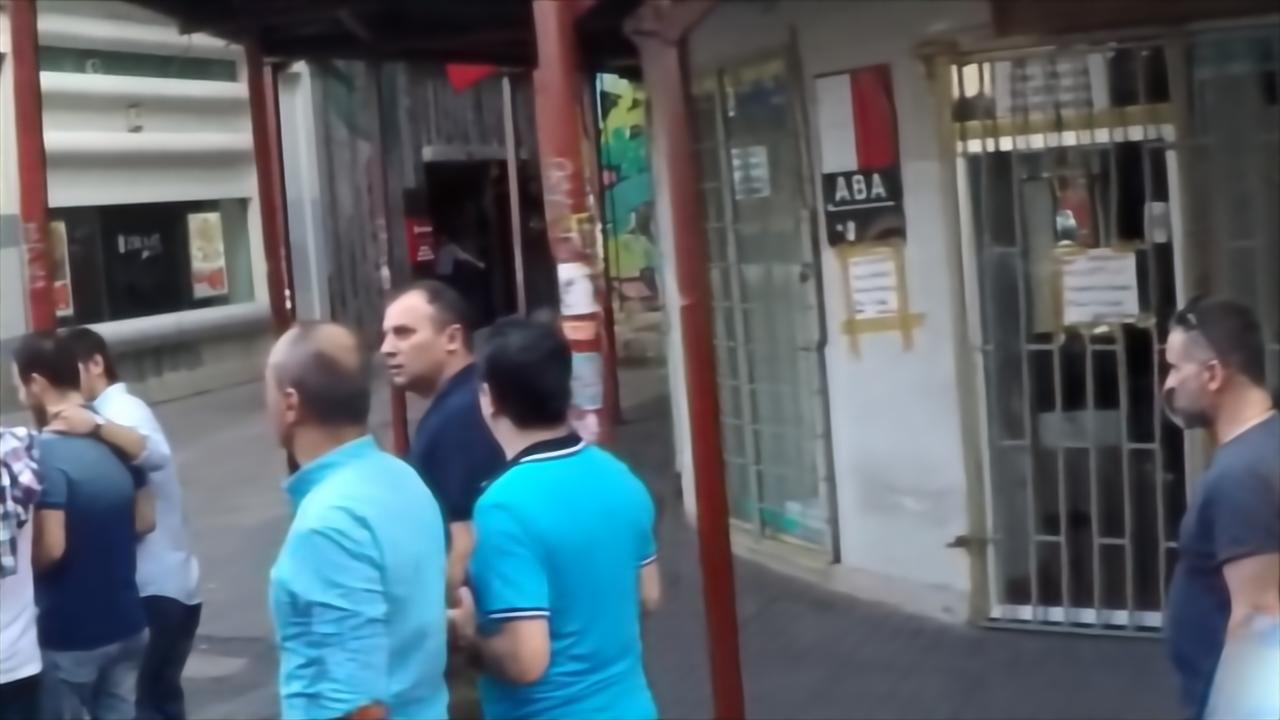} &
		\includegraphics[width = 0.6\columnwidth]{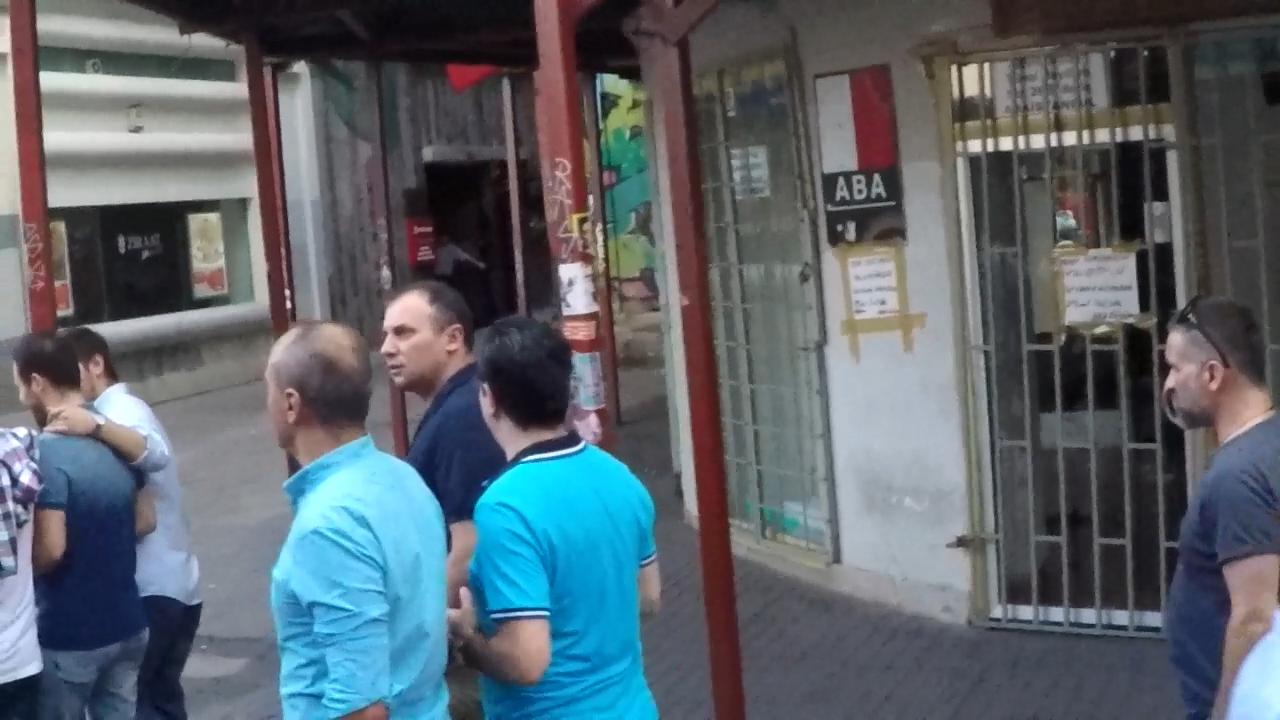}\\
		
		{\small{(c) Ref rec., $(26.5, 0.91)$}}&
		{\small{(d) Our rec., $(30.3, 0.94)$}} &
		{\small{(e) GT}}\\

	\end{tabular}
	\caption{\textbf{Test set results:} A visual motion deblurring example with a sequence of 9 frames and a noise level of $\sigma=1$. (a) Conventional motion blur, (b) Our image (color-coded motion), (c) Deblurring results using (a) and \cite{Nah_deepDeblur_gopro}, (d) our deblurring results and (e) Ground truth image.}
	\label{fig:testSet_4}
\end{figure*}

\section{Quantitative comparison statistics} \label{stat}

In Section 5.2 of the main paper, a quantitative comparison between our method and the blind deblurring method of Nah \etal \cite{Nah_deepDeblur_gopro} is presented. The test scenes are created using the test-set of the GoPro dataset, presented in \cite{Nah_deepDeblur_gopro}. Sequences of 7-13 frames are used. To simulate our camera, each image is blurred with its corresponding color-coded blur kernel (according to its relevant time in the sequence), and then all the frames are summed-up. The input for the blind deblurring method of \cite{Nah_deepDeblur_gopro} is just the sum of the consecutive frames, as performed in \cite{Nah_deepDeblur_gopro} (in both cases, the proper gamma-related transformation is applied, as discussed in \cite{Nah_deepDeblur_gopro}). The corresponding inputs (i.e. the scenes created using the same frame sequence) are deblurred using the relevant algorithm, after adding AWGN with $\sigma=[0,3]$ on a $[0,255]$ scale. As mentioned in the paper and in the previous section, in our case an additional diffraction-related spatial blur is added, therefore our method handles a more difficult task, and still achieves better performance. 

The global average performance over all the noise levels is presented in the paper. The per-noise level statistics is presented in Tables~\ref{Tab:goProStats_sig0}-\ref{Tab:goProStats_sig3}. Note that our advantage over the method presented in \cite{Nah_deepDeblur_gopro} is similar for all the tested noise levels.
As can be clearly seen, the deviation from the global average is not significant, which means that the noise sensitivity of both methods is not high in the $\sigma=[0,3]$ domain (which is relatively low, and therefore simulates good light conditions). Our model converged well for higher levels of noise (up to $\sigma=9$). However, since the model of Nah \etal was not trained for higher noise levels than $\sigma=2$, comparison in these noise levels is not fair. 

\begin{table}
\begin{center}
\begin{tabular}{|l|c|c|}
\hline
$N_{frames}$ & Nah \etal & Ours \\
\hline\hline

$N=7$ & $28.1/0.93$ & $\bf{30.9}/\bf{0.95}$ \\
$N=9$ & $27/0.91$ & $\bf{30}/\bf{0.94}$ \\
$N=11$ & $26/0.89$ & $\bf{28.9}/\bf{0.92}$ \\
$N=13$ & $24.9/0.88$ & $\bf{28}/\bf{0.91}$ \\

\hline
\end{tabular}
\end{center}
\caption{\textbf{Quantitative comparison to blind deblurring:} PSNR/SSIM comparison between the method presented in \cite{Nah_deepDeblur_gopro} and our method, for various lengths of motion ($N_{frames}$) and no noise} \label{Tab:goProStats_sig0}
\end{table}

\begin{table}
\begin{center}
\begin{tabular}{|l|c|c|}
\hline
$N_{frames}$ & Nah \etal & Ours \\
\hline\hline

$N=7$  & $28.1/0.93$ & $\bf{30.9}/\bf{0.95}$ \\
$N=9$  & $27/0.91$   & $\bf{30}/\bf{0.94}$ \\
$N=11$ & $25.9/0.89$ & $\bf{29}/\bf{0.92}$ \\
$N=13$ & $24.9/0.87$ & $\bf{28}/\bf{0.91}$ \\

\hline
\end{tabular}
\end{center}
\caption{\textbf{Quantitative comparison to blind deblurring:} PSNR/SSIM comparison between the method presented in \cite{Nah_deepDeblur_gopro} and our method, for various lengths of motion ($N_{frames}$) and noise level of $\sigma=1$} \label{Tab:goProStats_sig1}
\end{table}

\begin{table}
\begin{center}
\begin{tabular}{|l|c|c|}
\hline
$N_{frames}$ & Nah \etal & Ours \\
\hline\hline

$N=7$ & $28/0.93$ & $\bf{30.9}/\bf{0.95}$ \\
$N=9$ & $27/0.91$ & $\bf{30}/\bf{0.93}$ \\
$N=11$ & $26/0.89$ & $\bf{29}/\bf{0.92}$ \\
$N=13$ & $24.9/0.86$ & $\bf{28}/\bf{0.91}$ \\

\hline
\end{tabular}
\end{center}
\caption{\textbf{Quantitative comparison to blind deblurring:} PSNR/SSIM comparison between the method presented in \cite{Nah_deepDeblur_gopro} and our method, for various lengths of motion ($N_{frames}$) and noise level of $\sigma=2$} \label{Tab:goProStats_sig2}
\end{table}

\begin{table}
\begin{center}
\begin{tabular}{|l|c|c|}
\hline
$N_{frames}$ & Nah \etal & Ours \\
\hline\hline

$N=7$  & $28/0.92$   & $\bf{30.8}/\bf{0.94}$ \\
$N=9$  & $26.9/0.9$ & $\bf{29.9}/\bf{0.93}$ \\
$N=11$ & $25.9/0.88$ & $\bf{28.9}/\bf{0.92}$ \\
$N=13$ & $24.9/0.86$ & $\bf{27.9}/\bf{0.91}$ \\

\hline
\end{tabular}
\end{center}
\caption{\textbf{Quantitative comparison to blind deblurring:} PSNR/SSIM comparison between the method presented in \cite{Nah_deepDeblur_gopro} and our method, for various lengths of motion ($N_{frames}$) and noise level of $\sigma=3$} \label{Tab:goProStats_sig3}
\end{table}

\section{Experimental setup description} \label{expSetup}
    As discussed in Section~3 of the paper, the required spatiotemporal coding is achieved using two components: (i) the passive component, which is a phase-mask containing concentric phase rings, similar to the mask used in \cite{IEEE_depth}; and (ii) the active component, which is a gradual change in the focus setting during exposure. 

The phase-mask contains two phase rings. The first ring with normalized radii of ${\bf{r_1}}=[0.55, 0.8]$ and phase-shift $\phi_1=6.5[rad]$, and the second ring with radii of ${\bf{r_2}}=[0.8, 1]$ and phase-shift $\phi_2=13.2[rad]$ (the phase-shifts are for the blue wavelength of $\lambda=455[nm]$, which serves as the reference for both $\phi$ and $\psi$). As this mask has been designed for depth reconstruction \cite{IEEE_depth}, it achieves a strong chromatic separation between different defocus conditions. By analyzing the results presented in \cite{IEEE_depth}, the strongest chromatic separation is achieved in the defocus domain between $\psi=[0,8]$ (see Section 3 of the paper for the definition of $\psi$). 

\begin{figure}[ht]
\begin{center}
\includegraphics[width=0.9\linewidth]{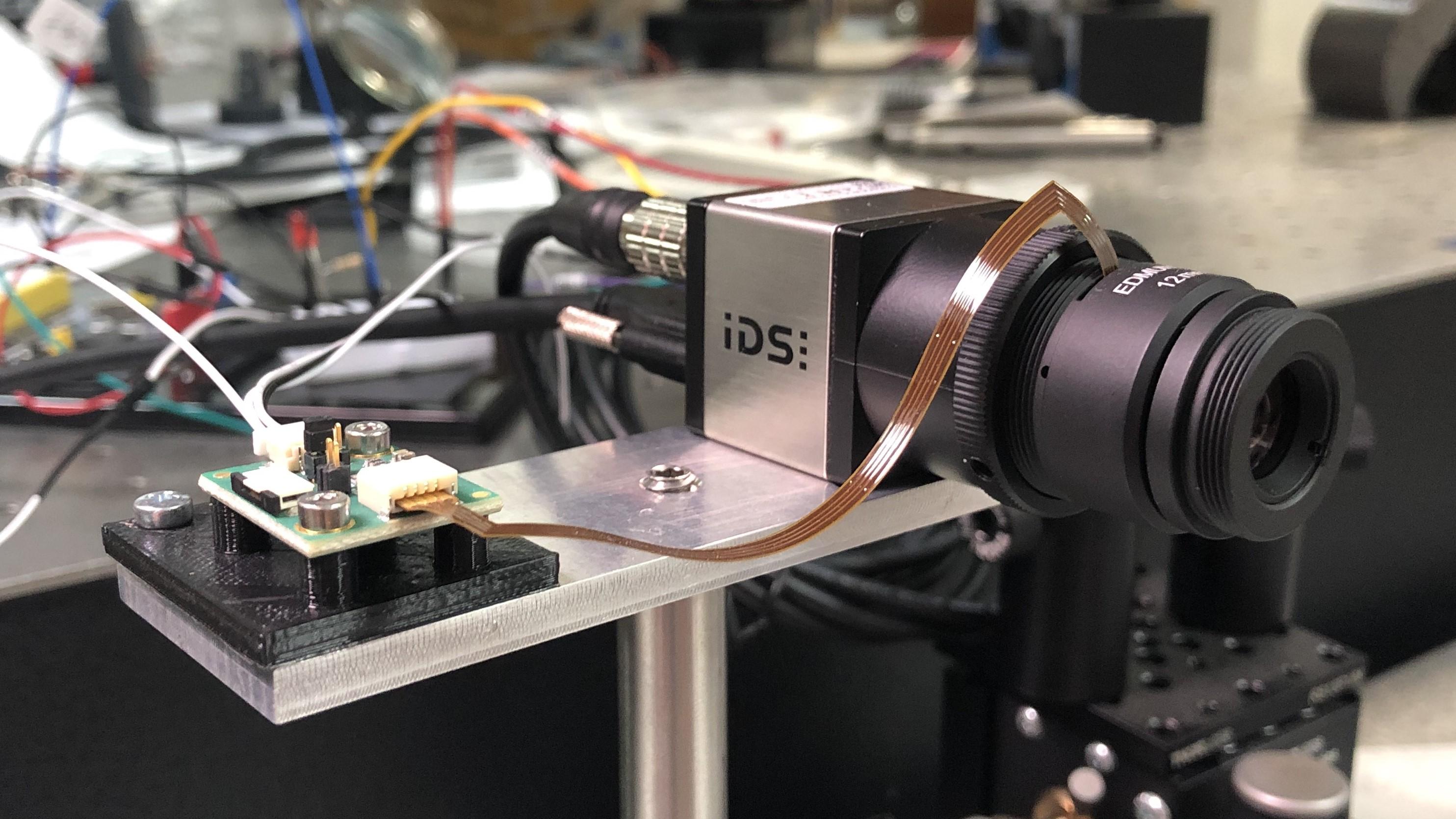}
\end{center}
   \caption{\textbf{The table-top experimental setup:} The C-mount lens is incorporated with both the phase-mask and the focusing lens. A micro-controller synchronizes the focus variation to the frame exposure using the camera flash signal.}
\label{fig:setup_supp}
\end{figure}

\begin{figure}[ht]
\begin{center}
\includegraphics[width=0.9\linewidth]{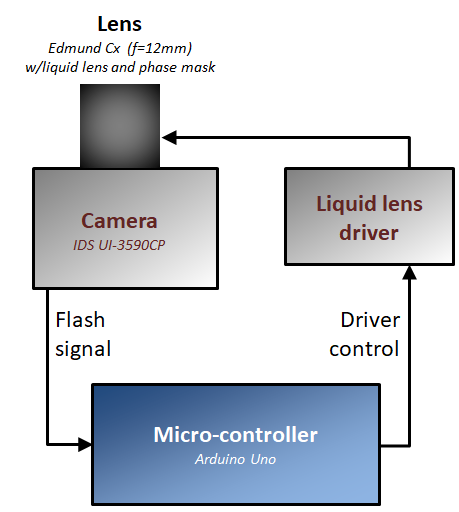}
\end{center}
   \caption{\textbf{Block diagram of the experimental setup:} Following the setup photo presented in Fig~\ref{fig:setup_supp}, the block diagram describes the setup structure and inter-connections.}
\label{fig:blkDgrm}
\end{figure}

The implementation of the designed spatiotemporal coding is illustrated in Figs.~\ref{fig:setup_supp}-\ref{fig:blkDgrm}. The phase-mask is incorporated in a lens with $f=12mm$ (Edmund Cx C-mount lens \#33-632). Following the phase-mask and $\psi$ domain selection, the second component (the dynamic focus variation) is set to achieve the proper changes. The C-mount lens is equipped with a liquid focusing lens (Corning Varioptic A-25H0 lens), which allows electronic focus setting without any moving parts. Calibration is performed to find the proper liquid lens settings that result in the desired $\psi=[0,8]$ focus variation of the C-mount lens. Then the loop is closed using a micro-controller (Arduino Uno), which is connected to the camera and also to the liquid lens driver. The camera outputs a flash signal, indicating the start of the exposure. This signal is used to trigger the liquid lend to start the focus variation. The focus variation is calibrated so that it takes place gradually during the entire exposure time.

\section{Additional experimental results} \label{expRes}
    \subsection{Comparison to other coding methods}

\begin{figure}[tb]
    \def\spkSz{0.4}
	\centering
	\begin{tabular}{ c c c}
	
		\includegraphics[width = \spkSz\columnwidth]{Figures/spkWArr.png} &
		\\
		{\small{(a) Rotating target}}\\
		
		\includegraphics[width = \spkSz\columnwidth]{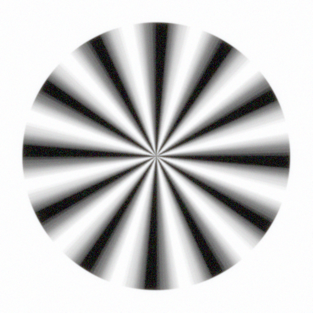} &
		\includegraphics[width = \spkSz\columnwidth]{Figures/flt_spk_rec_rs.png}\\
		{\small{(b) Fluttered-shutter img.}}&
		{\small{(c) Fluttered-shutter rec.}}\\
		
		\includegraphics[width = \spkSz\columnwidth]{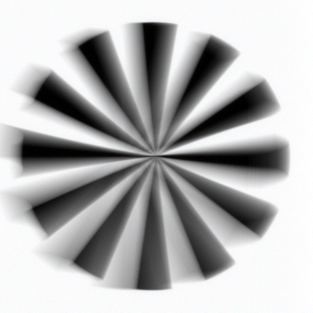} &
		\includegraphics[width = \spkSz\columnwidth]{Figures/prb_spk_rec.png}\\
		{\small{(d) Parabolic motion img.}}&
		{\small{(e) Parabolic motion rec.}}\\
		
		\includegraphics[width = \spkSz\columnwidth]{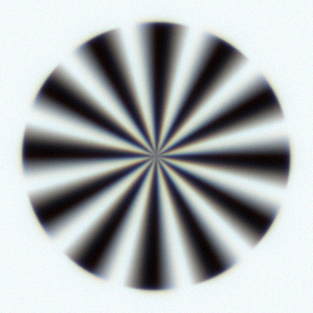} &
		\includegraphics[width = \spkSz\columnwidth]{Figures/our_spk_Rec.png}\\
		{\small{(f) Our img.}}&
		{\small{(g) Our rec.}}\\

	\end{tabular}
	\caption{\textbf{Simulation results of rotating target:} (a) rotating target and the intermediate images vs. reconstruction results for (b,c) fluttered-shutter, (d,e) parabolic motion camera and (f,g) our method.}
	\label{fig:sim_res_full}
\end{figure}

Following the comparison to other coding methods presented in Section~5.1 of the paper, the full results (including the intermediate images) are presented in Fig.~\ref{fig:sim_res_full}.

\subsection{Outdoor experimental results}

\begin{figure}[tb]
    
	\centering
	\includegraphics[width = 1\columnwidth]{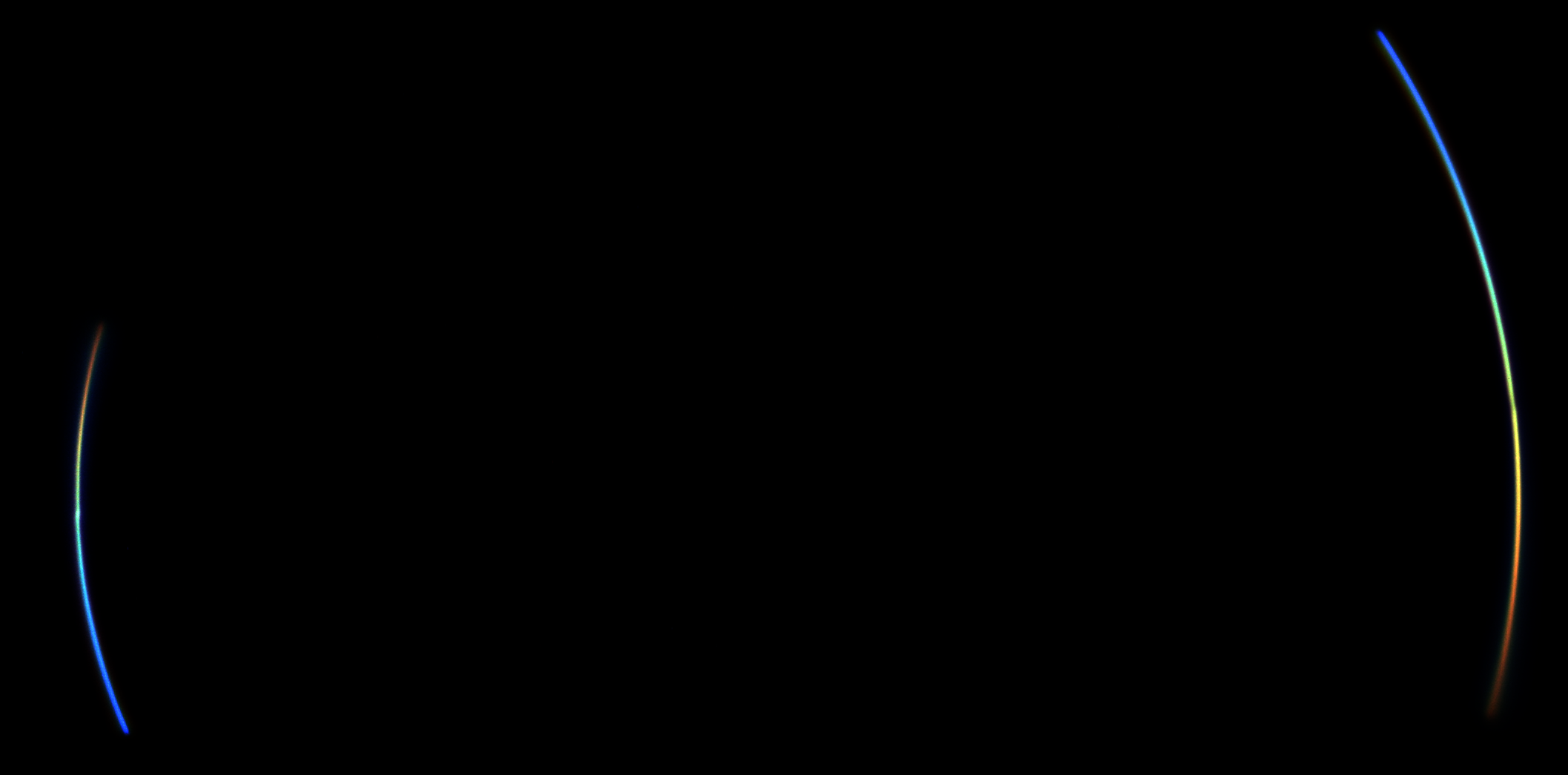} 
	
	\caption{\textbf{PSF encoding validation experiment:} two rotating white LEDs simulating point sources captured using our camera. The color coded motion trace indicates both direction and velocity.}
	\label{fig:LEDs}
		
\end{figure}

A table-top experimental setup that demonstrates our spatiotemporal aperture coding method is built (see details in the main paper and previous section). As mentioned in Section~5.3 of the paper, an experimental validation of the PSF encoding is performed, using a rotating wheel with two white LEDs (simulating point sources). The full image of the LEDs is presented in Fig.~\ref{fig:LEDs}. The color coded motion of each LED is clearly visible; the order of the colors indicates the direction, and the extent of the color trace is a cue for the object velocity. 

\begin{figure*}[tb]
    \def\outSz{0.85}
	\centering
	\begin{tabular}{c c}
		
		\includegraphics[width = \outSz\columnwidth]{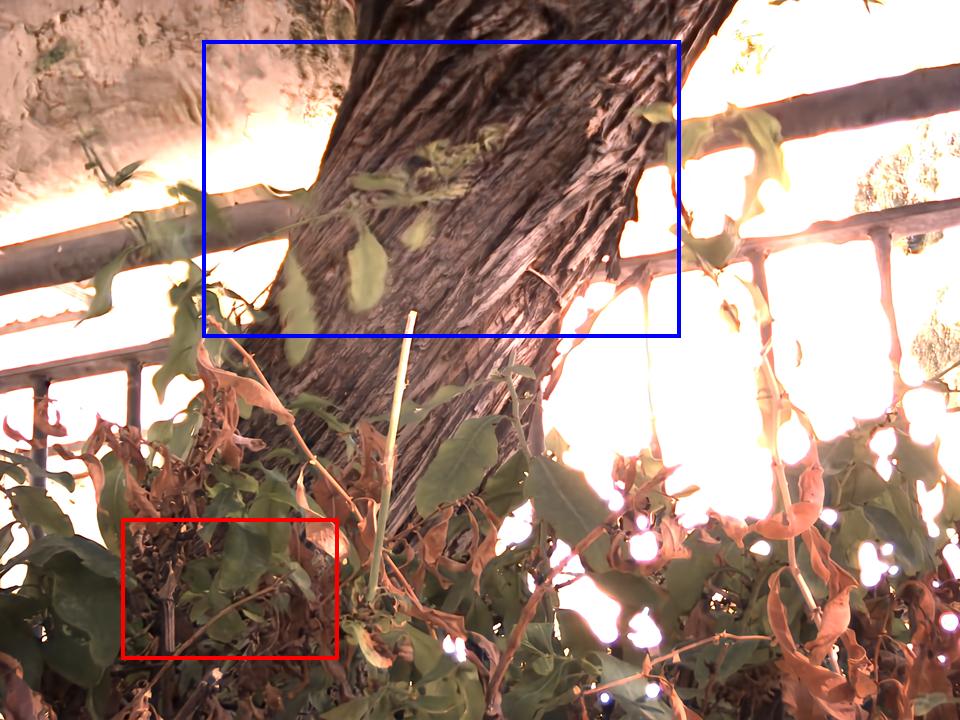}&
		\includegraphics[width = 1\columnwidth]{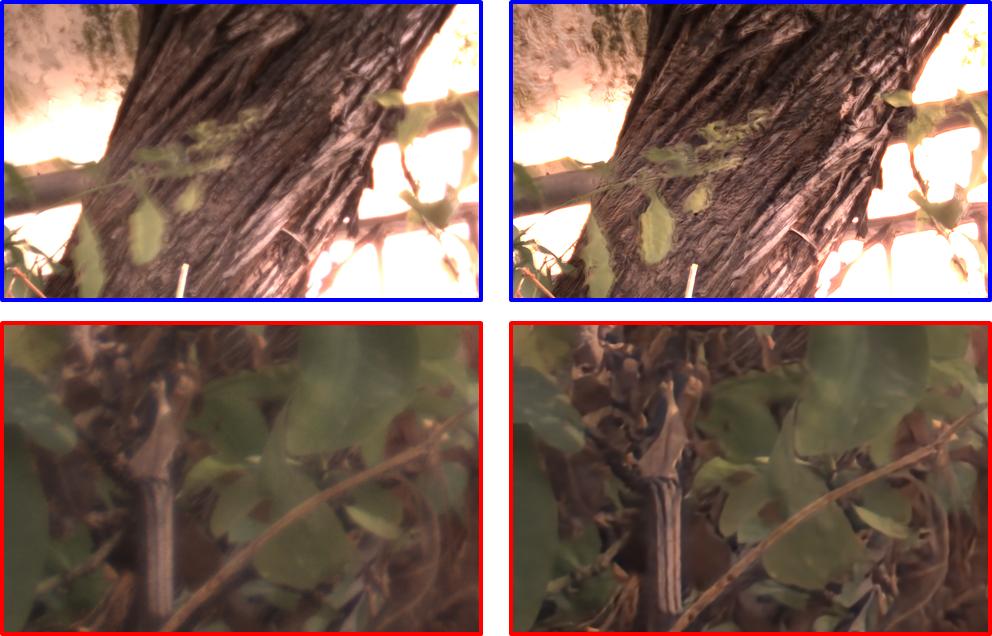}\\
		
		\includegraphics[width = \outSz\columnwidth]{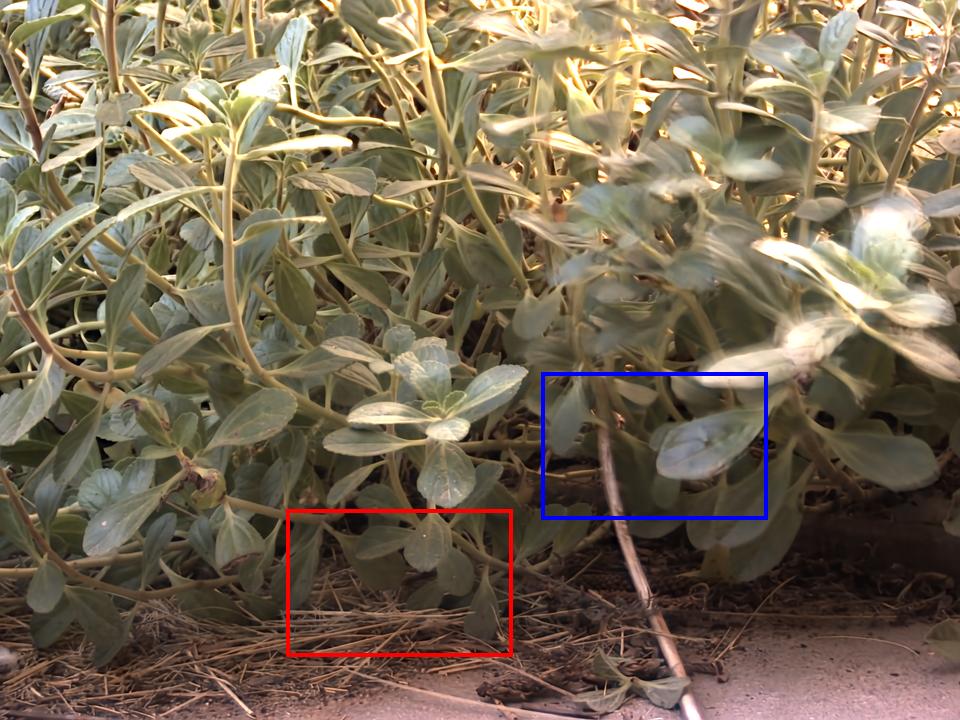}&
		\includegraphics[width = 1\columnwidth]{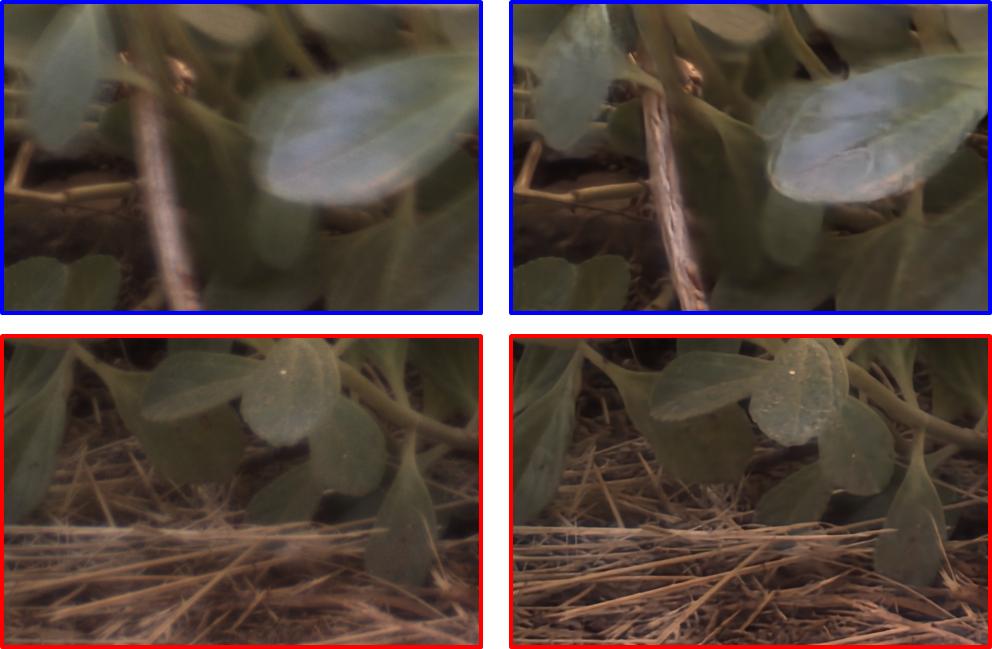}\\

	\end{tabular}
	\caption{\textbf{Outdoor experiment:} (left) full outdoor scenes with marks on magnified areas, and zoom-ins on (middle) intermediate image and (right) reconstruction results.}
	\label{fig:out_1}
\end{figure*}

\begin{figure*}[tb]
    \def\fullSz{1.7}
	\centering
	\begin{tabular}{c c}
		
		\includegraphics[width = \fullSz\columnwidth]{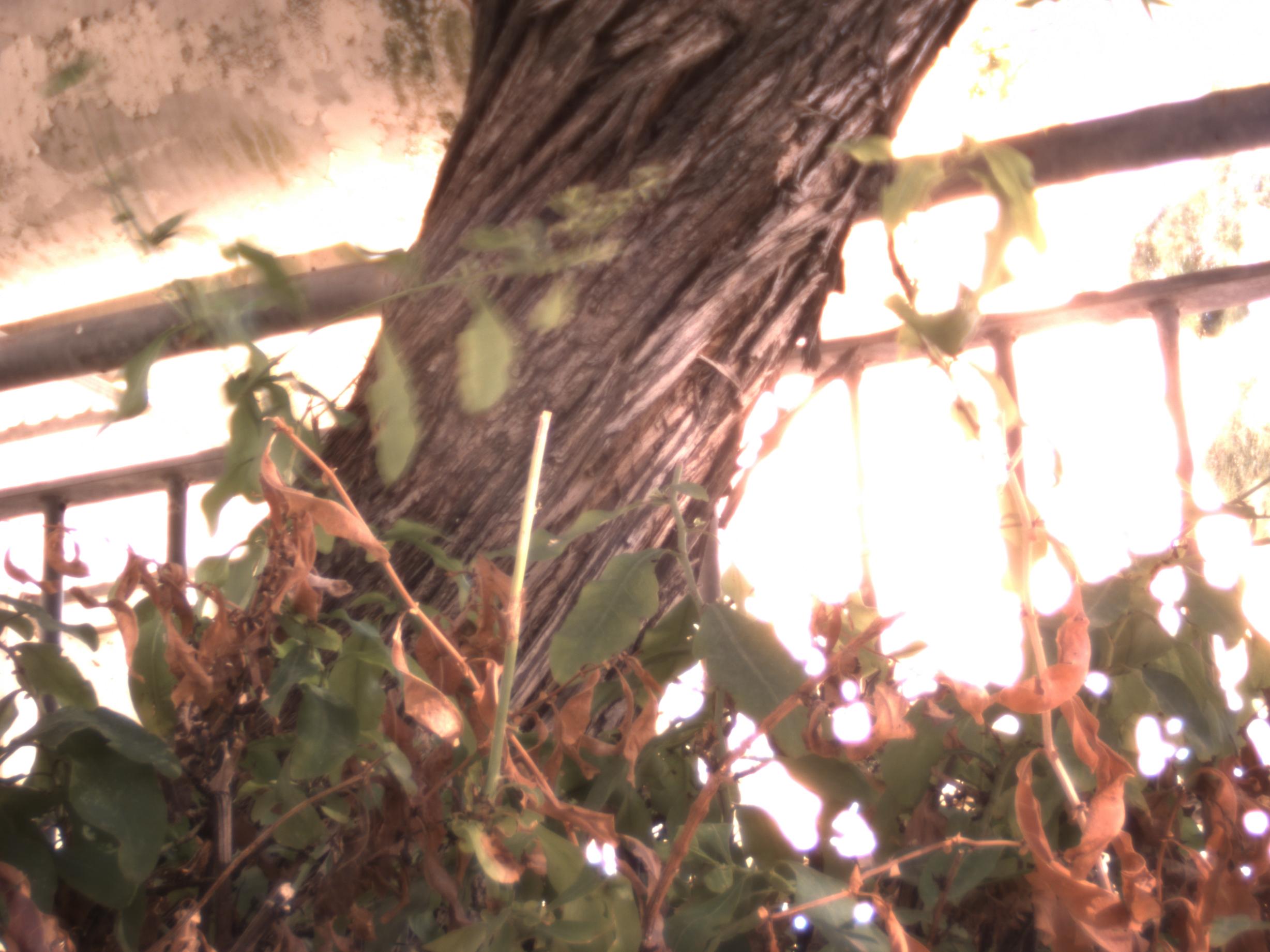}\\
		\includegraphics[width = \fullSz\columnwidth]{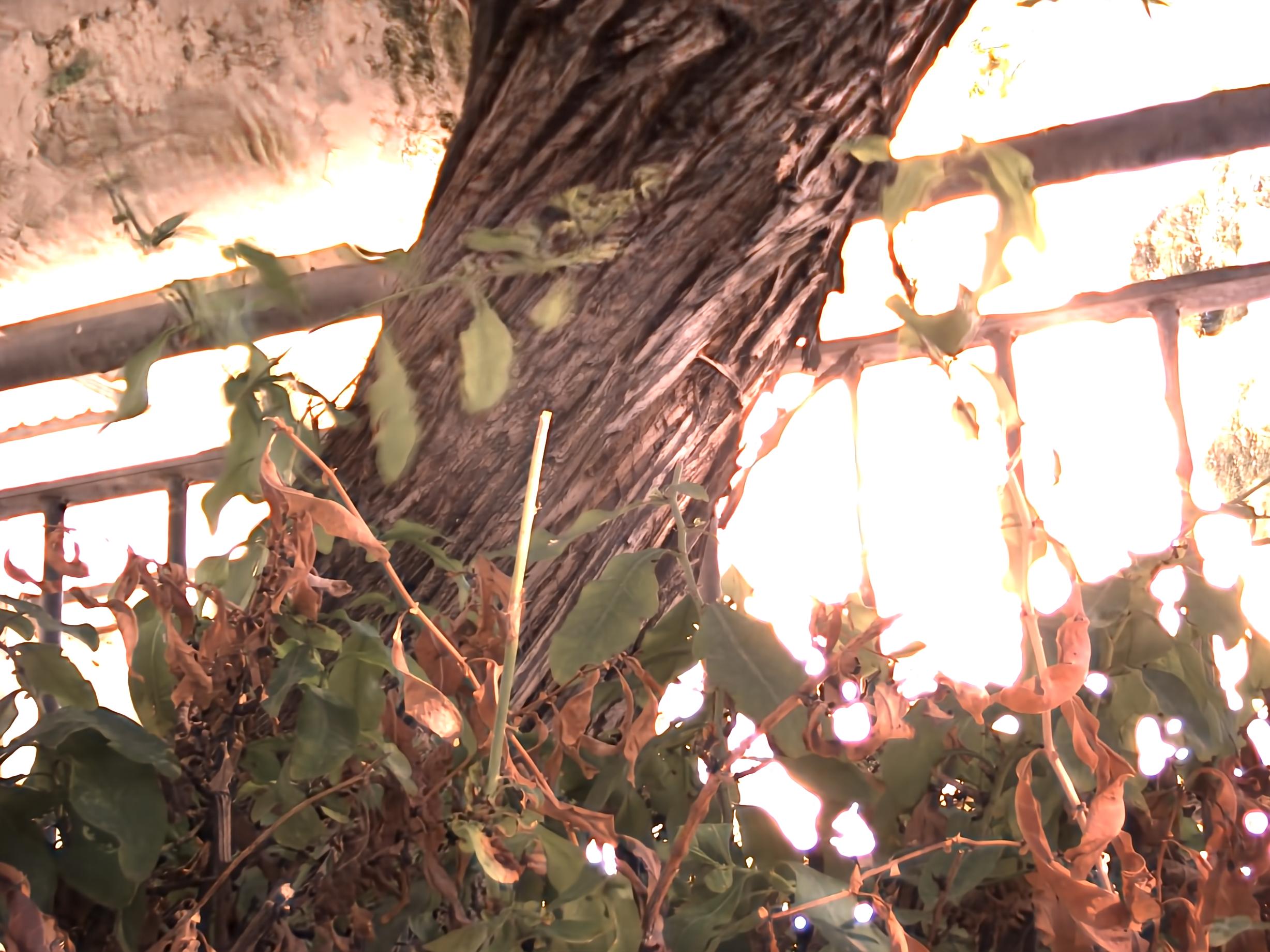}\\

	\end{tabular}
	\caption{\textbf{Outdoor experiment, full images:} (top) intermediate image and (bottom) reconstruction results.}
	\label{fig:out_2}
\end{figure*}

\begin{figure*}[tb]
    \def\fullSz{1.7}
	\centering
	\begin{tabular}{c c}
		
		\includegraphics[width = \fullSz\columnwidth]{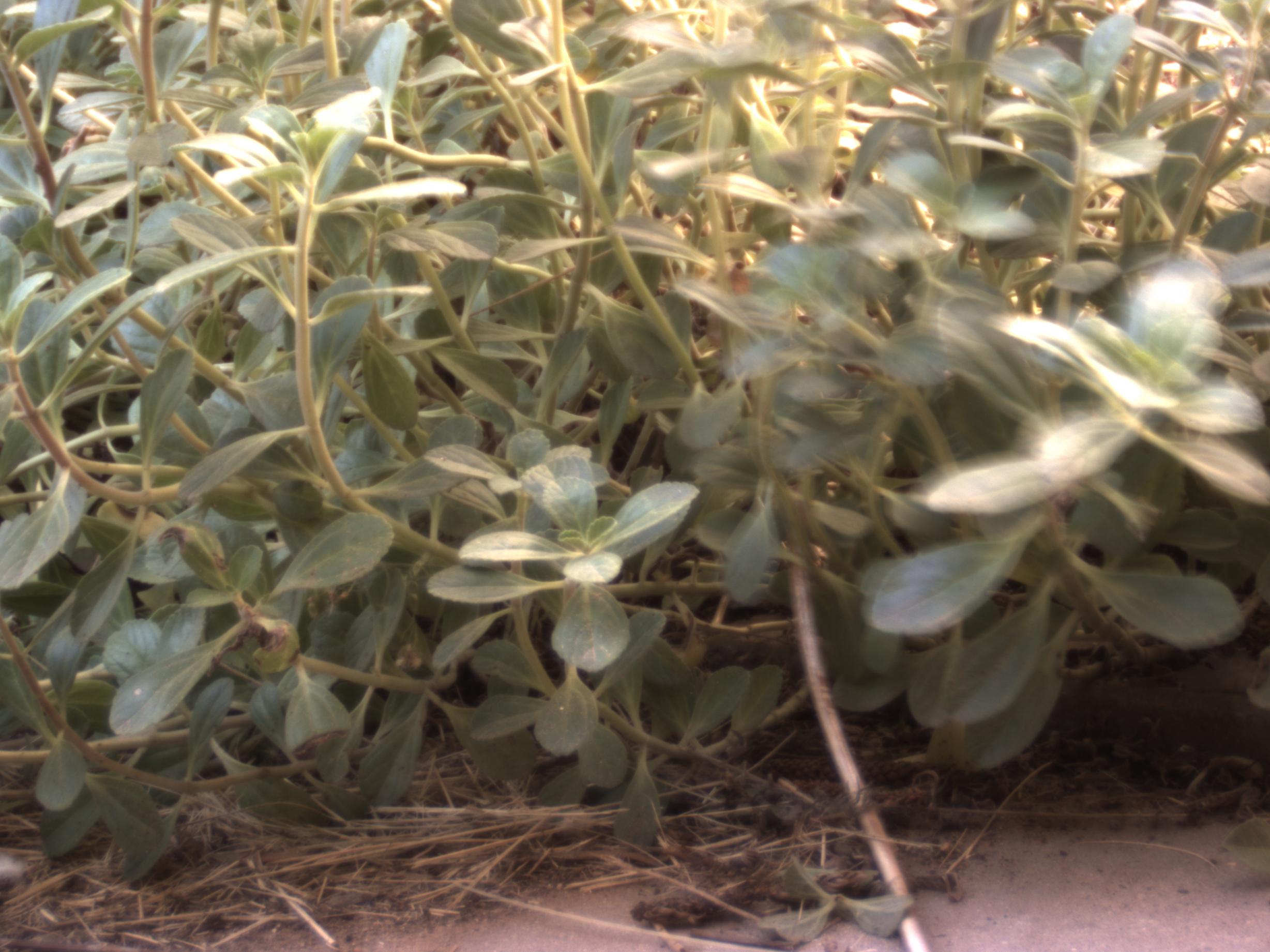}\\
		\includegraphics[width = \fullSz\columnwidth]{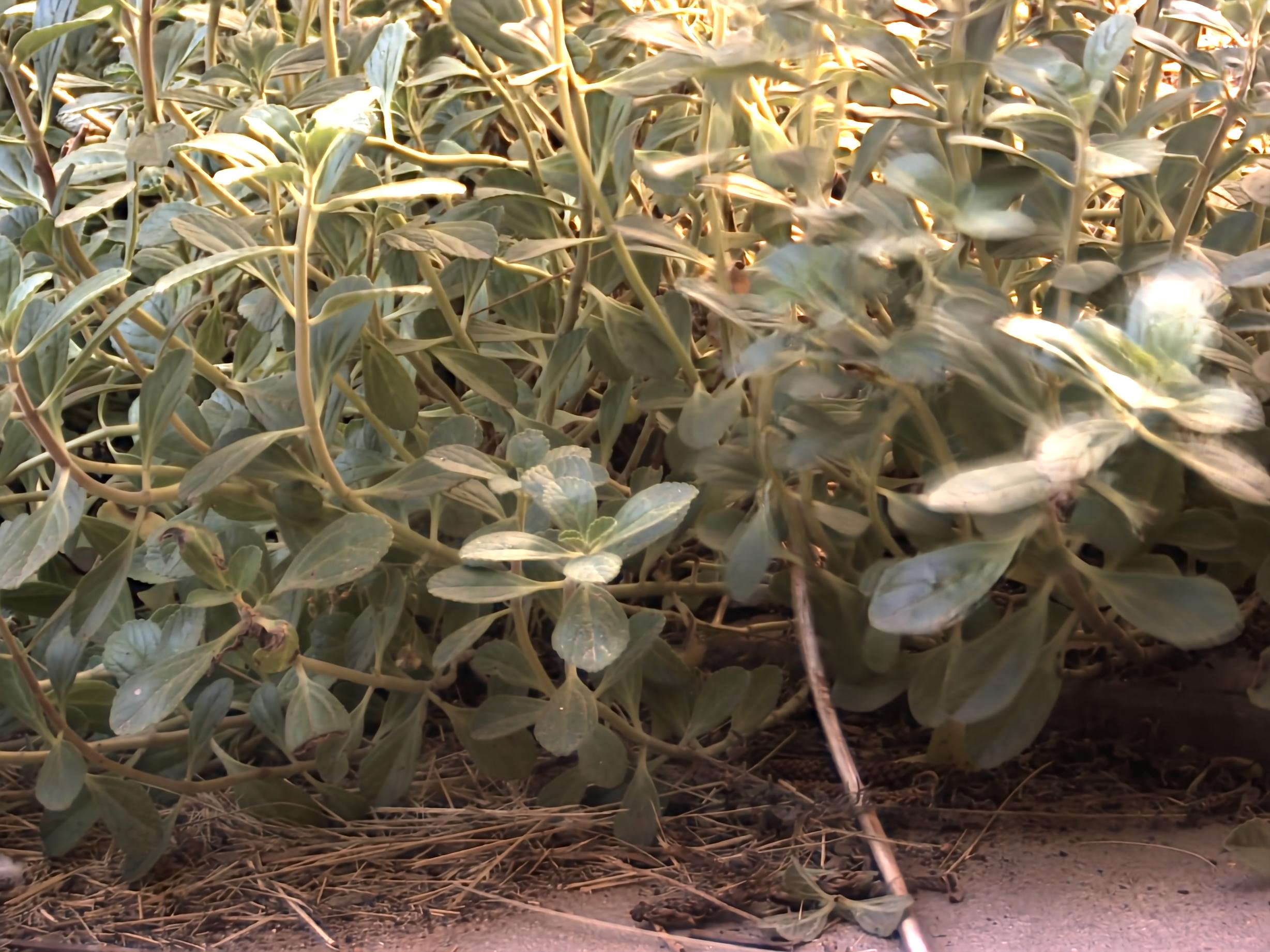}\\

	\end{tabular}
	\caption{\textbf{Outdoor experiment, full images:} (top) intermediate image and (bottom) reconstruction results.}
	\label{fig:out_3}
\end{figure*}

Following the results presented in the paper, additional outdoor scenes (of plants moving in the wind) are presented here (see Fig.~\ref{fig:out_1}-\ref{fig:out_3})\footnote{Note that the experimental camera sensor size is $4912\times3684$ pixels, while the test set images (of the GoPro dataset) are $1280\times720$ pixels, therefore the motion extent scale in pixel terms is different.}. As it is very complex to generate controlled motion 'in the wild', only our method is used, without a reference conventional camera. In such scenes, the motion occurs in every direction, and in various velocities. One can see (especially in the zoom-ins on Fig.~\ref{fig:out_1}) that our CNN is able to reconstruct the objects moving in different velocities and directions, while also deblurring the static parts (which also get some blur in the coding process). Notice that we do not re-train the reconstruction network, but rather use the same used in all other experiments (that was trained on the GoPro dataset with the color-coded motion simulation). Thus, in areas where the motion extent is quite large, the reconstruction performance decrease. The reason is that our model is trained with data that contains a limited velocities range; inside this range, the reconstruction results are good, while in areas with motion beyond this limit, the reconstruction ability is limited. This is an inherent trade-off in our method- as the velocities range in the dataset gets larger, the CNN can handle a longer motion extent. However, the cost is that some compromise is done in the reconstruction performance of the slower motion range.

\end{document}